\magnification=1000
\documentstyle{amsppt}
\baselineskip 20pt
\input amssym.def
\input amssym.tex

\NoBlackBoxes
\topmatter

\title  SPATIAL STRUCTURE IN LOW DIMENSIONS FOR\\
DIFFUSION LIMITED TWO-PARTICLE REACTIONS\endtitle

\rightheadtext{DIFFUSION LIMITED REACTIONS}
\leftheadtext{MAURY BRAMSON AND JOEL L. LEBOWITZ}

\endtopmatter

\vskip .3cm

\centerline{BY MAURY BRAMSON\footnote{Partially supported by NSF Grant
DMS-96-26196.} AND JOEL L. LEBOWITZ\footnote{Partially supported
by NSF Grant DMR-95-23266.  J.L.L. would also like to thank
DIMACS and its supporting agencies, the NSF under contract STC-91-19999 and
the N.J. Commission on Science and Technology.\hfil\hfil}}

\vskip .5truein
\centerline{\it University of Minnesota and Rutgers University}
\vskip .5truein
Consider the system of particles on ${\Bbb Z}^d$ where particles are
of two types, $A$ and $B$, and execute simple random walks in
continuous time.  Particles do not interact with their own type, but
when a type $A$ particle meets a type $B$ particle, both disappear.
Initially, particles are assumed to be distributed according to
homogeneous Poisson random fields, with equal intensities for the two
types.  This system serves as a model for the chemical reaction
$A+B\to inert$.  In [BrLe91a], the densities of the two types of
particles were shown to decay asymptotically like $1/t^{d/4}$ for
$d<4$ and $1/t$ for $d\geq 4$, as $t\to\infty$. This change in
behavior from low to high dimensions corresponds to a change in
spatial structure.  In $d<4$, particle types segregate, with only one
type present locally.  After suitable rescaling, the process converges
to a limit, with density given by a Gaussian process.  In $d>4$, both
particle types are, at large times, present locally in concentrations
not depending on the type, location or realization. In $d=4$, both
particle types are present locally, but with varying concentrations.
Here, we analyze this behavior in $d<4$; the behavior for $d\geq 4$
will be handled in a future work [BrLe99].
\vskip .5truein
\noindent{\bf AMS 1991 subject classification.} 60K35.
\vskip .3cm
\noindent{\bf Key words:} Diffusion limited reaction; annihilating
random walks; 
asymptotic densities; spatial structure

\newpage

\noindent{\bf Table of Contents}
\vskip .3cm
\noindent 1. Introduction\dotfill 2

\noindent 2. Summaries of the proofs of Theorems 1 and 2\dotfill 6

\noindent 3. Approximation of $\xi_t$ by earlier conditional
expectations\dotfill 10

\noindent 4. Approximation of $\xi_s*K_{t-s}$ by $\xi_0*N_t$\dotfill 17

\noindent 5. Approximation of $\xi_t$ by $\xi_0*N_t$\dotfill 23

\noindent 6. Upper bounds on $E[\xi^m_t(\cdot)]$\dotfill 27

\noindent 7. Approximation of $\xi^A_t$ and $\xi^B_t$\dotfill 33

\noindent 8. Convergence of $^T\hat\xi_t$ to $(2\lambda)^{1/2}(\Phi
*N_t)$\dotfill 43

\noindent 9. Local behavior of $\xi_t$\dotfill 53

\noindent References\dotfill 60
\vskip .5truein
\noindent{\bf 1. Introduction}
\vskip .3cm
Consider a system of particles of two types on ${\Bbb Z}^d$, $A$ and $B$,
which execute simple random walks in continuous time at rate $d$.  That is,
the motion of different particles is independent, and a particle at site
$x$ will jump to a given one of its $2d$ nearest neighbors at rate $1/2$.
Particles are assumed not to interact with their own type -- multiple $A$
particles or multiple $B$ particles can occupy a given site.  However, when
a particle meets a particle of the opposite type,
both disappear.  (When a particle simultaneously meets more than one
particle of the opposite type, it will cause only one of these particles to
disappear.)

We assume that particles are initially distributed according to
independent homogeneous Poisson random fields, with intensity
$\lambda$ for each type of particle. That is, the probability of there
being $j_1$ type $A$ particles and $j_2$ type $B$ particles at a given
site $x$ is $e^{-2\lambda}\lambda^{j_1+j_2}/j_1!j_2!$.  If there are
initially both $A$ and $B$ particles at a site $x$, they immediately
cancel each other out as much as possible.  We denote by $\xi^A_t(x)$
and $\xi^B_t(x)$ the number of $A$ particles, respectively, the number
of $B$ particles at site $x$, and by $\xi^A_t(E)$ and $\xi^B_t(E)$ the
number of such particles in a finite set $E\subset {\Bbb Z}^d$.  Also,
set $\xi^\#_t(E) = \xi^A_t(E) +
\xi^B_t(E)$, for the total number of particles in $E$. We associate with
each $A$ particle the value $-1$ and with each $B$ particle
the value 1, and denote by $\xi_t(x)$ the signed number of particles at
$x\in {\Bbb Z}^d$, i.e., $\xi_t(x) = \xi^B_t(x) -\xi^A_t(x)$.  Similarly,
$\xi_t(E) = \xi^B_t(E)-\xi^A_t(E)$.  We denote by $\xi_t$, $\xi_t\in ({\Bbb
Z}^2_+)^{{\Bbb Z}^d}$, the random state of the system at time $t$, where
${\Bbb Z}_+$ designates the nonnegative integers; the first
coordinate at each site corresponds to the number of $A$ particles there,
and the second coordinate to the number of $B$ particles.  We write
$\xi_{0-}$ for the initial state before $A$ and
$B$ particles originally at the same site have annihilated one another.

The above two-particle annihilating random walk can serve as a model
for the irreversible chemical reaction $A+B\to inert$, where both
particle types are mobile.  $A$ and $B$ can also represent matter and
antimatter.  There has been substantial interest in this model in the
physics literature over the last two decades following papers by
Ovchinnikov and Zeldovich [OvZe78], Toussaint and Wilczek [ToWi83],
and Kang and Redner (KaRe85); see [BrLe91a] for a more complete set of
references, and [LeCa95] and [LeCa97] for more recent work.

Let $\rho_A(t)$ and $\rho_B(t)$ denote the densities of $A$ and $B$
particles at the origin, i.e.,
$$
\eqalign{\rho_A(t) &= E[\#A\ \text{particles at}\ 0\ \text{at time}\ t],\cr
\rho_B(t) &= E[\#B\ \text{particles at}\ 0\ \text{at time}\ t].\cr}
\leqno(1.1)
$$
(In the paper, $E$ will be used for both expectations and sets.)
Since $\xi_{0-}$ is translation invariant, its densities do not depend on the
site $x$.  The difference $\rho_B(t)-\rho_A(t)$ remains constant for all
$t$, because particles annihilate in pairs.  Since $\rho_A(0-) =\rho_B(0-)
=\lambda$, one has $\rho_A(t) =\rho_B(t)$ for all $t$, which we will denote
by $\rho (t)$.  In [BrLe91a], it was shown that
$$
\eqalign{c_d\lambda^{1/2}/t^{d/4} &\leq \rho(t)\leq
c'_d\lambda^{1/2}/t^{d/4}\hskip .5in\text{for}\ d<4,\cr
c_4(\lambda^{1/2}\vee 1)/t &\leq \rho(t)\leq c'_4(\lambda^{1/2}\vee 1)/t
\quad\quad\hskip .15cm \text{for}\ d=4,\cr
c_d/t &\leq \rho(t)\leq c'_d/t\hskip .9in\text{for}\ d>4,\cr}
\leqno(1.2)
$$
for large $t$ and appropriate positive constants $c_d$ and $c'_d$;
here, $a\vee b = \max (a,b)$. (Bounds were also derived when the
initial densities are 
unequal.)  The asymptotic power laws in (1.2) were previously
obtained, in [ToWi83] and [KaRe85], 
using heuristic arguments.

The asymptotics of $\rho(t)$ in (1.2) are tied to the spatial structure of
$\xi_t$, which also depends on $d$.  The slow rate of decay for $d<4$
corresponds to the presence locally, at large times, of only one type of
particle, typically.  This behavior is a consequence of the random
fluctuations in the numbers of $A$ and $B$ particles locally in the initial
state, and the tendency for particles of the local minority type to be
annihilated before they can be replenished by the arrival of particles from
outside the region.
In particular, the random walks executed by these particles are diffusive,
which imposes limitations on the rate of mixing of particles.  {}For
$d>4$, $A$ and $B$ particles remain sufficiently mixed so that the behavior
is different, and mean field reasoning gives the correct asymptotics.
Namely, assuming that $d\rho_A(t)/dt$ is proportional to
$-\rho_A(t)\rho_B(t) = -(\rho_A(t))^2$, then $\rho (t) = \rho_A(t)$ will
decay like a multiple of $1/t$, which is the right answer.  In this
latter setting, the limiting density does not depend on the initial densities.
The dimension $d=4$ is a hybrid of the previous two cases, with both
mechanisms playing a role.

It is the purpose of this paper to analyze the spatial structure of $\xi_t$
in $d<4$.  The behavior of $\xi_t$ in $d\geq 4$ will be covered in
[BrLe99].  Our main results are Theorems 1 and 2.  Theorem 1 gives the
macroscopic limiting behavior of the process.
It says, in essence, that $\xi_t$, under diffusive scaling, converges to a
limit which is the convolution of white noise with the normal kernel.
Regions where this convolution is positive correspond to regions where only
$B$ particles are present, with negative regions corresponding to the
presence of $A$ particles.

By white noise, we mean the stochastic process $\Phi$ whose domain is the
set of finite rectangular solids $D\subset {\Bbb R}^d$, with sides
parallel to the coordinate axes, such that any linear combination
$\sum^n_{j=1} a_j\Phi (D_j)$ is normally
distributed with mean $0$, and, for any $D_1$ and $D_2$,
$$
E[\Phi (D_1)\Phi (D_2)] = \vert D_1\cap D_2\vert .
\leqno(1.3)
$$
Loosely speaking, $<\Phi (x)\Phi (y)> = \delta (x-y)$ for $x,y\in {\Bbb
R}^d$, i.e., $\Phi (x)$ is a Gaussian field with a $\delta$-function
covariance, in physics terminology. (One can alternatively define
$\Phi$ as the linear functional 
on the Schwartz space of rapidly decreasing functions $f:{\Bbb R}^d\to
{\Bbb R}$, where $\Phi (f)$ is normally distributed with mean $0$, and
$E[\Phi (f_1)\Phi (f_2)] = \int_{{\Bbb R}^d} f_1(x)f_2(x)dx$.)  These rules
specify a generalized Gaussian
random field on ${\Bbb R}^d$.  We will assume that $\Phi (D)$ is, for
each realization, continuous in the coordinates of $D$; a version of the
process exists for which this holds.  White noise is closely connected with
Brownian sheet, and the above definition is motivated by this relationship.
(More detail will be given in Section 8.)

We will write $N_t(\cdot)$ for the density of a normal random variable with
mean 0 and covariance matrix $tI$, i.e.,
$$
N_t(x) = (2\pi t)^{-d/2}e^{-\vert x\vert^2/2t}\quad \text{for}\ x\in {\Bbb
R}^d ,
\leqno(1.4)
$$
where $\vert\cdot\vert$ is the Euclidean norm.  We write $N_t(E) =
\int_EN_t(x)dx$ for measurable $E\subset {\Bbb R}^d$.  Let $\Phi
*N_t$ denote the convolution of $\Phi$ with $N_t$, that is,
$$
(\Phi *N_t)(D) =\int_{{\Bbb R}^d}\Phi (D-x)N_t(x)dx ,
\leqno(1.5)
$$
where $D+y$ designates $D$ translated by $y$.  Since $\Phi *N_t$ is
the average of translates of a generalized Gaussian random field, 
$\Phi *N_t$ is also Gaussian.  Because of the smoothness of $N_t$, $\Phi
*N_t$ will have a density $(\Phi *N_t)(x)$;  it is Gaussian with variance
$(4\pi t)^{-d/2}$ at each point.  We will write $(\Phi
*N_t)^-(D)$ and $(\Phi *N_t)^+(D)$ for the integrals of the negative and
positive parts of $(\Phi *N_t)(x)$ over $D$.

To state Theorem 1, we need to normalize $\xi_t$.  The notation
$^T\hat\xi_t$ (respectively, $^T\hat\xi^A_t$ and $^T\hat\xi^B_t$) will
denote $\xi_t$ (respectively, $\xi^A_t$ and $\xi^B_t$) after scaling time
by $T$, space by $T^{1/2}$ in each direction, and the weight of individual
particles by $T^{d/4}$.  That is,
$$
^T\hat\xi_t(E) = \xi_{Tt}(T^{1/2}E)/T^{d/4} ,
\leqno(1.6)
$$
where $E\subset {\Bbb Z}^d_{T^{1/2}}$, which is ${\Bbb Z}^d$ scaled by
$T^{1/2}$ in each direction. The factor $T^{d/4}$ is mandated by the first
line of (1.2). {}For $E\subset {\Bbb R}^d$, we set
$^T\hat\xi_t(E)=\!^T\hat\xi_t(E\cap {\Bbb Z}^d_{T^{1/2}})$.

Theorem 1 gives the limiting macroscopic behavior of $\xi_t$.  It states
that $(^T\hat\xi^A_t(D)$,\hfill\break $^T\hat\xi^B_t(D)$) converges weakly,
on ${\Bbb R}^2$, to $(2\lambda)^{1/2}((\Phi *N_t)^-(D)$, $(\Phi
*N_t)^+(D)$), where $\lambda$ is the initial density for the $A$ and $B$
particles.  Here
and elsewhere in the paper, unless stated otherwise, rectangular solids $D$
will be of the form $\prod^d_{j=1} (y_j,x_j]$.  They will be called
``rectangles'' for short.
\vskip .3cm
\noindent{\bf Theorem 1.} {\it Let $\Phi$ and $N_t$ be defined as above,
with $t>0$, and let $D\subset {\Bbb R}^d$ be any finite rectangle.  Then,
for $d<4$,
$$
(^T\hat\xi^A_t(D),\! ^T\hat\xi^B_t (D))\Rightarrow ((2\lambda)^{1/2}(\Phi
*N_t)^-(D),(2\lambda)^{1/2}(\Phi *N_t)^+(D))
\leqno(1.7)
$$
as $T\to\infty$.}
\vskip .3cm
A more general version of Theorem 1, Theorem 4, is demonstrated in Section
8.  There, it is shown that $^T\hat\xi^A_t(D)$ and $^T\hat\xi^B_t(D)$ are
uniformly well approximated by $(2\lambda)^{1/2}(\Phi *N_t)^-(D)$ and
$(2\lambda)^{1/2}(\Phi *N_t)^+(D)$ over all $t\in [1/M,M]$, $M>1$, and
all $D$ in a fixed cube.  The rectangles $D$ in both theorems can easily be
generalized, although one needs the cardinality of the collection of sets
employed in Theorem 4 not to be too large, in order to avoid the piling up
of small probability events where either $^T\hat\xi^A_t(E)$ or
$^T\hat\xi^B_t(E),\ 
E\subset {\Bbb R}^d$, is badly behaved.

In the course of demonstrating Theorem 4, one obtains estimates that
give the asymptotic behavior of $\rho (t)$.  It is shown at the end of
Section 6, that
$$
\lim_{t\to\infty} t^{d/4} \rho (t) = (\lambda /\pi)^{1/2}(4\pi)^{-d/4}
\leqno(1.8)
$$
when $d<4$.  This strengthens the first line of (1.2).  These limits were
given in [ToWi83]. We also note it follows immediately from (1.7), that
$$
^T\hat\xi_t(D)\Rightarrow (2\lambda)^{1/2}(\Phi *N_t)(D)\quad
\text{as}\ T\to\infty . 
\leqno(1.9)
$$
Conversely, (1.7) will follow from (1.9), if one also knows that the
particle types segregate.

In order to understand the spatial structure of $\xi_t$, one also needs to
know its behavior on the microscopic scale.  By (1.8), the correct spatial
scaling will be $t^{1/4}$ in each direction, and so we set
$$
\check\xi_t (E) =\xi_t(t^{1/4}E),
\leqno(1.10)
$$
with $E\subset {\Bbb Z}^d_{T^{1/4}}$, and $\check\xi^A_t$ and
$\check\xi^B_t$ being defined analogously. 
One can also guess at the limiting spatial structure of $\check\xi_t$.
Particles will, at large times, only be annihilated occasionally.  This
allows particles the time to mix locally, without interaction.  They should
therefore be independently distributed locally, as $t\to\infty$.  This
produces a Poisson random field, after conditioning on the intensity at
$x=0$, $(2\lambda)^{1/2}(\Phi *N_1)(0)$, with the type of particle
present depending on the sign of $(\Phi *N_1)(0)$.  The random variable
$(\Phi *N_1)(0)$ is normally distributed, with mean $0$ and variance
$(4\pi)^{-d/2}$.

With this behavior in mind, we denote by ${\Cal P}_c$ the Poisson random
field of $A$ particles with intensity $c^-$ if $c\leq 0$, and the Poisson
random field of $B$ particles with intensity $c^+$ if $c>0$.  We interpret
${\Cal P}_c$ as a vector, with the first component corresponding to $A$
particles and the second coordinate to $B$ particles.  Also, for $F$ a
probability distribution function on ${\Bbb R}$, set
$$
{\Cal P}_F =\int {\Cal P}_cdF(c),
\leqno(1.11)
$$
i.e., ${\Cal P}_F$ is the convex combination of homogeneous Poisson random
fields with intensities weighted according to $F$.

In Theorem 2, $\Rightarrow$ denotes weak convergence with respect to
the Borel measures 
on ${\Bbb R}^d$ having finite mass on all compact subsets.  (The space of
measures is assumed to be equipped with the topology of vague convergence
on ${\Bbb R}^d$, i.e., integration is against $f\in C^+_c({\Bbb R}^d)$, where
$C^+_c({\Bbb R}^d)$ is the set of nonnegative continuous functions on
${\Bbb R}^d$ with 
compact support.)
\vskip .3cm
\noindent{\bf Theorem 2.} {\it {}For $d<4$,
$$
(\check\xi^A_t,\check\xi^B_t)\Rightarrow {\Cal P}_F\quad \text{as}\
t\to\infty ,
\leqno(1.12)
$$
where $F$ is the distribution of a normal random variable with mean $0$ and
variance $2\lambda (4\pi)^{-d/2}$}.
\vskip .3cm
In this paper, we will demonstrate Theorem 2 and Theorem 4, the more
general version of Theorem 1 mentioned earlier.  Versions of these results
were summarized in [BrLe91b].  An outline of the main steps leading to these
results will be given in the next section.

Certain features of the asymptotic behavior of the model considered here are
shared by two simpler systems, coalescing random walk and annihilating random
walk.  Both cases consist of particles on ${{\Bbb Z}^d}$, of a single
type, which execute independent simple random walks.  In the first case, when
two particles meet, they coalesce into a single particle, whereas, in the
second case, they annihilate one another.  The two models can be interpreted
in terms of the chemical reactions $A+A\to A$ and $A+A\to inert$,
respectively.  {}For both models, it is natural to assume that all sites are
initially occupied.

The asymptotic behavior of both models is known.  {}For the coalescing random
walk, the density is asymptotically $1/\sqrt{\pi t}$ in $d=1$, $(\log t)/\pi
t$ in $d=2$, and $1/\gamma_d t$, for appropriate $\gamma_d$, in $d\geq 3$
([BrGr80]).  The asymptotic density of annihilating random walk is, in each
case, one half as great ([Ar81]).  Scaling, so as to compensate for
the decrease in 
density, produces analogs of Theorem 2.  {}For $d\geq 2$, the limiting measure
is again Poisson ([Ar81]), but, in $d=1$, it is not ([Ar79]).

Recent work [KeVa98] considers a generalization of the above coalescing
random walk.  There, coalescence is not automatic when two particles meet,
and occurs with a probability that depends on the number of particles
present at a site.  Results are obtained for $d\geq 6$.
\vskip .5truein
\noindent{\bf 2. Summaries of the Proofs of Theorems 1 and 2}
\vskip .3cm
In this section, we summarize the proofs of Theorems 1 and 2.  We
present the main 
steps, providing motivation in each case.
Proofs of the individual steps are given in the remaining sections.

Rather than directly show Theorem 1, our approach will be to first show
Theorem 3, which is given below.  This result is a more concrete analog of
Theorem 1, which compares $\xi_t$ with $\xi_0*N_t$ along individual sample
paths, instead of showing weak convergence of $^T\hat\xi_t$ to
$(2\lambda)^{1/2}(\Phi *N_t)$.  The
result also shows that the distribution of particles for
$\xi_t$, at large times, is essentially deterministic if $\xi_0$ is known.
Error bounds for the corresponding estimates, in (2.2) and (2.3), are
given in terms of powers 
of $T$; one has $T^{d/4-1/9,000}$ inside of $P(\cdot)$, and $T^{-1/9,000}$
on the right side of the inequalities.  (The exact values of the small
constants are not important, but show that convergence occurs at least at a
polynomial rate.)

Here and later on in the paper, ${\Cal D}_R$ will denote the set of all
rectangles contained in $D_R = \prod^d_{j=1}(-R/2,R/2]$, the semiclosed
cube of length $R$
centered at the origin.  (Recall that rectangles are always assumed to be
of the form $\prod^d_{j=1}(y_j,x_j]$.) Since the particles in $\xi_t$
reside on ${\Bbb Z}^d$, we will implicitly interpret such rectangles as
subsets of ${\Bbb Z}^d$, when there is no risk of ambiguity; $\vert D\vert$
will denote the number of sites in $D\cap {\Bbb Z}^d$.  Since $\xi_0$ is
discrete, the convolution $\xi_0*N_t$ will be defined by summing over
${\Bbb Z}^d$, 
i.e.,
$$
(\xi_0*N_t)(x) =\sum_{y\in {\Bbb Z}^d}\xi_0(x-y)N_t(y)\quad \text{for}\ x\in
{\Bbb Z}^d .
\leqno(2.1)
$$
This contrasts with the convolution in (1.5), where one integrates over ${\Bbb
R}^d$.  Throughout the paper, the initial density $\lambda$ of $A$ and $B$
particles will be considered to be fixed, with $\lambda > 0$.  As always,
$f(x)^+ = f(x)\vee 0$ and $f(x)^- = -f(x)\vee 0$.
\vskip .3cm
\noindent {\bf Theorem 3.} {\it {}For $d<4$ and given $M>1$,
$$
\align
P\bigl( \sup_{t\in [T/M,MT]}\sup_{D\in {\Cal D}_{MT^{1/2}}}\bigr.
\bigm|\xi^A_t(D)-\sum_{x\in D}(\xi_0 &*N_t)(x)^-\bigm|
\bigl.\geq T^{d/4-1/9,000}\bigr) \\
&\leq T^{-1/9,000}\tag2.2
\endalign
$$
and
$$
\align
P\bigl( \sup_{t\in [T/M,MT]}\sup_{D\in {\Cal D}_{MT^{1/2}}}\bigr.
\bigm|\xi^B_t(D)-\sum_{x\in D}(\xi_0&*N_t)(x)^+\bigm|
\bigl.\geq T^{d/4-1/9,000}\bigr)\\
&\leq T^{-1/9,000}\tag2.3
\endalign
$$
hold for sufficiently large $T$.}
\vskip .3cm
\noindent In Section 8, we will derive Theorem 4, and hence Theorem 1, from
Theorem 3.  The basic procedure will be to show that $\xi_0$, when scaled as
in (1.6), converges weakly to white noise $\Phi$, and then to use Theorem 3
and the continuity of $*$ to obtain (1.7).

In order to demonstrate Theorem 3, we first demonstrate the following
analog for $\xi_t(D)$, with $t\in [T/M,T]$ and $D\in {\Cal D}_{T^{1/2}}$.
(Rescaling $T$ will allow us to extend $[T/M,T]$ to $[T/M,MT]$, and ${\Cal
D}_{T^{1/2}}$ to ${\Cal D}_{MT^{1/2}}$, when convenient later on.)
\vskip .3cm
\noindent{\bf Proposition 2.1.} {\it {}For $d<4$ and $M>1$,
$$
P\Bigl( \sup_{t\in [T/M,T]} \sup_{D\in {\Cal D}_{T^{1/2}}}
\vert\xi_t(D)-(\xi_0*N_t)(D)\vert \geq T^{d/4-1/80}\Bigr)\leq
\exp\{-T^{1/42}\}
\leqno(2.4)
$$
holds for sufficiently large $T$.}
\vskip .3cm
\noindent This bound is considerably weaker than those in (2.2) and (2.3),
in that it only measures the imbalance between the numbers of $A$ and $B$
particles locally, rather than their absolute numbers.  (The exponential
bound on the right side of the inequality is, of course, stronger.)
Proposition 2.1 will be shown in Section 5. 

In order to derive (2.2) and (2.3), one also needs to know that, locally,
the number of particles of the ``minority type'' is negligible.  {}For this,
it will be sufficient to show that the expected number of such particles is
small at specific times that are not too far apart.  One will then be able
to fill in the behavior at intermediate times, and apply Markov's inequality
to the expectation.  {}For these purposes, we will employ Proposition 2.2.
Together with Proposition 2.1, it will be used to demonstrate Theorem 3, in
Section 7.  Throughout the paper, we will employ the notation
$$
\xi^m_t (E) = \xi^A_t(E)\wedge\xi^B_t(E) ,
\leqno(2.5)
$$
for the number of particles of the minority type in $E$, where $E\subset
{\Bbb Z}^d$.  Here and later on, we use $C_1,C_2,\dots$ for positive
constants whose exact values do 
not concern us.
\vskip .3cm
\noindent{\bf Proposition 2.2.} {\it Let $d<4,\ M>1$, and choose $R_T$ so
that $R_T =\delta_1(T)T^{1/2}$, where $\delta_1(T)\geq T^{-d/48}$ and
$\delta_1(T)\to 0$ as $T\to\infty$. {}For sufficiently large $T$, there exist
$K$ and $t_1<t_2<...<t_K$, with $t_k-t_{k-1}\leq \delta_1(T)T$,
$[t_1,t_K]\supset [T/M,T]$, $[t_1,t_{K-1}]\subset [T/2M,T]$, and $C_1$
(depending on 
$\lambda$ and $M$), so that}
$$
E[\xi^m_{t_k}(D_{R_T})]\leq C_1\delta_1(T)T^{-d/4}(R_T)^d\quad \text{for}\
k=1,\dots ,K .
\leqno(2.6)
$$
The cube $D_{R_T}$ contains approximately $(R_T)^d$ sites, and so, by
(1.2), will contain of order of magnitude $T^{-d/4}(R_T)^d$ particles.
Inequality (2.6) implies that $\xi^m_{t_k}(D_{R_T})$ is, on the average,
much smaller than this, for $R_T$ chosen as above.

Proposition 2.2 will follow from machinery introduced in
[BrLe91a].  The basic idea is that, if $E[\xi^m_t(D_{R_T})]$ is large for too
long a stretch of time, enough annihilation will occur to contradict the
bounds on $\rho (t)$ in (1.2).  The bound (2.6) will also enable us to derive
precise asymptotics on $\rho (t)$, for $d<4$, in Section 6.  These are an
improvement of the upper and lower bounds on $\rho (t)$ in (1.2).

The proof of Proposition 2.1 employs two main results.  To state these, we
need to introduce some additional notation.  Let $K_t(x)$ denote the
probability that a simple rate-$d$ continuous time random walk in ${\Bbb
Z}^d$, starting at the origin, is at $x$ at time $t$.  Denote by
$_s\eta_t$, $s>0$, the stochastic process in $t$ that is identical
to $\xi_t$ up until time $s$, and for which, starting at time $s$, the existing
particles continue to execute independent simple random walks as before,
but without annihilation. We let $\eta_t$ denote the process of independent
random walks with initial state $\eta_0 =\xi_{0-}$, the initial
configuration of $\xi_t$ before $A$ and $B$ particles at the same site have
annihilated one another.  We also set $_0\eta_t = \eta_t$.  The
processes $\xi_t$ and $_s\eta_t$, $s\geq 0$, can all be constructed on the
same probability space, so that they are all adapted to the same family of
increasing $\sigma$-algebras ${\Cal F}_t$, $t\geq 0$.  To do this, we specify
an arbitrary ranking of all of the particles initially in the system, with
the rule that when
more than two particles of opposite types meet, the highest ranked $A$ and
$B$ particles are the ones which are annihilated.  Then,
${\Cal F}_t$ is defined to be the $\sigma$-algebra generated by the
labeled random walks 
corresponding to $\eta_r$, for $r\leq t$.  Later on, we will also employ the
$\sigma$-algebras ${\Cal F}^\xi_t\subset {\Cal F}_t$, where ${\Cal
F}^\xi_t$ is generated by $\xi_r$, for $r\leq t$.

It is easy to see that for any finite set $E\subset {\Bbb Z}^d$ and $s\leq
t$,
$$
E[\xi_t(E)\mid {\Cal F}_s] = E[_s\eta_t (E)\mid {\Cal F}_s]
= (\xi_s*K_{t-s})(E),
\leqno(2.7)
$$
where $*$ is defined as in (2.1).  (In (2.7), the outer $E$ stands for
expectation, whereas the inner $E$ is a subset of ${\Bbb Z}^d$.) The following
proposition says that,
for large $t$, $(\xi_{t^{1/4}}*K_{t-t^{1/4}})(E)$ is a good
approximation of $\xi_t(E)$.  The reasons for this are basically that (1)
there are few enough particles locally at time $t^{1/4}$, and therefore
little enough randomness, so that, up to an error which is of smaller order
than $\vert E\vert /t^{d/4}$,
$_{t^{1/4}}\eta_t (E)$ can be replaced by its conditional expectation, and
(2) the annihilation of pairs of $A$ and $B$ particles over $[t^{1/4},t]$
reduces this randomness still further, and so $\xi_t(E)$ can also be replaced
by the same conditional expectation.  Using (2.7), one can then substitute
$(\xi_{t^{1/4}}* K_{t-t^{1/4}})(E)$ for this conditional expectation.
Since, for our applications, $\epsilon \vert E\vert$ will not be much less
than $t^{d/2}$ and $\epsilon$ will not be too small, the bound on the
right side of (2.8) will be quite small. This result is demonstrated
in Section 3. 
Here and later on, we use the abbreviation $v_t(\epsilon) =\epsilon\wedge
t^{3/16}$.
\vskip .3cm
\noindent{\bf Proposition 2.3.} {\it {}For $d<4$ and sufficiently large $t$,
$$
\eqalign{P(\vert \xi_t(E)-(\xi_{t^{1/4}}&*K_{t-t^{1/4}})(E)\vert\geq
\epsilon\vert E\vert t^{-d/4})\cr
&\leq 6\exp\{-((\epsilon v_t(\epsilon)\vert E\vert
t^{-9d/20})\wedge t^{1/8})\}\cr}
\leqno(2.8)
$$
holds for all $\epsilon$ and $E\subset {\Bbb Z}^d$, with $\vert E\vert\leq
t^d$.}
\vskip .3cm
The other estimate needed for Proposition 2.1 is a comparison of
$\xi_0*N_t$ with $\xi_{t^{1/4}}*K_{t-t^{1/4}}$.  These two quantities will
typically be close since particles do not wander far by time $t^{1/4}$, and
since $K_{t-t^{1/4}}$ can be approximated by $N_t$, by using an appropriate
version of the local central limit theorem.  The desired result,
Proposition 2.4, is demonstrated in Section 4.
\vskip .3cm
\noindent{\bf Proposition 2.4.}  {\it {}For any $d$, let $t$ be sufficiently
large and $s\leq t^{1/4}$.  Then, for all $\epsilon\geq 0$,}
$$
P(\vert (\xi_0*N_t)(0)-(\xi_s*K_{t-s})(0)\vert\geq\epsilon t^{-d/4})\leq
4\exp\{-(\epsilon^2\wedge 1)t^{1/4}\} .
\leqno(2.9)
$$

\noindent Note that since $\xi_0$ is translation invariant, the analog of
(2.9) holds at all $x\in {\Bbb Z}^d$.  Corresponding bounds therefore hold
for finite $E\subset {\Bbb Z}^d$, when factors of $\vert E\vert$ are inserted
for both inequalities.

Propositions 2.3 and 2.4 are employed in Section 5 to show Proposition 2.1.
After combining the two results, we still need to show that the bounds hold
simultaneously over all $t\in [T/M,T]$ and $D\in {\Cal
D}_{T^{1/2}}$.  Since the probabilities of the exceptional sets in (2.8)
and (2.9) will be exponentially small, they can be summed over a fine lattice
of elements in $[T/M,T]\times {\Cal D}_{T^{1/2}}$, while maintaining such
bounds.  One can then ``fill in'' the events corresponding to the values
between the lattice points, to produce the desired uniformity over
$[T/M,T]\times {\Cal D}_{T^{1/2}}$, as in (2.4) of Proposition 2.1.

Retracing the steps taken so far in this section, we have just discussed
Propositions 2.3 and 2.4, which are the main steps in showing Proposition
2.1.  As we discussed earlier, Proposition 2.1, together with
Proposition 2.2, is 
employed to derive Theorem 3.  By rescaling the process $\xi_t$ in Theorem
3, one then obtains Theorem 1.

We still need to discuss Theorem 2. 
The additional work required to demonstrate the theorem from the previous
results is done in Section 9.  The basic reasoning
is that, over intervals of time ending in $t$ that are short relative to
$t$, relatively little annihilation occurs, because
of the smooth decrease in the density in (1.8).  {}For
the space scale of interest to us for $\check\xi_t$, namely $t^{1/4}$, this is
long enough for the local particles (typically of only one type) to
thoroughly mix.  Such a mixed state will, for a typical realization, be
nearly Poisson for large $t$.  Its intensity near $0$ will be given by
$t^{d/4}\vert (\xi_0*N_t)(0)\vert$.  Laplace transforms are
employed to carry out the proof.

As mentioned earlier, the behavior of $\xi_t$, for $d\geq 4$, will be
handled in the future paper [BrLe99].  The behavior will be different than
that considered here, for $d<4$, since both types of particles will
co-exist locally.  This leads to a different rate of decay for $\rho (t)$,
which is given in (1.2).  There are certain similarities, though, and the
analogs of the results from Sections 3-5, for $d<4$, will also be stated for
$d\geq 4$ at the end of their respective sections.  They will be
applied in [BrLe99]. 
\vskip .5truein
\noindent{\bf 3. Approximation of $\xi_t$ by Earlier Conditional
Expectations}
\vskip .3cm
In this section, we demonstrate Proposition 2.3, which states that for
finite sets $E\subset {\Bbb Z}^d$ in $d<4$, $\xi_t(E)$ is approximated by
$(\xi_{t^{1/4}}*K_{t-t^{1/4}})(E)$ with high probability.  As
outlined in Section 2, we do this by analyzing $_{t^{1/4}}\eta_t$, where, we
recall, $_s\eta_r$, $r\in [0,t]$, is the process that evolves like $\xi_r$
up until time $s$, but where, over $(s,t]$, the annihilation of particles
is quenched.  The main goal of this section will be to show the following
analog of Proposition 2.3, with $_{t^{1/4}}\eta_t$ substituted for $\xi_t$.
\vskip .3cm
\noindent{\bf Proposition 3.1.} {\it {}For $d<4$ and sufficiently large $t$,
$$
\eqalign{P(\vert _{t^{1/4}}\eta_t(E)-(\xi_{t^{1/4}}&*K_{t-t^{1/4}})(E)\vert
\geq\epsilon\vert E\vert t^{-d/4})\cr
&\leq 3\exp\{-((\epsilon v_t(\epsilon)\vert E\vert t^{-9d/20})\wedge
t^{1/8})\}\cr}
\leqno(3.1)
$$
holds for all $\epsilon$ and $E\subset {\Bbb Z}^d$, with $\vert
E\vert\leq t^d$.} 
\vskip .3cm
\noindent (The exponent $1/4$ in the subscript $t^{1/4}$ is not crucial
here; other choices would require a modification of the term $t^{9d/20}$.)

It is not difficult to deduce Proposition 2.3 from Proposition 3.1.  The
main step is given by the following lemma.
\vskip .3cm
\noindent{\bf Lemma 3.1.} {\it {}For all $d$ and $s\leq t$,
$$
\eqalign{&P(_s\eta_t(E)- \xi_t(E)\geq 0\ \mid\ {\Cal F}^\xi_t)\geq
1/2,\cr
&P(_s\eta_t(E)-\xi_t(E)\leq 0\ \mid\ {\Cal F}^\xi_t)\geq
1/2,\cr}
\leqno(3.2)
$$
hold a.s. for all $E\subset {\Bbb Z}^d$, with $\vert E\vert < \infty$.}
\vskip .3cm
\noindent{\it Proof}.  Both parts of (3.2) follow from the symmetric
behavior of $A$ and $B$ particles.  Two particles of types $A$ and $B$,
which meet at some $\tau\in (s,t]$ under $\xi_r$, continue to evolve
as independent simple 
random walks, $Y^A(r)$ and $Y^B(r)$, on $[\tau ,t]$, under $_s\eta_t$.
The difference of indicator functions $1(Y^B(r)\in 
E)-1(Y^A(r)\in E)$ is symmetric, and is independent of all other such pairs
of random walks, when conditioned on ${\Cal F}^\xi_t$.  The sum of all such
differences equals $_s\eta_t(E)-\xi_t(E)$, and will again be symmetric when
conditioned on ${\Cal F}^\xi_t$.  This implies (3.2).\hfill //
\vskip .3cm
Using Lemma 3.1, Proposition 2.3 follows immediately from Proposition 3.1.
Setting $s = t^{1/4}$, one sees that at least half of the time when the
exceptional event in (2.8) holds, the same is true for the event in (3.1).  So,
the upper bound on the probability on the event in (3.1) implies that in
(2.8).

We now turn our attention to demonstrating Proposition 3.1.  Most of the
work required for the proposition is to show that $(\vert\xi_{t^{1/4}}\vert
*K_{t-t^{1/4}})(E)$ is typically not
too large.  One then uses this bound in conjunction with a large deviation
estimate.  The desired bound on this convolution is given by the
following result. 
\vskip .3cm
\noindent{\bf Proposition 3.2.} {\it Let $d<4$.  {}For given $\delta > 0$,
suppose that $s$ is sufficiently large, and that $t\geq s^4$.  Then,
$$
P((\vert\xi_s\vert * K_{t-s})(x)\geq 2s^{-(d/4-\delta)})\leq e^{-s^{2/3}}
\leqno(3.3)
$$
for all $x$.}
\vskip .3cm
We first demonstrate Proposition 3.2.  This will require Lemmas 3.2-3.4,
which are given below.  We will then show how Proposition 3.1 follows from
Proposition 3.2.

We will employ moment generating functions to show (3.3).  Rather than
analyzing $\vert\xi_s\vert *K_{t-s}$ directly, we will look at
$\vert^x\xi_s\vert *K_{t-s}$, for $x\in {\Bbb Z}^d$. The process
$^x\xi_s$ denotes the 
analog of $\xi_s$, but where the initial state is restricted to ${\Cal
A}_s(x)$, i.e., for $\vert E\vert < \infty$, $^x\xi_0(E) = \xi_0(E\cap {\Cal
A}_s(x))$.  The set ${\Cal A}_s(x)$ is defined by
$$
{\Cal A}_s(x) =\{y\in {\Bbb Z}^d:\vert y-x\vert_\infty\leq s/3\} ,
\leqno(3.4)
$$
where $\vert\cdot\vert_\infty$ denotes the sup norm.  We construct $^x\xi_s$
on the same space as
$\xi_s$, by assigning the same random walk paths to corresponding particles
as was done before (2.7). Let ${\Cal C}_s$ denote
the set of all $x\in {\Bbb Z}^d$ for which each coordinate is a multiple of
$\lfloor s\rfloor$, the integer part of $s$.  {}For the first part of the
argument, we will restrict $x$ to ${\Cal C}_s$.  There, we will use the
independence of the processes $^{x_1}\xi_r,\!^{x_2}\xi_r,\dots$, for
$x_j\in {\Cal C}_s,\ j=1,2,\dots$, and $r\in [0,t]$. Then, when we analyze
$\vert\xi_s\vert *K_{t-s}$, we will also consider translates of ${\Cal C}_s$.

Lemma 3.2 states that, for any $x$, $\xi_s(x)$ and $^x\xi_s(x)$ are
close in expectation. 
This is not surprising. The initial states $\xi_0$ and $^x\xi_0$ only
differ at sites further than $s/3$ from $x$.  By time $s$, the probability
will be small that this difference will have worked its way to $x$.
\vskip .3cm
\noindent{\bf Lemma 3.2.} {\it {}For all $d$ and sufficiently large $s$,
$$
E\left[\left\vert\xi_s(x)-\,^x\xi_s(x)\right\vert\right] \leq e^{-C_2s}
\leqno(3.5)
$$
for all $x$ and appropriate $C_2>0$.}
\vskip .3cm
Together with (1.2), (3.5) gives the following bound on
$E[\vert^x\xi_s(x)\vert ]$.  Lemma 3.2 will also be used in (3.19).
\vskip .3cm
\noindent{\bf Corollary 3.1.} {\it {}For $d<4$ and sufficiently large $s$,
$$
E[\vert^x\xi_s(x)\vert ]\leq C_3/s^{d/4}
\leqno(3.6)
$$
for all $x$ and appropriate $C_3$.}
\vskip .3cm
\noindent The corollary will be used in the proof of Lemma 3.3.
\vskip .3cm
\noindent{\it Proof of Lemma 3.2}. We consider the ``discrepancy'' $^x{\Cal
G}_r$ between $\xi_r$ and $^x\xi_r$, at each time $r$.  This set is defined
as those $A$ and $B$ particles, from either $\xi_r$ or $^x\xi_r$, which
still exist by time $r$ for one process but not for the other.  (The
initial discrepancy consists of particles of $\xi_0$ lying outside ${\Cal
A}_s(x)$.)  In order for $\xi_r(y)\ne\!^x\xi_r(y)$, $^x{\Cal G}_r$ must
contain a particle at $y$.  One can check that the particles in $^x{\Cal
G}_r$ execute independent random walks except when $A$ and $B$ particles
from the same process meet one another, and at least one of them is in
$^x{\Cal G}_{r-}$.  If both are in $^x{\Cal G}_{r-}$, the particles
annihilate one another, and both disappear from
$^x{\Cal G}_r$.  If only one is in $^x{\Cal G}_{r-}$,
then, upon annihilation, this particle is replaced in the discrepancy by a
particle of the opposite type (e.g., $B$ instead of $A$) at the same site,
which belongs to the opposite process.  So, $^x{\Cal G}_r$ is dominated by
a set of random walks, whose initial positions are given by the initial
positions of the $A$ and $B$ particles outside ${\Cal A}_s(x)$.

The distance between $x$ and ${\Cal A}_s(x)^c$ is $s/3$.  The initial
positions of $A$ and $B$ particles for $\xi_r$ are given by Poisson random
fields with intensity $\lambda$.  It is therefore not difficult to show,
using moment generating functions, that, for large $s$, the expected number
of random walks at $x$ at time $s$, which were originally in ${\Cal
A}_s(x)^c$, is bounded above by
$$
\lambda\sum_{z = \lfloor s/3\rfloor +1}z^{d-1}e^{-C_4z}\leq\lambda e^{-C_4s/4}
$$
for appropriate $C_4>0$ (see, e.g., the proof of Lemma 7.3).  Since
this dominates 
$E[\vert\xi_s(x)-\,^x\xi_s(x)\vert ]$, (3.5) follows.\hfill //
\vskip .3cm
In order to derive (3.3) of Proposition 3.2, we will need the following
bound on the moment generating function of $\vert^x\xi_s(x)\vert$.
\vskip .3cm
\noindent{\bf Lemma 3.3.} {\it Let $d<4$, and fix $\delta > 0$.  Suppose
that $\theta > 0$ is bounded, $s$ is sufficiently large, and $m\in {\Bbb
Z}^+$ is chosen so that $m!\geq s^{d/4}$.  Then,
$$
E[e^{\theta\vert^x\xi_s(x)\vert}]\leq 1+C_5 (e^{\theta
m}-1)/s^{d/4-\delta}
\leqno(3.7)
$$
for all $x$ and appropriate $C_5$.}
\vskip .3cm
\noindent{\it Proof}. {}For given $m\in {\Bbb Z}^+$ and $\theta > 0$, it
is easy to check that
$$
E[e^{\theta\vert^x\xi_s(x)\vert}] \leq 1+(e^{\theta m}-1)P(^x\xi_s(x)\ne 0) +
E[e^{\theta\vert^x\xi_s(x)\vert}-1; \vert^x\xi_s(x)\vert > m] .
\leqno(3.8)
$$
By (3.6), the second term on the right is bounded above by
$$
C_4(e^{\theta m}-1)/s^{d/4} .
\leqno(3.9)
$$
Denote by $\eta^\#_s(x)$ the total number of particles at $x$ for the
process $\eta_s$. To handle the third term, we note that it is at most
$$
E[e^{\theta\eta^\#_s(x)}-1;\ \eta^\#_s(x)>m] .
\leqno(3.10)
$$
Since $\eta^\#_s(x)$ is Poisson with mean $2\lambda$, (3.10) is
$$
\leq e^{-2\lambda}\sum_{j>m}(e^{\theta j}-1)(2\lambda)^j/j! .
$$
{}For given $\delta > 0$, bounded $\theta$, $m!\geq s^{d/4}$ and large enough
$s$, this is
$$
\leq (2\lambda)^m(e^{\theta m}-1)/m!\leq (e^{\theta m}-1)/s^{d/4-\delta} .
\leqno(3.11)
$$
This provides an upper bound on the third term.  Together, (3.8)-(3.11)
give the bound in (3.7).\hfill //
\vskip .3cm
Using Lemma 3.3, we derive Lemma 3.4.  It is the analog of Proposition 3.2,
but with $\vert^y\xi_s(y)\vert$ in place of $\vert\xi_s(y)\vert$, and the
sum being over $y\in {\Cal C}_s$ rather than $y\in {\Bbb Z}^d$.  In
the proof, we 
will use the local central limit bound
$$
\limsup_{t\to\infty} \{t^{d/2}\sup (K_{t-s}(x):x\in {\Bbb Z}^d,\ s\leq
t/2)\} < \infty ,
\leqno(3.12)
$$
as well as
$$
\limsup_{t\to\infty} \Bigl\{ s^d\sup \Bigl( \sum_{y\in {\Cal C}_s}
K_{t-s}(x-y):x\in {\Bbb Z}^d,\ s\leq t^{1/2}\Bigr)\Bigr\} < \infty ,
\leqno(3.13)
$$
which hold for all $d$. These bounds follow from (4.5) and (4.8),
respectively, together with 
simple bounds on the normal kernel $N_{t-s}$.  In Section 4, we
will go into greater detail on such bounds.
\vskip .3cm
\noindent {\bf Lemma 3.4.} {\it Let $d<4$.  {}For given $\delta >0$,
suppose that $s$ is sufficiently large, and that $t\geq s^2$.  Then,
$$
P\Bigl( \sum_{y\in {\Cal C}_s}\vert^y\xi_s(y)\vert K_{t-s}(x-y)\geq
s^{-(5d/4-\delta)}\Bigr) \leq \exp \{-t^{d/2}/s^{5d/4}\}
\leqno(3.14)
$$
for all $x$.}
\vskip .3cm
\noindent{\it Proof}.  Set $\theta = \theta_y = t^{d/2}K_{t-s}(x-y)/m$ in
(3.7), where $m$ is the smallest integer satisfying $m! \geq s^{d/4}$.  By
(3.12), $m\theta_y$ is bounded for large $s$ and $t\geq 2s$.  So, by Lemma
3.3, one has, for given $\delta' > 0$,
$$
\eqalign{E[\exp\{t^{d/2} \vert^y\xi_s(y)\vert &K_{t-s}(x-y)/m\}]\cr
&\leq 1 + C_5(\exp \{t^{d/2}K_{t-s}(x-y)\}-1)/s^{d/4-\delta'}\cr
&\leq 1 + C_6t^{d/2}K_{t-s}(x-y)/s^{d/4-\delta'}\cr
&\leq \exp\{C_6t^{d/2}K_{t-s}(x-y)/s^{d/4-\delta'}\}\cr}
\leqno(3.15)
$$
for all $x$ and $y$, and appropriate $C_6$.  One also has $s^{\delta'}\geq m$
for large $s$, and $m$ chosen as above.  So, by (3.15),
$$
E[\exp\{ t^{d/2}\vert^y\xi_s(y)\vert K_{t-s}(x-y)/s^{\delta'}\}]
\leq\exp\{C_6t^{d/2}K_{t-s}(x-y)/s^{d/4-\delta'}\}.
\leqno(3.16)
$$

Now, for $y_j\in {\Cal C}_s$,\ $j=1,2,\dots$, the processes
$^{y_1}\xi_r,^{y_2}\xi_r,\dots$ are independent, as was mentioned below
(3.4). Setting $r=s$, it therefore follows from (3.16), that
$$
\eqalign{E\Bigl[ \exp\Bigl\{ t^{d/2}\sum_{y\in {\Cal C}_s}&\vert^y\xi_s(y)\vert
K_{t-s}(x-y)/s^{\delta'}\Bigr\}\Bigr]\cr
&\ \leq\exp\Bigl\{ (C_6t^{d/2}/s^{d/4-\delta'})\sum_{y\in {\Cal
C}_s}K_{t-s}(x-y)\Bigr\} .\cr}
$$
By (3.13), this is
$$
\leq\exp\{ C_7t^{d/2}/s^{5d/4-\delta'}\}
$$
for large $s$ with $t\geq s^2$, and appropriate $C_7$.  So, by Chebyshev's
inequality,
$$
P\Bigl( \sum_{y\in {\Cal C}_s} \vert^y\xi_s(y)\vert K_{t-s}(x-y)\geq
s^{-(5d/4-\delta)}\Bigr)\leq\exp\{ (C_7s^{\delta'} - s^{\delta-\delta'})
t^{d/2}/s^{5d/4}\} .
\leqno(3.17)
$$
If we set $\delta' = \delta /3$, this is at most $\exp\{
-t^{d/2}/s^{5d/4}\}$ for large $s$, which implies (3.14).\hfill\break

\hfill //
\vskip .3cm
Using Lemmas 3.2 and 3.4, we demonstrate Proposition 3.2.
\vskip .3cm
\noindent{\it Proof of Proposition 3.2}.  The bound (3.14) holds
independently of $x$.  Consequently, it also holds if one instead sums over
translates ${\Cal C}_s+z$ of ${\Cal C}_s$, with $z\in {\Bbb Z}^d$.  By
summing over all such $\lfloor s\rfloor^d$ translates, one obtains,
for given $\delta > 0$, 
$$
P\Bigl( \sum_{y\in {\Bbb Z}^d}\vert^y\xi_s(y)\vert K_{t-s}(x-y)\geq
s^{-(d/4-\delta)}\Bigr) \leq s^d\exp\{-t^{d/2}/s^{5d/4}\}
\leqno(3.18)
$$
for all $x$, and for large $s$ with $t\geq s^2$.  On the other hand, since
$\sum_{y\in {\Bbb Z}^d} K_{t-s}(x-y) = 1$, it follows from Lemma 3.2 and
Chebyshev's inequality, that for large $s$,
$$
P\Bigl( \sum_{y\in {\Bbb Z}^d} (\vert\xi_s(y)\vert -\vert^y\xi_s(y)\vert
)K_{t-s}(x-y)\geq s^{-(d/4-\delta)}\Bigr) \leq s^{d/4-\delta}e^{-C_2s}
\leqno(3.19)
$$
for all $x$.  Together, (3.18) and (3.19) imply that for large $s$ and
$t\geq s^4$,
$$
P\Bigl( \sum_{y\in {\Bbb Z}^d}\vert \xi_s(y)\vert K_{t-s}(x-y) \geq
2s^{-(d/4-\delta)}\Bigr) \leq e^{-s^{2/3}} .
$$
This is equivalent to (3.3), and completes the proof of Proposition
3.2.\hfill //
\vskip .3cm
We now proceed to prove Proposition 3.1.  This will complete our proof of
Proposition 2.3.  The argument consists of applying
moment generating functions to $_{t^{1/4}}\eta_t$ conditioned on ${\Cal
F}_{t^{1/4}}$. Since the conditioned process $_{t^{1/4}}\eta_r$ evolves
according to independent simple
random walks over $(t^{1/4},t]$, and since $_{t^{1/4}}\eta_{t^{1/4}} =
\xi_{t^{1/4}}$ is known, the computations are fairly explicit.
Proposition 3.2 supplies the main technical estimate needed to bound the
right side of (3.23).
\vskip .3cm
\noindent{\it Proof of Proposition 3.1}. We abbreviate the left side of
(3.1) by setting
$$
\bar\eta =\,_{t^{1/4}}\eta_t(E)-(\xi_{t^{1/4}}*K_{t-t^{1/4}})(E) .
\leqno(3.20)
$$
We also set $\bar K(x) = K_{t-t^{1/4}}(E-x)$.

We proceed to estimate the moment generating function of $\bar\eta$,
conditioned on ${\Cal F}_{t^{1/4}}$.  Since over $[t^{1/4},t]$,\
$_{t^{1/4}}\eta_r$ evolves according to independent simple random walks,
one can, for given $\theta$, write
$$
\eqalign{E[e^{\theta\bar\eta} \mid {\Cal F}_{t^{1/4}}] = \prod_{x\in
{\Bbb Z}^d} 
&[\bar K(x)e^{-\theta (1-\bar K(x))}+(1-\bar K(x))e^{\theta\bar
K(x)}]^{\xi^A_{t^{1/4}}(x)}\cr
&\cdot [\bar K(x)e^{\theta (1-\bar K(x))}+ (1-\bar K(x))e^{-\theta\bar
K(x)}]^{\xi^B_{t^{1/4}}(x)}.\cr}
\leqno(3.21)
$$
{}For small $\theta$, this is
$$
\eqalign{&\leq \prod_x [1+\theta^2\bar K(x)(1-\bar
K(x))]^{\vert\xi_{t^{1/4}}(x)\vert}\cr
&\leq \exp\Bigl\{ \theta^2\sum_x \vert\xi_{t^{1/4}}(x)\vert \bar K(x)\Bigr\}\cr
&= \exp\bigl\{ \theta^2(\vert\xi_{t^{1/4}}\vert *K_{t-t^{1/4}})(E)\bigr\}.\cr}
\leqno(3.22)
$$
It follows from (3.21)-(3.22) and Chebyshev's inequality (applied to both
$\theta >0$ and $\theta <0$), that, for $\gamma >0$,
$$
P(\vert\bar\eta\vert\geq\gamma \mid {\Cal F}_{t^{1/4}})\leq
2\exp\{\theta^2(\vert\xi_{t^{1/4}}\vert *K_{t-t^{1/4}})(E)-\gamma\theta\} .
\leqno(3.23)
$$

Let $G$ denote the set where
$$
(\vert\xi_{t^{1/4}}\vert *K_{t-t^{1/4}})(E) < 2\vert E\vert t^{-(d-4\delta)/16}
,
$$
for given $\delta > 0$.  By Proposition 3.2, applied to each $x\in E$ with
$s=t^{1/4}$,
$$
P(G^c)\leq\vert E\vert e^{-t^{1/6}}\leq e^{-t^{1/8}}
\leqno(3.24)
$$
for sufficiently large $t$ and $\vert E\vert\leq t^d$.  (Of course, any
power of $t$ suffices.)

On the other hand, on $G$, the right side of (3.23) is at most
$$
2\exp\{(2\theta^2\vert E\vert t^{-(d-4\delta)/16})-\gamma\theta\} ,
\leqno(3.25)
$$
which provides an upper bound on $P(\vert\bar\eta\vert \geq \gamma\mid
{\Cal F}_{t^{1/4}})$, for small $\theta$.  {}For a given $\epsilon >0$, let
$$
\gamma =\epsilon\vert E\vert t^{-d/4},\quad \theta = {1\over
4}t^{-\delta /4}((\epsilon t^{-3d/16})\wedge 1) .
\leqno(3.26)
$$
Setting $v_t(\epsilon) = \epsilon\wedge
t^{3d/16}$, one can write $\theta = {1\over 4}t^{-(3d+4\delta)/16}v_t
(\epsilon)$.
Substitution of $\gamma$ and $\theta$ into (3.23) and (3.25), with
$\delta < d/20$, implies that, on 
$G$,
$$
\eqalign{P(\vert\bar\eta\vert\geq\epsilon \vert E\vert t^{-d/4}\mid {\Cal
F}_{t^{1/4}}) &\leq 2\exp\Bigl\{ -{\epsilon v_t(\epsilon)\vert E\vert\over
8t^{(7d+4\delta)/16}}\Bigr\}\cr
&\leq 2\exp\{-\epsilon v_t(\epsilon)\vert E\vert t^{-9d/20}\}\cr}
\leqno(3.27)
$$
for large $t$.  It follows from (3.24) and (3.27), that for large $t$,
$$
P(\vert\bar\eta\vert\geq\epsilon\vert E\vert t^{-d/4})\leq
3\exp\{-((\epsilon v_t(\epsilon)\vert E\vert t^{-9d/20})\wedge t^{1/8})\},
$$
which implies (3.1).\hfill //
\vskip .3cm
As mentioned at the end of Section 2, we will want to employ the higher
dimensional analogs of results, such as Proposition 2.3, in [BrLe99].  The
modifications required for Proposition 2.3 are straightforward to make.
The restriction to $d<4$ was needed for Corollary 3.1, which employed
(1.2).  If one 
instead employs the corresponding bound for $d\geq 4$, one obtains
$$
E[\vert^x\xi_s(x)\vert ]\leq C_3/s
\leqno(3.28)
$$
for large $s$ and all $x$, in place of (3.6).  The bound in (3.6) was used
in (3.9); replacing the term $s^{d/4}$ by $s$ there gives the corresponding
bound for $d\geq 4$.  This leads to the analog of Lemma 3.3, with $s$
replacing $s^{d/4}$ in both places, and to the analog of Lemma 3.4, with
$s^{d+1}$ replacing $s^{5d/4}$ in both places.  This last change
requires us to replace 
the bound $s^{d/4-\delta}$ with $s^{1-\delta}$ in Proposition 3.2.  In
the proof of Proposition 3.1, one replaces $t^{(d-4\delta)/16}$ with
$t^{(1-\delta)/4}$ in the bound defining $G$.  This leads to the upper
bound, for large $t$,
$$
P(\vert\bar\eta\vert\geq \epsilon\vert E\vert t^{-1}\mid {\Cal
F}_{t^{1/4}})\leq 2\exp\{-\epsilon v_t(\epsilon)\vert E\vert t^{-9/5}\}
\leqno(3.29)
$$
corresponding to (3.27), where now $v_t(\epsilon) = \epsilon\wedge
t^{3/4}$, and to the analogous bounds corresponding to (3.1).  Application
of Lemma 3.1, as before, then produces the following analog of Proposition
2.3 for $d\geq 4$.
\vskip .3cm
\noindent{\bf Proposition 3.3.} {\it {}For $d\geq 4$ and sufficiently large
$t$,}
$$
\eqalign{P(\vert\xi_t(E)-(\xi_{t^{1/4}}*K_{t-t^{1/4}})&(E)\vert\geq\epsilon
\vert E\vert t^{-1})\cr
&\leq 6\exp\{-(\epsilon v_t(\epsilon)\vert E\vert t^{-9/5})\wedge
t^{1/8})\}\cr}
\leqno(3.30)
$$
holds for all $\epsilon$ and $E\subset {\Bbb Z}^d$, with $\vert
E\vert\leq t^d$. 
\vskip .5truein
\noindent{\bf 4. Approximation of $\xi_s*K_{t-s}$ by $\xi_0*N_t$}
\vskip .3cm
In this section, we demonstrate Proposition 2.4, which states that
$(\xi_s*K_{t-s})(0)$ is approximated by $(\xi_0*N_t)(0)$, with high
probability, when $s\leq t^{1/4}$.  In order to demonstrate the
proposition, it is enough to verify the following two results.
\vskip .3cm
\noindent{\bf Proposition 4.1.} {\it {}For any $d$, let $t$ be sufficiently large
and $s\leq t^{1/4}$.  Then, for all $\epsilon\in [0,1]$,}
$$
P(\vert (\xi_s*N_t)(0) - (\xi_s*K_{t-s})(0)\vert\geq\epsilon t^{-d/4})\leq
2\exp\{ -4\epsilon^2t^{1/2}\} .
\leqno(4.1)
$$
\vskip .3cm
\noindent{\bf Proposition 4.2.} {\it {}For any $d$, let $t$ be sufficiently
large and $s\leq t^{1/4}$.  Then, for all $\epsilon\in [0,1]$,}
$$
P(\vert (\xi_0*N_t)(0) - (\xi_s*N_t)(0)\vert\geq \epsilon t^{-d/4})\leq
2\exp\{ -4\epsilon^2t^{1/4}\} .
\leqno(4.2)
$$
\vskip .3cm
{}For these propositions, we will need estimates on $K_{t-s}$, which follow
from a standard local central limit
theorem.  Related bounds are also used
in Sections 3 and 9.  We employ the references [BhRa86] and [Pe75] for
these purposes.

Assume that $d=1$.  By Theorem 16, on page 207 of [Pe75],
$$
(1+(x/t^{1/2})^3)(N_t(x)-K_t(x)) = o(t^{-1})\quad \text{as}\ t\to\infty ,
\leqno(4.3)
$$
holds uniformly in $x\in {\Bbb Z}$.  (Because $K_t$ is symmetric, its
cumulant of order 3 will be 0, which gives the simplified form in (4.3).)
One can also employ Theorem 22.1, on page 231 of [BhRa86].  (Both results
are stated for discrete times, although the derivation for continuous $t$
is, of course, the same.)

Assume now that $d$ is arbitrary.  Since the evolution of $K_t$ in
different coordinates is independent, one can apply (4.3) to conclude that
$$
(1+(\vert x\vert /t^{1/2})^{3d})(N_t(x)-K_t(x)) = o(t^{-(d+1)/2}) .
\leqno(4.4)
$$
The next two bounds follow quickly from (4.4):
$$
\sup_{x\in {\Bbb Z}^d}\vert N_t(x)-K_t(x)\vert\leq C_8t^{-(d+1)/2}
\leqno(4.5)
$$
and
$$
\sum_{x\in {\Bbb Z}^d} \vert N_t(x)-K_t(x)\vert\leq C_8t^{-1/2} ,
\leqno(4.6)
$$
for appropriate $C_8$ and large $t$.  {}From these bounds, one also obtains
that
$$
\sum_{x\in {\Bbb Z}^d} (N_t(x)-K_t(x))^2\leq C_9t^{-d/2-1} ,
\leqno(4.7)
$$
for appropriate $C_9$.

Suppose that ${\Bbb R}^d$ is partitioned into sets $E_j$, $j=1,2,\dots$,
such that each $E_j$ contains a cube of length $M$ with $M\leq
C_{10}t^{1/2}$, for a given $C_{10}$.  Again using (4.4), one can generalize
(4.6) so that, for $x_j\in E_j$,
$$
\sum_j \vert N_t(x_j) - K_t(x_j)\vert\leq C_{11}M^{-d}t^{-1/2}
\leqno(4.8)
$$
holds for some $C_{11}$ independently of the choice of $x_j$ and the
partition $\{E_j\}$.  Using this
and simple estimates on $\sum_jN_t(x_j)$, it is not difficult to derive
(3.13), which was employed in the proof of Lemma 3.4.

We will also need some basic estimates on $N_t$.  Set $g(s,r) =
N_{t-s}(x)$, where $r = \vert x\vert$ and $x\in {\Bbb R}^d$.  It is easy to
check that, for appropriate $C_{12}$,
$$
\left\vert {\partial g\over\partial s}(s,r)\right\vert\leq
C_{12}t^{-(d/2+1)}\left( 1+{r^2\over t}\right) e^{-r^2/4t}
\leqno(4.9)
$$
for all $s\leq t/2$ and $x$.  So, for all $x\in {\Bbb Z}^d$,
$$
\vert N_t(x)-N_{t-s}(x)\vert\leq C_{12}st^{-(d/2+1)}\left( 1+{r^2\over
t}\right) e^{-r^2/4t} .
\leqno(4.10)
$$
With a little work, it follows from this that
$$
\sum_{x\in {\Bbb Z^d}} (N_t(x)-N_{t-s}(x))^2\leq C_{13}s^2t^{-(d/2+2)}
\leqno(4.11)
$$
for appropriate $C_{13}$.

One can also check that
$$
\left\vert {\partial g\over\partial r}(0,r)\right\vert \leq
C_{14}t^{-(d/2+1)}re^{-r^2/2t} .
\leqno(4.12)
$$
{}For $\vert x-x'\vert\leq M\leq t^{1/2}$, one can use this to show
$$
\vert N_t(x) -N_t(x')\vert\leq C_{15}t^{-(d/2+1)}(\vert x\vert
+M)Me^{-\vert x\vert^2/4t} ,
\leqno(4.13)
$$
for appropriate $C_{15}$.  With a little work, one can then show that, for
appropriate $C_{16}$ and any $y\in {\Bbb R}^d$,
$$
\sum_{x\in {\Bbb Z}^d} \max_{x'} \{(N_t(x-y)-N_t(x'-y))^2: \vert x-x'\vert\leq
M\}\leq C_{16}t^{-(d/2+1)}M^2.
\leqno(4.14)
$$

In order to show Proposition 4.1, we first show its analog, where
$\xi_s$ is replaced by $\xi_0$.  In Proposition 4.3 and all following results
in this section, all dimensions $d$ are allowed.
\vskip .3cm
\noindent{\bf Proposition 4.3.} {\it Let $t$ be sufficiently large and
$s\leq t^{1/2}$.  Then, for all $\epsilon\in [0,1]$,}
$$
P(\vert (\xi_0*N_t)(0) - (\xi_0*K_{t-s})(0)\vert\geq \epsilon t^{-d/4})\leq
\exp\{ -4\epsilon^2t^{1/2}\} .
\leqno(4.15)
$$
\vskip .3cm
\noindent{\it Proof}. To obtain (4.15), we compute an upper bound on the
corresponding moment generating function, and then apply Chebyshev's
inequality.  First, recall that $\xi_0(x)$, at each site $x$, is the
difference of two Poisson random variables, each with intensity $\lambda$.
Since these random variables are independent at different sites, one has
that, for given $\theta$,
$$
\eqalign{E[\exp\{\theta ((\xi_0*N_t)(0) &- (\xi_0*K_{t-s})(0))\}]
= \prod_{x\in {\Bbb Z}^d} E[\exp\{\theta R(x)\xi_0(x)\}]\cr
&= \exp \Bigl\{\lambda\sum_{x\in {\Bbb Z}^d} (\exp \{\theta R(x)\} + \exp
\{-\theta R(x)\} - 2)\Bigr\} ,\cr}
\leqno(4.16)
$$
where $R(x) = N_t(-x)-K_{t-s}(-x)$.

Together, (4.5) and (4.10) imply that, for appropriate $C_{17}$,
$$
\vert R(x)\vert\leq C_{17}t^{-(d+1)/2}
$$
for large $t$ and all $x$, since $s\leq t^{1/2}$.  So, for
$\vert\theta\vert\leq C_{18}t^{(d+1)/2}$ and appropriate $C_{18}$, (4.16)
is at most
$$
\exp\{2\lambda\theta^2\sum_x (R(x))^2\} .
$$
By (4.7) and (4.11), this is
$$
\leq\exp\{C_{19}\lambda\theta^2t^{-(d/2+1)}\} ,
\leqno(4.17)
$$
for appropriate $C_{19}$.  Combining the inequalities from (4.16) through
(4.17), it follows that
$$
E[\exp\{\theta ((\xi_0*N_t)(0) - (\xi_0*K_{t-s})(0))\}]
\leq \exp\{ C_{19}\lambda\theta^2t^{-(d/2+1)}\}.
$$
Applying Chebyshev's inequality for both $\theta > 0$ and $\theta < 0$, one
obtains
$$
P(\vert (\xi_0*N_t)(0)-(\xi_0*K_{t-s})(0)\vert\geq\epsilon
t^{-d/4})
\leq 2\exp\{C_{19}\lambda\theta^2t^{-(d/2+1)}-\vert\theta\vert\epsilon
t^{-d/4}\}.
$$
Setting $\vert\theta\vert = 5\epsilon t^{(d+2)/4}$, one has, for large
$t$, the upper bound $\exp\{-4\epsilon^2t^{1/2}\}$, which implies
(4.15).\hfill //
\vskip .3cm
In order to obtain Proposition 4.1 from Proposition 4.3, we need to replace
$\xi_0$ by $\xi_s$.  The following lemma will enable us to do that.
\vskip .3cm
\noindent{\bf Lemma 4.1.} {\it Let $f(\cdot)$ be any nonrandom function.
{}For all $s$,}
$$
P\Bigl( \sum_{x\in {\Bbb Z}^d} f(x)(\eta_s(x)-\xi_s(x)) \geq 0 \mid {\Cal
F}^\xi_s\Bigr) \geq 1/2 .
\leqno(4.18)
$$
\vskip .3cm
\noindent The statement in (4.18) is similar to that in (3.2) of Lemma 3.1.
The reasoning that is required is analogous, with the point being that
exchanging the random walk motions of $A$ and $B$ particles in $\eta_t$, after
annihilation occurs in the corresponding process $\xi_t$, does not change
the law of $\sum_x f(x)(\eta_s(x)-\xi_s(x))$, conditioned on ${\Cal
F}^\xi_s$.

By first setting $f(x) = N_t(-x)-K_{t-s}(-x)$ and then $f(x) =
K_{t-s}(-x)-N_t(-x)$, one obtains the following corollary of Lemma 4.1.
\vskip .3cm
\noindent{\bf Corollary 4.1.} {\it {}For all $s\leq t$ and $\epsilon$,}
$$
\eqalign{P(\vert (\xi_s*N_t)(0)-(\xi_s &*K_{t-s})(0)\vert\geq\epsilon
t^{-d/4})\cr
&\leq 2P(\vert (\eta_s*N_t)(0) - (\eta_s*K_{t-s})(0)\vert\geq \epsilon
t^{-d/4}).\cr}
\leqno(4.19)
$$

The distribution of $\eta_s(x),\ x\in {\Bbb Z}^d$, for each $s$, is
given by the difference of 
two Poisson random fields, each with intensity $\lambda$.  So, the
distribution of $\eta_s(x)$ is constant over $s$, and one may substitute
$\eta_0$ for $\eta_s$ on the right side of (4.19).  This, in turn, may be
replaced by $\xi_0$, since the joint distributions of $\xi_0(x)$ and
$\eta_0(x)$, 
over $x\in {\Bbb Z}^d$, are the same.  Consequently, one obtains the
following result. 
\vskip .3cm
\noindent{\bf Corollary 4.2.} {\it {}For all $s\leq t$ and $\epsilon$,}
$$
\eqalign{P(\vert (\xi_s*N_t)(0)-(\xi_s&*K_{t-s})(0)\vert\geq \epsilon
t^{-d/4})\cr
&\leq 2P(\vert (\xi_0*N_t)(0) - (\xi_0*K_{t-s})(0)\vert\geq \epsilon
t^{-d/4}).\cr}
\leqno(4.20)
$$
\vskip .3cm
\noindent Proposition 4.1 is an immediate consequence of Proposition 4.3
and Corollary 4.2.

We now turn our attention to showing Proposition 4.2.  Our first step is to
replace $\xi_0$ by $\eta_0$ and $\xi_s$ by $\eta_s$ in (4.2).  {}For this, we
apply Lemma 4.1 again, this time with $f(x) = N_t(-x)$.  Since $\xi_0(x) =
\eta_0(x)$ for all $x$, and $\xi_0\in {\Cal F}^\xi_s$, we obtain the
following result.
\vskip .3cm
\noindent{\bf Corollary 4.3.} {\it {}For all $s\leq t$ and $\epsilon$,}
$$
\eqalign{P(\vert (\xi_0*N_t)(0)-(\xi_s &*N_t)(0)\vert \geq\epsilon
t^{-d/4})\cr
&\leq 2P(\vert (\eta_0*N_t)(0) -(\eta_s*N_t)(0)\vert\geq \epsilon
t^{-d/4}).\cr}
\leqno(4.21)
$$

On account of Corollary 4.3, in order to show Proposition 4.2, it suffices
to demonstrate the following variant.
\vskip .3cm
\noindent{\bf Proposition 4.4.} {\it Let $t$ be sufficiently large and
$s\leq t^{1/4}$.  Then, for all $\epsilon\in [0,1]$,}
$$
P(\vert (\eta_0*N_t)(0) - (\eta_s*N_t)(0)\vert\geq \epsilon t^{-d/4})\leq
\exp\{ -4\epsilon^2t^{1/4}\} .
\leqno(4.22)
$$

In order to demonstrate (4.22), it is more convenient to instead focus on
the motion of the individual particles corresponding to $\eta_s$, which are
undergoing rate-$d$ simple random walks on ${\Bbb Z}^d$.  We will show, in
effect, that for $s\leq t^{1/4}$, only a negligible number of particles
will have moved far enough by time $s$ to alter $N_t(\cdot)$ by more than a
negligible amount from its initial value.  We will employ the following
notation.  Label the positions at time $s$ of the $\eta^\#_0(x)
=\eta^A_0(x)+\eta^B_0(x)$ particles initially at $x$ by $X_s(x,j),\
j=1,\dots ,\eta^\#_0(x)$, where the ordering is chosen independently of the
type of particle; $\eta^\#_0(x),\ x\in {\Bbb Z}^d$, are independent
mean-$2\lambda$ Poisson random variables.  Set
$$
{\Cal J} =\{(x,j):1\leq j\leq \eta^\#_0(x)\} ,
$$
and let sgn$(x,j) = 1$ whenever the corresponding particle is a $B$
particle, and sgn$(x,j) = -1$ whenever it is an $A$ particle.  Also, for
$(x,j)\in {\Cal J}$, set
$$
Y(x,j) = \text{sgn}(x,j)(N_t(x)-N_t(X_s(x,j))) .
$$
(Since $s$ and $t$ are thought of as being fixed here, they are suppressed
in $Y(x,j)$.)  Using the above notation, we can rewrite (4.22) as
$$
P\Bigl(\Bigl\vert\sum_{(x,j)\in {\Cal J}}Y(x,j)\Bigr\vert\geq\epsilon
t^{-d/4}\Bigr) \leq \exp\{ -4\epsilon^2t^{1/4}\}.
\leqno(4.23)
$$

We break the demonstration of (4.23) into two steps.  {}For the first
step, Lemma 4.2, we set
$$
Y(x) = W(N_t(x)-N_t(X_s(x))),
$$
where $X_s(x)$ is a rate-$d$ simple random walk on ${\Bbb Z}^d$ starting at
$x$, and $W$ is an independent random variable taking values 1 and $-1$
with equal probability.  We introduce the quantities
$$
\eqalign{\psi_1(x) &= \max_{x'}\{(N_t(x)-N_t(x'))^2:\vert x-x'\vert\leq
t^{1/4}\},\cr
\psi_2(x) &= N_t(0)\sum_{\vert y\vert >t^{1/4}}K_s(y)(N_t(x)+N_t(x-y)),\cr}
\leqno(4.24)
$$
with $\psi (x) = \psi_1(x)+\psi_2(x)$.
\vskip .3cm
\noindent{\bf Lemma 4.2.} {\it Let $t$ be sufficiently large.  Then, for
all $s,x$ and $\vert\theta\vert\leq C_{20}t^{d/2}$,
$$
E[e^{\theta Y(x)}]\leq \exp\{C_{21}\theta^2\psi (x)\}
\leqno(4.25)
$$
for appropriate $C_{21}$ (depending on $C_{20}$).}
\vskip .3cm
\noindent{\it Proof}. Note that $Y(x)$ is symmetric, and that $\vert Y(x)
\vert\leq t^{-d/2}$ for all $s,t$ and $x$.  So, for
$\vert\theta\vert\leq C_{20}t^{d/2}$,
$$
\eqalign{E[e^{\theta Y(x)}] &= E\Bigl[ \sum^\infty_{k=0}(\theta
Y(x))^{2k}/(2k)!\Bigr] \leq 1+C_{21}\theta^2 E[(Y(x))^2]\cr
&\leq\exp\{ C_{21}\theta^2E[(Y(x))^2]\},\cr}
\leqno(4.26)
$$
for appropriate $C_{21}$.

Let $G(x)$ denote the event on which $\vert X_s(x)-x\vert\leq t^{1/4}$.
Then, on $G(x)$,
$$
(Y(x))^2\leq\psi_1(x),
\leqno(4.27)
$$
where $\psi_1(x)$ is given in (4.24).  Also, one has that
$$
E[(Y(x))^2;G^c(x)] \leq \sum_{\vert y\vert
>t^{1/4}}K_s(y)(N_t(x)-N_t(x-y))^2\leq \psi_2(x).
\leqno(4.28)
$$
Together, (4.26)-(4.28) imply that
$$
E[e^{\theta Y(x)}]\leq \exp\{ C_{21}\theta^2\psi (x)\}
$$
which is (4.25).\hfill //
\vskip .3cm
Conditioned on ${\Cal J}$, the random variables $Y(x,j),\ (x,j)\in {\Cal
J}$, are independent, and, for each $x$ and $j$, are distributed like
$Y(x)$.  Letting 
$Z(x),\ x\in {\Bbb Z}^d$, denote independent mean-$2\lambda$ Poisson random
variables, Lemma 4.2 therefore implies the following result.
\vskip .3cm
\noindent{\bf Corollary 4.4.} {\it Let $t$ be sufficiently large.  Then,
for all $s$ and $\vert\theta\vert\leq C_{20}t^{d/2}$,
$$
E\Bigl[ \exp \Bigl\{ \theta\sum_{(x,j)\in {\Cal
J}}Y(x,j)\Bigr\}\Bigr]
\leq E\Bigl[ \exp \Bigl\{ C_{21}\theta^2\sum_{x\in {\Bbb Z}^d}\psi
(x)Z(x)\Bigr\}\Bigr]
\leqno(4.29)
$$
holds a.s.}
\vskip .3cm
We now demonstrate (4.23).  We do this by bounding the right side of
(4.29), and applying Chebyshev's inequality.
\vskip .3cm
\noindent{\it Proof of (4.23).} We first note that, by (4.14) and (4.24),
$$
\sum_{x\in {\Bbb Z}^d} \psi_1(x)\leq C_{16}t^{-(d+1)/2} .
\leqno(4.30)
$$
Also, using $s\leq t^{1/4}$, it follows from a standard large deviation
estimate on $K_s$, that
$$
\sum_x \psi_2(x) = 2N_t(0)\sum_{\vert y\vert >t^{1/4}} K_s(y)\leq
e^{-C_{22}t^{1/4}}
\leqno(4.31)
$$
for large $t$ and appropriate $C_{22}$.  So, by (4.30) and (4.31),
$$
\sum_x\psi(x)\leq 2C_{16}t^{-(d+1)/2}
\leqno(4.32)
$$
for large $t$.

Since $Z(x)$ are independent mean-$2\lambda$ Poisson random variables,
$$
E\Bigl[ \exp\Bigl\{C_{21}\theta^2\sum_{x\in {\Bbb Z}^d} \psi
(x)Z(x)\Bigr\}\Bigr] = \exp\Bigl\{ 2\lambda\sum_x (e^{C_{21}
\theta^2\psi(x)}-1)\Bigr\}.
\leqno(4.33)
$$
{}For $\theta^2\leq C_{23}/\sup_x\psi (x)$, this is at most
$\exp\{C_{24}\lambda\theta^2\sum_x\psi(x)\}$ for appropriate $C_{24}$
(depending on $C_{23}$), which, by (4.32), is
$$
\leq \exp\{ C_{25}\lambda\theta^2t^{-(d+1)/2}\}
\leqno(4.34)
$$
for large $t$ and appropriate $C_{25}$.  So, by Corollary 4.4 and
(4.33)-(4.34),
$$
E\Bigl[ \exp\Bigl\{ \theta\sum_{(x,j)\in {\Cal J}} Y(x,j)\Bigr\}\Bigr] \leq
\exp\{C_{25}\lambda\theta^2t^{-(d+1)/2}\} .
$$
Setting $C_{23} = 50C_{16}$, $\theta = 5\epsilon t^{(d+1)/4}$ and
applying Chebyshev's 
inequality implies that
$$
P\Bigl( \sum_{(x,j)\in {\Cal J}} Y(x,j)\geq\epsilon
t^{-d/4}\Bigr) \leq\exp\{-5\epsilon^2(t^{1/4}-5C_{25}\lambda )\} .
$$
This gives (4.23) for large $t$.\hfill //
\vskip .5truein
\noindent{\bf 5. Approximation of $\xi_t$ by $\xi_0*N_t$}
\vskip .3cm
In this section, we demonstrate Proposition 2.1, which gives a uniform
bound on $\vert\xi_t(D)-(\xi_0*N_t)(D)\vert$ over rectangles $D\in {\Cal
D}_{T^{1/2}}$, for $t\in [T/M,T]$ and $M>1$, where
$T$ is large.  Our main tools for this are Propositions 2.3 and 2.4, which
bound $\vert\xi_t(D)-(\xi_{t^{1/4}}*K_{t-t^{1/4}})(D)\vert$ and $\vert
(\xi_0*N_t)(0)-(\xi_{t^{1/4}}*K_{t-t^{1/4}})(0)\vert$, respectively.  It is
easy to extend the latter estimate from $0$ to $D$.  After
combining these bounds, we will sum the exceptional probabilities over
$D\in {\Cal D}_{T^{1/2}}$ and $t\in {\Cal S}_T$, where ${\Cal S}_T$ is an
appropriate lattice in $[T/M-1,T]$. It is then not difficult to extend the
bounds to all $t\in [T/M,T]$.

We first note that by Proposition 2.4, for large $t$,
$$
P(\vert (\xi_0*N_t)(E)
- (\xi_{t^{1/4}}*K_{t-t^{1/4}})(E)\vert\geq\epsilon\vert E\vert t^{-d/4})
\leq 4\vert E\vert\exp\{ -(\epsilon^2\wedge 1)t^{1/4}\}
\leqno(5.1)
$$
for $\vert E\vert <\infty$ and $\epsilon\geq 0$.  Together with
Proposition 2.3, 
this implies the following result.  Recall that $v_t(\epsilon)
=\epsilon\wedge t^{3d/16}$.
\vskip .3cm
\noindent{\bf Lemma 5.1.} {\it {}For $d<4$ and sufficiently large $t$,
$$
\eqalign{P(\vert\xi_t(E) &-(\xi_0*N_t)(E)\vert\geq 2\epsilon \vert E\vert
t^{-d/4})\cr
&\leq 4\vert E\vert \exp\{ -(\epsilon^2\wedge 1)t^{1/4}\} +
6\exp\{-((\epsilon v_t(\epsilon)\vert E\vert t^{-9d/20})\wedge
t^{1/8})\}\cr}
\leqno(5.2)
$$
for all $\epsilon$ and $E\subset {\Bbb Z}^d$, with $\vert E\vert\leq t^d$.}
\vskip .3cm
In order to derive Proposition 2.1, we rephrase (5.2) so that the bound on
the right side does not depend on $E$.  {}For $E$ and $t$ in the range of
interest to us, the inequality simplifies to that given in (5.3) with a
little work.  
\vskip .3cm
\noindent{\bf Proposition 5.1.} {\it {}For $d<4$, $M>1$ and
sufficiently large $T$,
$$
P(\vert\xi_t(E)-(\xi_0*N_t)(E)\vert\geq\epsilon_1T^{d/4})
\leq 8\exp\{-C_{26}((\epsilon_1)^2\wedge 1)T^{1/20}\},
\leqno(5.3)
$$
for appropriate $C_{26}>0$ (depending on $M$), all $t\in [T/M,T]$,
$\epsilon_1$ and $E\subset {\Bbb Z}^d$, with $\vert E\vert\leq MT^{d/2}$.}
\vskip .3cm
\noindent{\it Proof}. Setting $\epsilon_1 = 2\epsilon\vert E\vert
(tT)^{-d/4}$, the left side of (5.2) can be written as
$$
P(\vert\xi_t(E)-(\xi_0*N_t)(E)\vert\geq\epsilon_1T^{d/4}).
$$
This is the left side of (5.3).  The first term on the
right side of (5.2) is at most $4\vert E\vert\exp\{-C_{26}((\epsilon_1)^2
\wedge 1)T^{1/4}\}$ for $t\geq T/M$, $\vert E\vert\leq MT^{d/2}$ and
appropriate $C_{26}>0$.  {}For large $T$ and $\epsilon_1\geq T^{-1/12}$, this
is at most $\exp\{ -C_{26}((\epsilon_1)^2\wedge 1)T^{1/12}\}$, which is
dominated by the right side of (5.3), with the factor 2 instead of 8; the
factor 2 there ensures that
the inequality is trivial for $\epsilon_1 < T^{-1/12}$.  One can also check
that
$$
\epsilon v_t(\epsilon)\vert E\vert t^{-9d/20}\geq {1\over
4M}\epsilon_1v_t(\epsilon_1)T^{d/20} .
$$
Using this, it is easy to
see that the second term on the right side of (5.2) is dominated by the
right side of (5.3), with the factor 6, for large $T$.\hfill //
\vskip .3cm
Let ${\Cal S}_T$ denote the set of all $t\in [T/M-1,T]$ that are integer
multiples of $b_T\buildrel def.\over = \exp\{-T^{1/41}\}$.  By setting
$\epsilon_1 = {1\over 3}T^{-1/80}$ and summing over the exceptional
probabilities obtained from (5.3), one obtains the following uniform bound
over times $t\in {\Cal S}_T$ and rectangles $D\in {\Cal D}_{T^{1/2}}$.
\vskip .3cm
\noindent{\bf Proposition 5.2.} {\it {}For $d<4$,
$$
P\Bigl( \sup_{t\in S_T}\ \sup_{D\in {\Cal D}_{T^{1/2}}}
\vert\xi_t(D)-(\xi_0*N_t)(D)\vert\geq {1\over 3}T^{d/4-1/80}\Bigr) \leq b_T
\leqno(5.4)
$$
for sufficiently large $T$.}
\vskip .3cm
In order to deduce Proposition 2.1 from Proposition 5.2, we need to extend
(5.4) to all $t\in [T/M,T]$.  {}For this, it is enough to show that
$\vert\xi_t(D)-\xi_{t'}(D)\vert$ and $\vert
(\xi_0*N_t)(D)-(\xi_0*N_{t'})(D)\vert$ will both, with high probability,
remain small simultaneously over all $\vert t-t'\vert < b_T$, for each
given $t\in {\Cal S}_T$ and $D\in {\Cal D}_{T^{1/2}}$.  Such bounds are
provided by Lemmas 5.2 and 5.3.
\vskip .3cm
\noindent{\bf Lemma 5.2.} {\it {}For all $d,t$ and $D\in {\Cal
D}_{T^{1/2}}$,
$$
P\Bigl(\sup_{t'\in [t,t+b_T]}\vert \xi_t(D)-\xi_{t'}(D)\vert\geq 2\Bigr)
\leq (b_T)^{3/2}
\leqno(5.5)
$$
for sufficiently large $T$.}
\vskip .3cm
Since Lemma 5.3 will also be used in Section 7, it is stated somewhat more
generally than needed here.  We set $b^\delta_T = \exp\{-T^\delta\}$, and
define ${\Cal S}^\delta_T$ correspondingly.
\vskip .3cm
\noindent{\bf Lemma 5.3.} {\it {}For all $d$, $M>1,\ \delta >0,\ t\in
[T/M,T]$ and $D\in {\Cal D}_{T^{1/2}}$,
$$
P\Bigl( \sup_{t'\in [t,t+b^\delta_T]} \vert (\xi_0*N_t)(D)-(\xi_0*N_{t'})
(D)\vert\geq {1\over 3} T^{d/4-1/80}\Bigr) \leq \exp\left\{
-e^{T^\delta /4}\right\}
\leqno(5.6)
$$
for sufficiently large $T$.}
\vskip .3cm
Summing up the exceptional probabilities in (5.5) and (5.6), for $\delta =
1/41$, $t\in {\Cal S}_T$ and $D\in {\Cal D}_{T^{1/2}}$, and combining the
resulting bound with (5.4) implies that, for each $d<4$ and $M$,
$$
P\Bigl( \sup_{t\in [T/M,T]}\  \sup_{D\in {\Cal D}_{T^{1/2}}} \vert
\xi_t(D)-(\xi_0*N_t)(D)\vert\geq T^{d/4-1/80}\Bigr) \leq \exp\{ -T^{1/42}\}
$$
for sufficiently large $T$.  This implies Proposition 2.1, as desired.

The conclusion in Lemma 5.2, that the probability of $\xi_{t'}(D)$
increasing or decreasing by more than 1 over a small time interval is
very small, is not surprising.  There are several steps that require a bit
of estimation.
\vskip .3cm
\noindent{\it Proof of Lemma 5.2}. The $A$ and $B$ particles in $D$, for
the state $\xi_t$, form a subset of the particles in $D$, for the state
$\eta_t$.  So, in order for at least two of these particles in $D$ to leave
$D$ during $[t,t+b_T]$, under the process $\xi_{t'}$ (excluding
annihilations), the same must be true under
$\eta_{t'}$.  The particles in $\eta_{t'}$ execute rate-$d$ random walks.
Since $\eta_t(D)$ is Poisson with mean $2\lambda\vert D\vert$,
$$
E[(\eta_t(D))^2] = 4\lambda^2\vert D\vert^2+2\lambda\vert
D\vert
\leq 4(\lambda^2+\lambda)T^d\buildrel def.\over = \beta .
\leqno(5.7)
$$
It is not difficult to see, using (5.7), that the probability of
at least 2 jumps occuring over $[t,t+b_T]$, for those $\eta_t(D)$ particles
starting in $D$, is at most $\beta (db_T)^2$.  So, this is an upper
bound on the probability of two particles of $\xi_t$ leaving $D$ by time
$t+b_T$.

We also need an upper bound on the probability of at least 2 particles of
$\xi_t$, in $D^c$, entering $D$ by time $t+b_T$.  If one restricts $D^c$ to
those sites within distance 2 (in the sum norm) of $D$, one obtains the
same bound as above.  On the other hand, the probability of a random walk
moving $k$ steps over this time period decays like $(db_T)^k/k!$.  So, the
expected number of particles starting from distance at least 3, which enter
$D$ by time $t+b_T$, is at most
$$
2\lambda T^{d/2} \sum^\infty_{k=3} k^{d-1}(db_T)^k/k!\ << \beta
(b_T)^2
\leqno(5.8)
$$
for large $T$.  Applying Markov's inequality to this expectation and adding
the resulting probability to the other two exceptional probabilities, we
see that, for large $T$, the probability that
$$
P\Bigl( \sup_{t'\in [t,t+b_T]} \vert\xi_t(D)-\xi_{t'}(D)\vert\geq 2\Bigr)
\leq 3\beta (b_T)^2 .
$$
{}For large $T$, this is less than $(b_T)^{3/2}$.  This implies (5.5).\hfill //
\vskip .3cm
To demonstrate Lemma 5.3, we use moment generating functions and the
independence of $\xi_0(x)$ at different $x$.  Here, we abbreviate, and set
$I^\delta_{t,T} = [t,t+b^\delta_T]$.
\vskip .3cm
\noindent{\it Proof of Lemma 5.3}. In order to demonstrate (5.6), it
suffices to show that for given $\gamma > 0$ and large enough $T$,
$$
P\Bigl(\sup_{t'\in I^\delta_{t,T}} \vert (\xi_0*N_t)(0) -
(\xi_0*N_{t'})(0)\vert\geq\gamma T^{-(d/4+1/80)}\Bigr)\leq \exp\left\{
-e^{T^\delta/2}\right\},
\leqno(5.9)
$$
since $\xi_0$ is translation invariant, and we can sum over $x\in D$.  We
will use the inequality
$$
\sup_{t'\in I^\delta_{t,T}}\vert (\xi_0*N_t)(0) -
(\xi_0*N_{t'})(0)\vert\leq\sum_{x\in {\Bbb Z}^d} \vert\xi_0(-x)\vert
\sup_{t'\in I^\delta_{t,T}} \vert N_t(x)-N_{t'}(x)\vert ,
\leqno(5.10)
$$
and the fact that $\vert\xi_0(-x)\vert$, $x\in {\Bbb Z}^d$, are dominated by
independent Poisson random variables with mean $2\lambda$.  It follows
that, for $\theta >0$,
$$
\align
E\bigl[ \exp \bigl\{ \bigr. \bigr.
&\bigl. \bigl. \theta \sup_{t'\in I^\delta_{t,T}}
\vert (\xi_0*N_t)(0) - (\xi_0*N_{t'})(0)\vert\bigr\}\bigr] \\
&\leq \exp\bigl\{ 2\lambda\sum_x \bigl( \exp \bigl\{ \theta
\sup_{t'\in I^\delta_{t,T}} \vert N_t(x)-N_{t'}(x)\vert
\bigr\}-1\bigr)\bigr\}.\tag5.11
\endalign
$$

By (4.10), for large $T$, $t\in [T/M,T]$ and $x\in {\Bbb Z}^d$,
$$
\sup_{t'\in I^\delta_{t,T}} \vert N_t(x)-N_{t'}(x)\vert\leq
C_{27}b^\delta_T T^{-(d/2+1)} \left( 1+{\vert x\vert^2\over T}\right)
e^{-\vert x\vert^2/4T}
\leqno(5.12)
$$
for appropriate $C_{27}$.  Summation of both sides of
(5.12) implies that
$$
\sum_x \sup_{t'\in I^\delta_{t,T}} \vert N_t(x)-N_{t'}(x)\vert \leq
b^\delta_T .
\leqno(5.13)
$$
The right side of (5.12) is at most $b^\delta_T$.  So, for $\theta =
1/b^\delta_T$, the right side of (5.11) is at most
$$
\exp\Bigl\{ 4\lambda (b^\delta_T)^{-1}\sum_x \sup_{t'\in I^\delta_{t,T}}\vert
N_t(x)-N_{t'}(x)\vert\Bigr\} ,
$$
which, by (5.13), is at most $e^{4\lambda}$.  By Chebyshev's inequality, one
obtains that for given $\gamma > 0$ and large $T$,
$$
P\Bigl( \sup_{t'\in I^\delta_{t,T}} \vert (\xi_0*N_t)(0) -
(\xi_0*N_{t'})(0)\vert\geq \gamma T^{-(d/4+1/80)}\Bigr)
$$
$$
\leq \exp\{4\lambda -\gamma (b^\delta_T)^{-1}T^{-(d/4+1/80)}\} ,
$$
which, for large enough $T$, is at most $\exp\{-e^{T^\delta /2}\}$.  This
implies (5.9), and hence (5.6).\hfill //
\vskip .3cm
We recall that the analog of Proposition 2.3, for $d\geq 4$, was given at
the end of Section 3, after some minor changes in the argument, as
Proposition 3.3.  The other main result that has been employed in
Section 5, Proposition 
2.4, does not depend on $d$.  By replacing Proposition 2.3 by
Proposition 3.3, but otherwise reasoning the same as through Proposition 5.2,
one obtains the analog of (5.4), with the bound ${1\over
3}T^{d/4-1/80}$ replaced 
by ${1\over 3}T^{d/2-81/80}$.  Since neither Lemma 5.2 nor Lemma 5.3 depends on
$d$, one thus obtains the following analog of Proposition 2.1, for $d\geq
4$.
\vskip .3cm
\noindent{\bf Proposition 5.3.} {\it {}For $d\geq 4$ and $M>0$,
$$
P\Bigl( \sup_{t\in [T/M,T]}\  \sup_{D\in {\Cal D}_{T^{1/2}}}
\vert\xi_t(D)-(\xi_0*N_t)(D)\vert \geq T^{d/2-81/80}\Bigr)
\leq \exp\{-T^{1/42}\}
\leqno(5.14)
$$
for sufficiently large $T$.}
\vskip .3cm
Proposition 5.3 will be employed in [BrLe99].  Note that (2.4) and (5.14)
are the same, if one formally sets $d=4$ in both cases.  (The arguments
leading to Proposition 2.1, in fact, all hold for $d=4$ as well.)
\vskip .5truein
\noindent{\bf 6. Upper Bounds on $E[\xi^m_t(\cdot)]$}
\vskip .3cm
In this section, we demonstrate Proposition 2.2, which gives an upper bound
on the expected number of particles of the minority type in cubes
$D_{R_T}$, where the length $R_T$ is chosen appropriately.  We will use
this bound in Section 7, together with Proposition 2.1, to obtain the
desired estimates for $\xi^A_t$ and $\xi^B_t$, which are given in Theorem
3.  We also use
Proposition 2.2 to compute the limiting density $\rho (t)$ as $t\to\infty$,
for $d<4$, in (1.8), which is an improvement of the upper and lower
bounds given in 
(1.2).

In order to show Proposition 2.2, we rely heavily on a slight modification
of Lemma 4.6 from [BrLe91a].  {}For this, we set
$$
R_T = \delta_1(T)T^{1/2}\quad \text{and}\quad r_T = T^{7/24},
\leqno(6.1)
$$
for appropriate $\delta_1(t)$ to be specified shortly.  The above exponent
$7/24$ is itself not crucial, but needs to be slightly larger than $1/4$.
The required analog of Lemma 4.6 is given by Lemma 6.1 below.  In
contrast to our 
usual convention, we drop the assumption in the lemma that $\xi_0$ be the
difference of two Poisson random fields.
\vskip .3cm
\noindent{\bf Lemma 6.1.} {\it Assume $d<4$, and that $\xi_0$ is
translation invariant with $E[\xi^m_0 (D_{R_T})]$\hfill\break $\geq
L_1$, where $L_1\geq 
C_{28}(R_T/r_T)^d$ for appropriate $C_{28}$.  Assume that $\delta_1(T)\geq
T^{-d/48}$.  Then, for appropriate $C_{29} > 0$ (not depending on
$\delta_1(\cdot)$) and large enough $T$,}
$$
E[\xi^\#_0(D_{R_T})] - E[\xi^\#_{R^2_T} (D_{R_T})]\geq C_{29} L_1 .
\leqno(6.2)
$$
\vskip .3cm
\noindent (We abbreviate $(R_T)^d$ by $R^d_T$ and $(r_T)^d$ by $r^d_T$, here
and later on.)

Lemma 6.1 says that if $E[\xi^m_0(D_{R_T})]$, the mean number of particles
of the
minority type in $D_{R_T}$, is not too small, then, on the average, the
total number of particles lost in $D_{R_T}$, over the time interval
$[0,R^2_T]$, must also be of this order of magnitude.  The lemma, for
$d<4$, is identical to Lemma 4.6 of [BrLe91a], if $R_T$ and $r_T$ are
replaced by
$$
R'_T = \delta_1T^{1/2}\quad \text{and}\quad r'_T = \delta_2T^{1/4} ,
\leqno(6.3)
$$
where $\delta_1$ and $\delta_2$ are constants.  The purpose of chosing
$r_T$ as in (6.1), with $r_T > > r'_T$, is to permit smaller values of
$L_1$ when Lemma 6.1 (rather than Lemma 4.6) is applied.  The condition
$\delta_1(T)\geq T^{-d/48}$ is used at the end of the following sketch, as
well as in the proof of Proposition 2.2.
\vskip .3cm
\noindent{\it Sketch of the proof of Lemma 6.1.} The proofs of the lemma and
of Lemma 6.4 in [BrLe91a] are almost the same.  We begin here at the point
where they diverge, referring the reader to [BrLe91a] for the earlier part
of the argument.

Denote the left side of (6.2) by $u_T$.  The proofs of the two lemmas are
identical up through (4.30) and (4.31) of [BrLe91a], which state that
$$
C_{30}u_T/L_1\geq\cases R^2_T/r^2_T, &d=1,\\
R^2_T/(r^2_T\log r_T), &d=2,\\
R^2_T/r^3_T, &d=3,\endcases
\leqno(6.4)
$$
for appropriate $C_{30}$ and large $T$.  Substitution for $R_T$ and $r_T$,
as in (6.1), gives the bounds
$$
C_{30}u_T/L_1 \geq \cases (\delta_1(T))^2T^{5/12}, &d=1,\\
(\delta_1(T))^2T^{5/12}/\log T, &d=2,\\
(\delta_1(T))^2T^{1/8}, &d=3.\endcases
\leqno(6.5)
$$
Since $\delta_1(T)\geq T^{-d/48}$ is assumed, this implies that $u_T\geq
(C_{30})^{-1}L_1$, which, setting $C_{29} = (C_{30})^{-1}$, gives
(6.2).\hfill //
\vskip .3cm
We now show Proposition 2.2.  In addition to the lower bound
$\delta_1(T)\geq T^{-d/48}$, where $R(T) = \delta_1(T)T^{1/2}$, we also
assume here that $\delta_1(T)\to 0$ as $T\to\infty$.  The reason for this
is to be able to show, as in (2.6) of Proposition 2.2, that
$(T^{-d/4}R^d_T)^{-1}\xi^m_{t_k}(D_{R_T})$ is negligible in the limit, for
appropriate $t_k$.  As we will see in Section 8, this will not be true if
$\delta_1(T)$ is instead bounded away from $0$.  The major ingredients for
the proof of Proposition 2.2 are Lemma 6.1 and (1.2).  The main point of
the argument is that the decrease in the expected number of particles given
by (6.2) restricts the frequency with which (6.6) can occur.
\vskip .3cm
\noindent{\it Proof of Proposition 2.2}. Assume that
$$
E[\xi^m_t(D_{R_T})] > C_1\delta_1(T)T^{-d/4}R^d_T
\leqno(6.6)
$$
for large $T$ and some $t$, where $C_1\geq 1$ is fixed and will be chosen
later.  Setting $L_1 = C_1\delta_1(T)T^{-d/4}R^d_T$, it is not difficult to
see that the assumptions of Lemma 6.1 are satisfied, if time $t$ is
replaced by $0$: Clearly $\xi_t$ is translation invariant.  Also,
since $\delta_1(T)\geq T^{-d/48}$,
$$
L_1 = C_1\delta_1(T)T^{-d/4}R^d_T\geq (R_T/r_T)^d .
$$
Therefore, by (6.2),
$$
E[\xi^\#_t(D_{R_T})] - E[\xi^\#_{t+R^2_T}(D_{R_T})]\geq
C_1C_{29}\delta_1(T)T^{-d/4} R^d_T
\leqno(6.7)
$$
for appropriate $C_{29}$.

Let $t'_\ell = T/2M+\ell R^2_T$, for $\ell = 0,1,2,\dots$.  One can select
the values $t_k$ appearing in (2.6), so that they form
a subset of these $t'_\ell$.  We
argue inductively, setting $t_0 = T/2M$, and assuming that $t_{k-1}$ has
already been chosen.  Let $n_k$ denote the number of $t'_\ell$ satisfying
$$
t'_\ell -t_{k-1}\in [R^2_T,\ \delta_1(T)T - R^2_T]
\leqno(6.8)
$$
where (6.6) holds, with $t = t'_\ell$.  Also, set $t^* =
t_{k-1}+\delta_1(T)T$.  Since $E[\xi^\#_t(D_{R_T})]$ is decreasing in $t$,
it follows from (6.7), that
$$
\eqalign{E[\xi^\#_{T/2M}(D_{R_T})] &\geq E[\xi^\#_{t_{k-1}}(D_{R_T})] -
E[\xi^\#_{t^*}(D_{R_T})]\cr
&\geq C_1C_{29}n_k\delta_1(T)T^{-d/4}R^d_T.\cr}
\leqno(6.9)
$$
On the other hand, it follows from (1.2), that
$$
E[\xi^\#_{T/2M}(D_{R_T})] = 2\vert D_{R_T}\vert
\rho(T/2M)\leq C_{31}T^{-d/4}R^d_T,
\leqno(6.10)
$$
where $C_{31}$ depends on $\lambda$ and $M$.  So, comparing (6.9) and
(6.10), one obtains that
$$
n_k\leq C_{31}(C_1C_{29}\delta_1(T))^{-1} .
\leqno(6.11)
$$

The number of $t'_\ell$ satisfying (6.8) is at least
$$
\delta_1(T)T/R^2_T-3 = (\delta_1(T))^{-1}-3 .
$$
{}For $C_1 \buildrel def.\over = (2C_{31}/C_{29})\vee 1$ and
$\delta_1(T) < 1/6$, this is 
strictly larger than $n_k$.  So, for this choice of $C_1$ and large $T$,
$$
E[\xi^m_{t'_\ell}(D_{R_T})]\leq C_1\delta_1(T)T^{-d/4}R^d_T
$$
holds for at least one $t'_\ell$ satisfying (6.8).  Setting $t_k$
equal to this $t'_\ell$ produces (2.6), as desired.

Since $\delta_1(T)\to 0$ as $T\to\infty$, it is easy to check that $t_1\leq
T/M$ for large $T$.  Also, since $t'_\ell - t'_{\ell -1} = R^2_T$, one has
$t_{K-1}\leq T$ and $t_K>T$ for some $K\leq T^{d/24}$.  So, for
this choice of $K$, $[t_1,t_K]\supset [T/M,T]$ and $[t_1,t_{K-1}]\subset
[T/2M,T]$, which completes the proof
of the proposition. \hfill //
\vskip .3cm
In the remainder of the section, we analyze the asymptotic density $\rho
(t)$, for $d<4$. Recall that the initial states of $A$ and $B$ are given by
Poisson random fields with intensity $\lambda$.  We already know, as in
(1.2), that $\lambda^{-1/2}t^{d/4}\rho(t)$ is bounded above and
below; here, we show convergence and identify the limit.  The basic
procedure is as follows: The density $\rho (t)$ is given by
$E[\xi^A_t(D)]/\vert D\vert = E[\xi^B_t(D)]/\vert D\vert$. Using Proposition
2.2, the latter quantities can be
approximated by $E[\vert\xi_t(D)\vert ]/2\vert D\vert$, at appropriate $t$,
when $D$ is comparatively small.
By applying Lemma 5.1 and some elementary tail estimates to $\xi_t(D)$ and
$(\xi_0*N_t)(D)$, one can show that $E[\vert\xi_t(D)\vert ]$ is
approximated by $E[\vert (\xi_0*N_t)(D)\vert ]$.  Consequently, in order to
compute the asymptotics of $\rho (t)$, it suffices to do the same for $E
[\vert (\xi_0*N_t)(D)\vert ]/\vert D\vert$. This can be done using the
central limit 
theorem.

We begin with several elementary estimates of $\xi_t$ and $\xi_0*N_t$.
\vskip .3cm
\noindent{\bf Lemma 6.2.}  {}For all $E\subset {\Bbb Z}^d$, with $\vert E\vert
<\infty$, and all $t\geq 1$,
$$
E[(\xi_t(E))^2]\leq (\lambda^2+\lambda)\vert E\vert^2
\leqno(6.12)
$$
and
$$
E[((\xi_0*N_t)(E))^2]\leq C_{32}\lambda t^{-d/2}\vert E\vert^2
\leqno(6.13)
$$
for appropriate $C_{32}$.  Suppose that $E_t\subset D_{\epsilon (t)}$,
where $\epsilon (t) = o(t^{1/2})$.  Then,
$$
\lim_{t\to\infty} \sigma^2((\xi_0*N_t)(E_t))/(t^{-d/2}\vert E_t\vert^2) =
2\lambda (4\pi )^{-d/2} .
\leqno(6.14)
$$
\vskip .3cm
\noindent{\it Proof}.  The random variable $\vert\xi_t(E)\vert$ is
dominated by $\eta^\#_t(E)$, which is Poisson with mean $2\lambda\vert
E\vert$.  This implies (6.12).  Since
$$
(\xi_0*N_t)(E) = \sum_{x\in {\Bbb Z}^d} \xi_0(x)N_t(E-x)
$$
is the sum of independent mean-0 random variables with variances $2\lambda
(N_t(E-x))^2$,
$$
\sigma^2((\xi_0*N_t)(E)) = 2\lambda\sum_x (N_t(E-x))^2.
\leqno(6.15)
$$
The right side is at most $2\lambda\vert E\vert^2\sum_x(N_t(x))^2$.
One can check 
that, for $t\geq 1$, this is at most $C_{32}\lambda t^{-d/2}\vert E\vert^2$
for appropriate $C_{32}$, which gives (6.13), since $(\xi_0*N_t)(E)$ has
mean 0.  With a bit of estimation, one can also show that the right side of
(6.15), with $E = E_t$, is asymptotically equal to
$$
2\lambda \vert E_t\vert^2\int_{{\Bbb R}^d}(N_t(x))^2dx\ \sim\ 2\lambda (4\pi
t)^{-d/2}\vert E_t\vert^2
$$
for large $t$, which implies (6.14).\hfill //
\vskip .3cm
The following result allows us to compare the expectations
of $\xi_t(E)$ and $(\xi_0*N_t)(E)$, for appropriate $\vert E\vert$.  Its
proof employs Lemma 5.1, (6.12) and (6.13).
\vskip .3cm
\noindent{\bf Lemma 6.3.} {\it {}For $d<4$, $E\subset {\Bbb Z}^d$ with $\vert
E\vert\in [t^{19d/40},t^d]$, and large $t$,}
$$
E[\vert\xi_t(E)-(\xi_0*N_t)(E)\vert ]\leq t^{-(d/4+1/200)}\vert E\vert .
\leqno(6.16)
$$
\vskip .3cm
\noindent{\it Proof}. Let $H_1$ denote the set where
$\vert\xi_t(E)\vert\geq t\vert E\vert$ and $H_2$ the set where $\vert
(\xi_0*N_t)(E)\vert \geq t\vert E\vert$.  Also, let $Z =
\vert\xi_t(E)-(\xi_0*N_t)(E)\vert$.  Then, one can check that
$$
E[Z;H_1\cup H_2]\leq
2E[\vert\xi_t(E)\vert ;H_1]+2E[\vert (\xi_0*N_t)(E)\vert ;H_2].
\leqno(6.17)
$$
By (6.12) and the definition of $H_1$,
$$
E[\vert\xi_t(E)\vert ;H_1]\leq (\lambda^2+\lambda)t^{-1}\vert E\vert .
\leqno(6.18)
$$
Similarly, by (6.13) and the definition of $H_2$,
$$
E[\vert (\xi_0*N_t)(E)\vert ;H_2]\leq C_{32}\lambda t^{-(d/2+1)}\vert E\vert
.
\leqno(6.19)
$$
So, by (6.17)-(6.19),
$$
E[Z;H_1\cup H_2]\leq 2(C_{32}+1)(\lambda^2+\lambda)t^{-1}\vert E\vert .
\leqno(6.20)
$$

Assume that $\vert E\vert\in [t^{19d/40},t^d]$.  Substitution of $\epsilon
= t^{-1/200}/6$ in Lemma 5.1, implies that
$$
P(Z\geq {1\over 3} t^{-(d/4+1/200)}\vert E\vert)\leq 7\exp\{-t^{d/80}\}
\leqno(6.21)
$$
for large $t$.  Let $H_3$ denote the random set in (6.21).  Then, since
$\vert Z\vert\leq 2t\vert E\vert$ on $H^c_1\cap H^c_2$,
$$
E[Z;H^c_1\cap H^c_2\cap H_3]\leq 14t\exp\{-t^{d/80}\}\vert E\vert .
\leqno(6.22)
$$
Also,
$$
E[Z;H^c_3]\leq {1\over 3}t^{-(d/4+1/200)}\vert E\vert .
\leqno(6.23)
$$
Together, (6.20),(6.22) and (6.23) imply that
$$
E[Z]\leq t^{-(d/4+1/200)}\vert E\vert ,
$$
which is the desired inequality.\hfill //
\vskip .3cm
By using (6.14) and the central limit theorem, we obtain the following
limiting behavior for $E[\vert (\xi_0*N_t)(E_t)\vert ]$ and small $\vert
E_t\vert$.
\vskip .3cm
\noindent{\bf Lemma 6.4.} {\it Suppose that $E_t\subset {\Bbb Z}^d$, with
$\phi\ne E_t\subset D_{\epsilon (t)}$ and $\epsilon (t) = o(t^{1/2})$.
Then, for all $d$,}
$$
\lim_{t\to\infty} (t^{-d/4}\vert E_t\vert )^{-1}E[\vert
(\xi_0*N_t)(E_t)\vert] = (4\lambda /\pi)^{1/2}(4\pi )^{-d/4} .
\leqno(6.24)
$$
\vskip .3cm
\noindent{\it Proof}. Recall that
$$
(\xi_0*N_t)(E_t) =\sum_{x\in {\Bbb Z}^d}\xi_0(x)N_t(E_t-x),
$$
where the summands are independent mean-0 random variables.  By (6.14), for
$E_t\subset D_{\epsilon (t)}$ with $\epsilon (t) = o(t^{1/2})$,
$$
\lim_{t\to\infty} \sigma^2((\xi_0*N_t)(E_t)/(t^{-d/4}\vert E_t\vert)) =
2\lambda (4\pi)^{-d/2} .
\leqno(6.25)
$$
Since the 3rd moment of a rate-$\lambda$ Poisson random variable is
$\lambda (\lambda^2+3\lambda +1)$, and $\xi_0(x)$ is the difference of two
such random variables,
$$
E[\vert\xi_0(x)N_t(E_t-x)\vert^3]\leq 40(\lambda^3+\lambda)(N_t(E_t-x))^3.
$$
Using this, one can check that for large $t$,

$$
\sum_x E[(\vert\xi_0(x)N_t(E_t-x)\vert /(t^{-d/4}\vert E_t\vert ))^3]\leq
40(\lambda^3+\lambda)t^{-d/4} ,
\leqno(6.26)
$$
which $\to 0$ as $t\to\infty$.

By the bounds, in (6.25) and (6.26), on the variances and 3rd
moments of $(t^{-d/4}\vert E_t\vert)^{-1}\xi_0(x)N_t(E_t-x)$, and
by the Liapunov central limit theorem (see, e.g., [Ch74], page 200), it follows
that
$$
(t^{-d/4}\vert E_t\vert )^{-1}(\xi_0*N_t)(E_t)\Rightarrow (2\lambda)^{1/2}
(4\pi)^{-d/4}Z_{0,1} ,
\leqno(6.27)
$$
where $Z_{0,1}$ is a normal random variable with mean 0 and variance 1, and
$\Rightarrow$ denotes weak convergence.  We also know from (6.13) (or
(6.14)), that the 2nd moments of $(t^{-d/4}\vert E_t\vert
)^{-1}(\xi_0*N_t)(E_t)$ are bounded in $t$.  It follows from this and
(6.27), that
$$
\align
\lim_{t\to\infty} (t^{-d/4}\vert E_t\vert )^{-1}E[\vert
(\xi_0*N_t)(E_t)\vert ] &= (\lambda /\pi)^{1/2}(4\pi)^{-d/4} \int_{\Bbb R}
\vert x\vert e^{-x^2/2}dx\\
&= (4\lambda /\pi)^{1/2}(4\pi )^{-d/4} ,\endalign
$$
which implies (6.24).\hfill //
\vskip .3cm
We note in passing, that the assumption $\epsilon = o(t^{1/2})$, in Lemma
6.4, was only used for the limit on the variance in (6.14).  {}For $E_t\subset
D_{t^{1/2}}$, it is not difficult to see, using (6.15), that a lower bound
on the 2nd moment of $(\xi_0*N_t)(E)$ corresponding to (6.13), but with the
inequality reversed, still holds.  Using this, and reasoning as in Lemma
6.4, one can check that $(t^{-d/4}\vert
E_t\vert)^{-1}(\xi_0*N_t)(E_t)$ is bounded away from 0 in distribution (and
not just in mean).  In particular, the error bounds given in Theorem 3 are
of smaller order of magnitude than the terms $\xi^A_t(D)$ and $\xi^B_t(D)$
there, for $D$ of the same order as $D_{T^{1/2}}$ and $t\in [T/M,MT]$.
This is what one would 
expect, because of (1.2).

Lemmas 6.3 and 6.4 imply the following limiting behavior for
$E[\vert\xi_t(E_t)\vert ]$, when $\vert E_t\vert$ is small.
\vskip .3cm
\noindent{\bf Corollary 6.1.}  {\it Suppose $d<4$, and that $E_t\subset {\Bbb
Z}^d\cap D_{\epsilon (t)}$, with $\vert E_t\vert\geq t^{19d/40}$ and
$\epsilon (t) = o(t^{1/2})$.  Then,}
$$
\lim_{t\to\infty} (t^{-d/4}\vert E_t\vert )^{-1} E[\vert\xi_t(E_t)\vert ] =
(4\lambda /\pi)^{1/2}(4\pi )^{-d/4} .
\leqno(6.28)
$$

Using Corollary 6.1 and Proposition 2.2, we now compute the limiting
density $\rho (t)$ for $d<4$, which was given in (1.8).
\vskip .3cm
\noindent{\bf Proposition 6.1.}  {\it Assume that the initial distributions of
$A$ and $B$ particles for the process $\xi_t$ are given by independent
Poisson random fields with intensity $\lambda$.  Then, for $d<4$,}
$$
\lim_{t\to\infty}t^{d/4}\rho (t) = (\lambda /\pi)^{1/2}(4\pi)^{-d/4} .
\leqno(6.29)
$$
\vskip .3cm
\noindent{\it Proof}. Since $\rho (t)$ is decreasing in $t$, it suffices to
show (6.29) along a subsequence of times $u_1<u_2< \dots ,$ with
$\lim_{j\to\infty} u_j 
=\infty$ and $u_j-u_{j-1} = o(u_j)$.  {}For all $t$ and nonempty finite
$E\subset {\Bbb Z}^d$, one has
$$
\rho (t) = E[\xi^A_t(E) + \xi^B_t(E)]/2\vert E\vert
= E[\vert\xi_t(E)\vert ] /2\vert E\vert + E[\xi^m_t(E)]/\vert E\vert .
\leqno(6.30)
$$
Our approach will be to use Proposition 2.2 to select $t=u_j$, so that
the second term 
in (6.30) can be dropped in the limit, after scaling by $t^{-d/4}$.  The
limit in (6.29) will then follow from (6.28).

{}For given $t$, we set $T = t$ and $R_T = T^{23/48}$ in Proposition 2.2.
By Proposition 2.2, one can then choose $s\in (t-t^{47/48},t]$ such
that, for large $t$, 
$$
s^{d/4} E[\xi^m_s(D_{s^{23/24}})]/\vert
D_{s^{23/24}}\vert\leq 2C_1s^{-1/48}
\leqno(6.31)
$$
(since $\xi^m_{s}(D_{s^{23/24}})\leq\xi^m_s(D_{t^{23/24}}))$.
Employing such $s$ and (6.31), it is easy to construct $u_1<u_2<\dots ,$
with $\lim_{j\to\infty} u_j = \infty$ and $u_j-u_{j-1} = o(u_j)$, so that
$$
\lim_{j\to\infty} u^{d/4}_j E[\xi^m_{u_j}(E_{u_j})]/\vert E_{u_j}\vert = 0 ,
\leqno(6.32)
$$
where $E_t \buildrel def.\over = D_{t^{23/24}}$.  Together with
(6.30), (6.32) implies that 
$$
\lim_{j\to\infty} u^{d/4}_j \rho (u_j) =\lim_{j\to\infty}
u^{d/4}_jE[\vert\xi_{u_j}(E_{u_j})\vert ]/2\vert E_{u_j}\vert .
\leqno(6.33)
$$
Along with (6.28), (6.33) implies that
$$
\lim_{j\to\infty} u^{d/4}_j \rho (u_j) = (\lambda /\pi)^{1/2}(4\pi)^{-d/4}
.
$$
The limit (6.29) follows from this and the comment at the beginning of the
proof.
\vskip .2cm
\hfill //
\vskip .3cm
\noindent{\bf 7. Approximation of $\xi^A_t$ and $\xi^B_t$}
\vskip .3cm
In this section, we demonstrate Theorem 3, which, in $d<4$, enables us to
approximate $\xi^A_t(D)$ and $\xi^B_t(D)$ by $\sum_{x\in
D}(\xi_0*N_t)(x)^-$ and 
$\sum_{x\in D}(\xi_0*N_t)(x)^+$, respectively.  The bounds given in
(2.2)-(2.3)
of the theorem hold simultaneously over all times $t\in [T/M,MT]$ and
rectangles $D\in {\Cal D}_{MT^{1/2}}$, where $M$ is fixed, off
of an event which is of small probability when $T$ is large.  The main
tools used in deriving Theorem 3 are Propositions 2.1 and 2.2.  The
statement in Proposition 2.1 is analogous to those in Theorem 3, except that
here one approximates $\xi_t(D)$ by $(\xi_0*N_t)(D)$.  In order to derive the
estimates in Theorem 3 from Proposition 2.1, one needs to show that,
locally, the two particle types segregate, with the number of the minority
type typically being negligible.  Proposition 2.2 is employed for this.

One can break the reasoning required for this argument into two main
parts.  In Proposition 7.1, we will approximate $(\xi_0*N_t)(D)^\pm$ by
$\sum_{x\in D}(\xi_0*N_t)(x)^\pm$ for small rectangles $D$.
The reasoning here is straightforward, and is based on an
estimate that shows $(\xi_0*N_t)(x)$ does not fluctuate much locally.

We also need to approximate $\xi^A_t(D)$ and $\xi^B_t(D)$ by
$(\xi_0*N_t)(D)^-$ and $(\xi_0*N_t)(D)^+$, again for small $D$.  This
involves bounding $\xi^m_t(D)$.  Here, one needs to be more careful, since
Proposition 2.2 only holds for certain times $t_k$, and bounds are given
only on the expectation of $\xi^m_{t_k}(D)$.  The probability estimates
obtained by applying Markov's inequality to this expectation, at each
$t_k$, are much weaker than the exceptional probabilities in Proposition 2.1,
and one needs to work to keep these estimates small when summing over
different events.  One also needs to control the migration of particles
over each interval $[t_k,t_{k+1})$.  These difficulties are taken care
of in the 
work leading up to Proposition 7.2.

Together, the reasoning from the last two paragraphs shows that
$\xi^A_t(D)$ and $\xi^B_t(D)$ can be approximated by $\sum_{x\in
D}(\xi_0*N_t)(x)^-$ and $\sum_{x\in D}(\xi_0*N_t)(x)^+$, for small $D$.
Taking unions of such rectangles $D$, one obtains the corresponding
estimates for all $D\in {\Cal D}_{MT^{1/2}}$, as desired.  As mentioned
above, one needs to keep the exceptional probabilities which crop up under
control.

Lemma 7.1 is the main technical estimate needed for Proposition 7.1.  It is
employed there and elsewhere in the section, with $t' = t$; it is employed
in Section 9 with $t'\ne t$, but with $x'=x$.  The
argument is a straightforward application of moment generating functions.
\vskip .3cm
\noindent{\bf Lemma 7.1.} {\it Let $\vert t-t'\vert\leq t^\alpha$ and
$\vert x-x'\vert\leq t^{\alpha/2}$, where
$\alpha\in [1/2,1)$.  Then, for appropriate $C_{33}>0$,
$$
P(\vert (\xi_0*N_t)(x)-(\xi_0*N_{t'})(x')\vert\geq \epsilon t^{-d/4})\leq
2\exp\{-C_{33}\epsilon^2t^{1-\alpha}\}
\leqno(7.1)
$$
for large enough $t$ and all $\epsilon\in [0,1]$.}
\vskip .3cm
\noindent{\it Proof}. Fix $t,t',x$ and $x'$, and set
$$
R(y) = N_t(x-y)-N_{t'}(x'-y) .
$$
One has, for given $\theta$,
$$
\eqalign{E[\exp\theta\{(\xi_0*N_{t})(x) &- (\xi_0*N_{t'})(x')\}]\cr
&=\exp\Bigl\{ \lambda\sum_{y\in {\Bbb Z}^d}\Bigl( e^{\theta R(y)} +
e^{-\theta R(y)}-2\Bigr)\Bigr\}.\cr}
\leqno(7.2)
$$
Since $\vert t-t'\vert\leq t^\alpha$ and $\vert x-x'\vert\leq
t^{\alpha/2}$, with $\alpha\leq 1$, it follows from (4.10) and (4.13), that
$$
\vert R(y)\vert\leq C_{34}t^{(\alpha-d-1)/2},
\leqno(7.3)
$$
and from (4.11) and (4.14), that
$$
\sum_{y\in {\Bbb Z}^d} (R(y))^2\leq C_{35}t^{\alpha -d/2-1} ,
\leqno(7.4)
$$
for appropriate $C_{34}$ and $C_{35}$.  By (7.3) and (7.4), (7.2) is, for
$\vert\theta\vert\leq C_{36}t^{(d+1-\alpha)/2}$ and given $C_{36}$, at most
$$
\exp\Bigl\{ C_{37}\lambda\theta^2\sum_y (R(y))^2\Bigr\} \leq\exp
\{C_{35}C_{37}\lambda\theta^2t^{\alpha -d/2-1}\},
\leqno(7.5)
$$
for appropriate $C_{37}$.  Since $\alpha\geq 1/2$ and $\epsilon\leq 1$,
$\theta = \pm (2\lambda C_{35}C_{37})^{-1}\epsilon t^{d/4+1-\alpha}$
satisfy the above bounds on $\vert\theta\vert$.  Chebyshev's
inequality, applied to (7.2) and (7.5) for both values of 
$\theta$, implies that
$$
P(\vert (\xi_0*N_{t})(x)-(\xi_0*N_{t'})(x')\vert\geq \epsilon
t^{-d/4})\leq 2\exp\{-(4C_{35}C_{37}\lambda)^{-1}\epsilon^2t^{1-\alpha}\}.
$$
This implies (7.1), with $C_{33} = (4C_{35}C_{37}\lambda)^{-1}$.\hfill //
\vskip .3cm
We would like to replace the bound in (7.1), with $t'=t$, by one which
simultaneously 
holds over $t\in [T/M,T]$ and $x\in D_{T^{1/2}}$, if $\epsilon$ is chosen
not too small.  Such an estimate follows directly from Lemma 7.1 and Lemma
5.3.  (In some applications, $t$ will remain fixed, and only $x$ will
be allowed to 
vary.)
\vskip .3cm
\noindent{\bf Lemma 7.2.}  {\it Let $\alpha\in [1/2,1)$ and $\beta =
(1-\alpha)/8$.  Then, for all $M>1$,
$$
\align
P\bigl( \sup_{t\in [T/M,T]}\ \sup_{x\in D_{T^{1/2}}}\ \sup_{\vert
x'-x\vert \leq T^{\alpha /2}} \vert(\xi_0*N_t)(x)\bigr. &- \bigl.
(\xi_0*N_t)(x')\vert\geq T^{-d/4-\beta}\Bigr)\\
&\leq\exp\{-T^\beta\}\tag7.6
\endalign
$$
for sufficiently large $T$.}
\vskip .3cm
\noindent{\it Proof}. Set $\epsilon = MT^{-\beta}$, where $\alpha\in
[1/2,1)$ and $\beta\in (0,(1-\alpha)/8)$.  One can show that the bound in
(7.6), with ${1\over 2}\exp\{-T^\beta\}$ instead of $\exp\{-T^\beta\}$, holds
for $t\in {\Cal S}^\beta_T,\ x\in D_{T^{1/2}}$ and $\vert x'-x\vert\leq
T^{\alpha/2}$, by summing over the probabilities in (7.1).  To extend the bound
to all $[T/M,T]$ as in (7.6), one applies Lemma 5.3 with $D =\{x\}$.\hfill //
\vskip .3cm
In Section 8, we will apply Lemma 7.2 in a somewhat different setting,
where $(\xi_0*N_t)(x) =\sum_{y\in {\Bbb Z}^d}\xi_0(y)N_t(x-y)$ has been
extended to $x\in {\Bbb R}^d$.  This slight generalization causes no
changes in the statement of Lemma 7.2 or its proof.

Let ${\Cal D}^r_R$ denote those rectangles contained in the cube
$D_R$, for which the lengths of all sides are at most $r$.  Using Lemma 7.2, we
are able to compare $(\xi_0*N_t)(D)^+$ with $\sum_{x\in D}(\xi_0*N_t)(x)^+$
over such rectangles.  Proposition 7.1 is the first main ingredient for
demonstrating Theorem 7.3.
\vskip .3cm
\noindent{\bf Proposition 7.1.} {\it Let $\alpha\in [1/2,1)$ and $\beta =
(1-\alpha)/8$.  Then, for all $M>1$,
$$
\eqalign{P\Bigl( \sup_{t\in [T/M,T]}\ \sup_{D\in {\Cal
D}^{T^{\alpha/2}}_{T^{1/2}}}\Bigl| 
(\xi_0*N_t)(D)^- &- \sum_{x\in D} (\xi_0*N_t)(x)^-\Bigr|\geq\vert D\vert
T^{-d/4-\beta}\Bigr)\cr
&\leq \exp\{-T^\beta\}\cr}
\leqno(7.7)
$$
and
$$
\eqalign{P\Bigl( \sup_{t\in [T/M,T]}\ \sup_{D\in {\Cal
D}^{T^{\alpha/2}}_{T^{1/2}}} 
\Bigl| (\xi_0*N_t)(D)^+ &-\sum_{x\in D} (\xi_0*N_t)(x)^+\Bigr|\geq\vert
D\vert T^{-d/4-\beta}\Bigr)\cr
&\leq\exp\{ -T^\beta\}\cr}
\leqno(7.8)
$$
for sufficiently large $T$.}
\vskip .3cm
\noindent{\it Proof}. We consider just (7.8), since the argument for (7.7)
is the same.  Clearly, $(\xi_0*N_t)(D)^+\leq\sum_{x\in
D}(\xi_0*N_t)(x)^+$ always holds.  {}For the other direction, we may assume
that $(\xi_0*N_t)(y)\geq 0$ for some $y\in D$.  Then, on the nonexceptional
set $G$ in Lemma 7.2,
$$
(\xi_0*N_t)(x) > -T^{-d/4-\beta}\quad \text{for}\ x\in D ,
$$
for $\alpha\in [1/2,1)$ and $\beta = (1-\alpha)/8$.
It follows from this that, on $G$,
$$
\eqalign{(\xi_0*N_t)(D)^+\geq\sum_{x\in D} (\xi_0*N_t)(x) &=\sum_{x\in
D}[(\xi_0*N_t)(x)^+ -(\xi_0*N_t)(x)^-]\cr
&\geq\sum_{x\in D} (\xi_0*N_t)(x)^+ -\vert D\vert T^{-d/4-\beta}.\cr}
$$
This implies (7.8). \hfill //
\vskip .3cm
The second main ingredient for demonstrating Theorem 3 is to show that, for
suitable small $D$, $\xi^A_t(D)$ and $\xi^B_t(D)$ can be approximated by
$(\xi_0*N_t)(D)^-$ and $(\xi_0*N_t)(D)^+$.  On account of Proposition 2.1,
it will be enough to show that $\xi^A_t(D)$ and $\xi^B_t(D)$ can be
approximated by $\xi_t(D)^-$ and $\xi_t(D)^+$.  It will suffice to show that
$\xi^m_t(D)$ is typically negligible.  {}For this, we
cover $D_{T^{1/2}}$ with disjoint cubes $D^i,\ i=1,\dots ,I$, each of
length $T^{\alpha/2}$, so that $\cup_iD^i\subset D_{6T^{1/2}}$ and $I\leq
(6T^{(1-\alpha)/2})^d$.  We will employ the following result.
\vskip .3cm
\noindent{\bf Proposition 7.2.} {\it Let $d<4$, $\alpha\in [1-2/1,000,\ 1)$,
$\beta = (1-\alpha)/16$, and choose $D^i$ as above.  Then, for all $M>1$,
$$
P\Bigl( \sup_{t\in [T/M,T]} \sum^I_{i=1} \xi^m_t (D^i)\geq
T^{d/4-\beta}\Bigr) \leq T^{-\beta}
\leqno(7.9)
$$
for large enough $T$.}
\vskip .3cm
The number of particles of minimum type, $\xi^m_t(E)$, is increasing with
the set $E\subset {\Bbb Z}^d$.  Also,
$$
\xi^m_t(E) =\xi^A_t(E)-\xi_t(E)^- = \xi^B_t(E)-\xi_t(E)^+
\leqno(7.10)
$$
always holds.  The following corollary is therefore an immediate
consequence of (7.9).
\vskip .3cm
\noindent{\bf Corollary 7.1.} {\it Let $\alpha\in [1-2/1,000,\ 1),\ \beta
= (1-\alpha)/16$, and choose $D^i$ as above.  Then, for all $M>1$,
$$
P\Bigl( \sup_{t\in [T/M,T]} \sum^I_{i=1}\sup_{E\subset D^i}
(\xi^A_t(E)-\xi_t(E)^-)\geq T^{d/4-\beta}\Bigr) \leq T^{-\beta}
\leqno(7.11)
$$
and
$$
P\Bigl( \sup_{t\in [T/M,T]} \sum^I_{i=1} \sup_{E\subset D^i}
(\xi^B_t(E)-\xi_t(E)^+)\geq T^{d/4-\beta}\Bigr) \leq T^{-\beta}
\leqno(7.12)
$$
for large enough $T$.}
\vskip .3cm
Corollary 7.1, together with Proposition 2.1 and Proposition 7.1, provides
a quick proof of Theorem 3.  We therefore first show Theorem 3, and
afterwards return to the argument for Proposition 7.2.
\vskip .3cm
\noindent{\it Proof of Theorem 3}. We will demonstrate the inequality for
$\xi^B_t$; the argument for $\xi^A_t$ is the same.  By rescaling $T$, it
suffices to show this over $t\in [T/M,T]$ and $D\in {\Cal
D}_{T^{1/2}}$ (after replacing the term $1/9,000$ in the exponents with any
larger power).

Let $G_1$ denote the nonexceptional set in (7.12) and $G_2$ the
nonexceptional set in (2.4) of
Proposition 2.1.  Also, let $\alpha = 1-2/1,000$ and $\beta =
(1-\alpha)/16 = 1/8,000$.  {}For $D\in {\Cal D}_{T^{1/2}}$, we set $E^i =
D\cap D^i$. The sets $E^i$ are always rectangles.  So, by (2.4)
$$
\sum^I_{i=1} \vert\xi_t(E^i)^+ - (\xi_0*N_t)(E^i)^+\vert <
IT^{d/4-1/80}\leq T^{d/4-\beta}
\leqno(7.13)
$$
holds on $G_2$, for large $T$ and $t\in [T/M,T]$.  One can, of course,
write $\xi^B_t(D)$ as 
$\sum^I_{i=1}\xi^B_t(E^i)$.  It therefore follows from (7.13) that, on
$G_1\cap G_2$,
$$
\Bigl\vert\xi^B_t(D)-\sum^I_{i=1} (\xi_0*N_t)(E^i)^+\Bigr\vert <
2T^{d/4-\beta} ,
\leqno(7.14)
$$
for large $T$.

Let $G_3$ denote the nonexceptional set in (7.8) of Proposition 7.1.
Setting $D = E^i$ there, for each $i$, and summing over $i$ implies that
$$
\Bigl\vert\sum^I_{i=1}(\xi_0*N_t)(E^i)^+ -\sum_{x\in D}
(\xi_0*N_t)(x)^+\Bigr\vert < 6^dT^{d/4-2\beta}
$$
on $G_3$.  It follows from this and (7.14) that, on $G_1\cap G_2\cap G_3$,
$$
\Bigl\vert\xi^B_t(D)-\sum_{x\in D} (\xi_0*N_t)(x)^+\Bigr\vert < 3T^{d/4-\beta}
\leqno(7.15)
$$
for large $T$.  By (7.12), (2.4) and (7.8),
$$
P((G_1\cap G_2\cap G_2)^c)\leq 3T^{-\beta}
\leqno(7.16)
$$
for large $T$.  The bounds (7.15) and (7.16) imply (2.3), which is the
desired result.

\hfill //
\vskip .3cm
In Section 9, we will employ the following result. It is an immediate
consequence 
of Theorem 3 and Lemma 7.2.  {}For $D\in {\Cal D}_{T^\delta}$, $\delta <
1/2$, it gives the behavior of $\xi^A_T(D)$ and
$\xi^B_T(D)$ in terms of $(\xi_0*N_T)(0)^\pm$.  ($\vert D\vert$ and
$\delta$ will be chosen large enough so that the term $\vert D\vert
T^{-(d+\epsilon)/4}$ in $h(T,D)$ is dominant.)
\vskip .3cm
\noindent{\bf Corollary 7.2.} {\it Let $d<4$, $\delta < 1/2$ and $\epsilon
= 1/2-\delta$.  Set $h(T,D) = T^{d/4-1/10,000} + \vert D\vert
T^{-(d+\epsilon)/4}$.  Then, for sufficiently large $T$,
$$
P(\vert\xi^A_T(D)-\vert D\vert (\xi_0*N_T)(0)^-\vert \geq h(T,D)\ \text{for
some}\ D\in {\Cal D}_{T^\delta})\leq T^{-1/10,000}
\leqno(7.17)
$$
and
$$
P(\vert\xi^B_T(D)-\vert D\vert
(\xi_0*N_T)(0)^+\vert\geq h(T,D)\ \text{for some}\ D\in {\Cal D}_{T^\delta})
\leq T^{-1/10,000} .
\leqno(7.18)
$$}

In Section 8, we will reformulate the approximations for $\xi^A_t$ and
$\xi^B_t$ in (2.2) and (2.3) of Theorem 3 in terms of the convolutions
$(\Phi*N_t)^-$ and $(\Phi *N_t)^+$, where $\Phi$ is white noise.  This will
lead to Theorem 4, which is a generalization of Theorem 1.  To employ
Theorem 3, we need to 
restate it in its scaled format. Recall (1.6), where $^T\hat\xi_t(x)$
is defined over $x\in {\Bbb 
Z}^d_{T^{1/2}}$. The convolution
$^T\hat\xi_0*N_t$, employed below, is also taken over ${\Bbb
Z}^d_{T^{1/2}}$ and 
is defined in the obvious manner.  Since $N_{Tt}(T^{1/2}x) = T^{-d/2} N_t(x)$
always holds, it is easy to check that
$$
(^T\hat\xi_0*N_t)(x) = T^{d/4}(\xi_0*N_{Tt})(T^{1/2}x) .
\leqno(7.19)
$$
One can therefore rewrite Theorem 3 as follows.  (We assume here that
$D\subset {\Bbb R}^d$, and explicitly write
$D\cap {\Bbb Z}^d_{T^{1/2}}$ in the summation to avoid any ambiguity.)
\vskip .3cm
\noindent{\bf Theorem $3'$.} {\it {}For $d<4$, and given $M>1$,
$$
\eqalign{P\Bigl( \sup_{t\in [1/M,M]}\ \sup_{D\in {\Cal D}_M} &\vert
^T\hat\xi^A_t(D)- T^{-d/2}\sum_{x\in D\cap {\Bbb Z}^d_{T^{1/2}}}
(^T\hat\xi_0*N_t)(x)^-\vert\cr
&\geq T^{-1/9,000}\Bigr)\leq T^{-1/9,000}\cr}
\leqno(7.20)
$$
and
$$
\eqalign{P\Bigl( \sup_{t\in [1/M,M]} \sup_{D\in {\Cal D}_M} &\vert
^T\hat\xi^B_t(D)- T^{-d/2}\sum_{x\in D\cap {\Bbb Z}^d_{T^{1/2}}}
(^T\hat\xi_0*N_t)(x)^+\vert\cr
&\geq T^{-1/9,000}\Bigr)\leq T^{-1/9,000}\cr}
\leqno(7.21)
$$
hold for sufficiently large $T$.}
\vskip .3cm
We point out that, here and in Theorem 3, it is possible to replace the
summation in the formulas by the corresponding integrals.  We avoid doing
this, since $\xi_0$ (and $^T\hat\xi_0$) are discrete, which makes the
definition of $(\xi_0*N_t)(x)$ for nonlattice $x$ less natural.

The remainder of the section is devoted to demonstrating Proposition 7.2.
The main step for this is Proposition 7.3 below.  We will first show how
Proposition 7.2 follows from Proposition 7.3, and will then prove Proposition
7.3.  We introduce the following notation.  The set $\bar D^i,\ i=1,\dots
,I$, will denote all points (in ${\Bbb Z}^d$) within distance ${1\over
2}T^{\alpha/2}$ of 
$D^i$ in the max norm.  That is, $\bar D^i$ is the rectangle
centered at the middle of $D^i$, and having length $2T^{\alpha/2}$.  The times
$t_k$, in Proposition 7.3 and later on, will be those
given in Proposition 2.2 for $R_T = \lfloor T^{3\alpha/2 -1}\rfloor$.
(Recall that $\lfloor x\rfloor$ denotes the integral part of $x$.)
\vskip .3cm
\noindent{\bf Proposition 7.3.} {\it Let $d<4$, $\alpha\in [1-2/1,000,\ 1)$,
$\beta = (1-\alpha)/16$, and choose $\bar D^i$ and $t_k$ as above.
Then, for all 
$M>1$ and all $k = 1,\dots ,K-1$,
$$
P\Bigl( \sum^I_{i=1} \xi^m_{t_k}(\bar D^i)\geq T^{d/4-\beta}\Bigr)
\leq T^{-20\beta} 
\leqno(7.22)
$$
for large enough $T$.}
\vskip .3cm
In order to show Proposition 7.2 from Proposition 7.3, we will show that
the probability is small that any particle moves from outside $\bar D^i$, at
time $t_k$, to inside $D^i$, over times $[t_k,t_k+T^\gamma]$, for any $i$.
(The constant $\gamma$ will be chosen so that $\gamma < \alpha$ and
$\cup^{K-1}_{k=1}[t_k,t_k+T^\gamma]\supset [T/M,T]$.) {}For this estimate, we let
$W^{i,k}$ denote the number of particles (of either type) which violate
this condition, for given $i$ and $k$.
\vskip .3cm
\noindent{\bf Lemma 7.3.} {\it {}For $\gamma\in (0,\alpha)$ and each $k$,}
$$
P\Bigl( \sum^I_{i=1} W^{i,k}\ne 0\Bigr) \leq
\exp\{-C_{38}T^{(\alpha-\gamma)\wedge(\alpha /2)}\}
\leqno(7.23)
$$
for sufficiently large $T$ and appropriate $C_{38} > 0$.
\vskip .3cm
\noindent{\bf Proof}. Let $X_t$ be a continuous time rate-$d$ simple random
walk in $d$ dimensions, with $X_0 = 0$.  Using moment generating functions
and the reflection principle, it is not difficult to show that
$$
P\Bigl( \sup_{t\leq u}\vert X_t\vert\geq\vert x\vert \Bigr) \leq
4d\exp\Bigl\{ -C_{39}\vert
x\vert \Bigl( {\vert x\vert\over u}\wedge 1\Bigr)\Bigr\}
\leqno(7.24)
$$
for each $u$ and appropriate $C_{39}$.  Moreover, at time $t_k$, each
type of particle is dominated by a Poisson random 
field with intensity $\lambda$.  Therefore, for given $i$ and $k$,
$$
P(W^{i,k}\ne 0)\leq E[W^{i,k}]\leq 2\lambda \sum_{\vert x\vert\geq
T^{\alpha/2}} 4d\exp\Bigl\{ -C_{39}{\vert x\vert\over 2}\Bigl( {\vert
x\vert\over 2T^\gamma}\wedge 1\Bigr)\Bigr\} .
$$
{}For given $\alpha$ and $\gamma$, with $\gamma <\alpha$, this is
$$
\leq \exp\{ -C_{40}T^{\alpha /2} (T^{\alpha/2-\gamma}\wedge 1)\} = \exp\{
-C_{40}T^{(\alpha-\gamma)\wedge(\alpha/2)}\}
$$
for large $T$ and appropriate $C_{40}$.  Since $I\leq
(6T^{(1-\alpha)/2})^d$, this gives (7.23), for $C_{38} < C_{40}$, after
summing over $i$.\hfill //
\vskip .3cm
We now demonstrate Proposition 7.2, assuming Proposition 7.3.
\vskip .3cm
\noindent{\it Proof of Proposition 7.2}.  Let $\gamma\buildrel def.\over =
(9/8)\alpha -1/8 = 1-18\beta$.  Since $\alpha < 1$, one has $\gamma <
\alpha$.  Lemma 7.3 implies that, for a given $k$, $k=1,\dots ,K-1$,
no particles move from 
outside $\bar D^i$ to inside $D^i$, over ${\Cal T}_k\buildrel def.\over =
[t_k,t_k+T^\gamma]$, with overwhelming probability.  Under this event,
$$
\sum^I_{i=1} \xi^m_t(D^i)\leq\sum^I_{i=1} \xi^m_{t_k}(\bar D^i)\quad\
\text{for}\ 
t\in [t_k,t_k+T^\gamma].
$$
One can check that $(\alpha -\gamma)\wedge (\alpha /2) = 2\beta$.  So, by
Proposition 7.3 and Lemma 7.3,
$$
P\Bigl( \sup_{t\in {\Cal T}_k} \sum^I_{i=1} \xi^m_t (D^i)\geq
T^{d/4-\beta}\Bigr)\leq
T^{-20\beta} + \exp\{-C_{38}T^{2\beta}\}
\leqno(7.25)
$$
for large enough $T$ and appropriate $C_{38}>0$.

We wish to extend the range of $t$ in (7.25) to $[T/M,T]$, in order to
obtain (7.9). We first note ${\Cal T}_k$ has length $T^\gamma$.  By
assumption, $R_T = \lfloor T^{3\alpha/2 -1}\rfloor$, and so, for $t_k$
chosen as in 
Proposition 2.2,
$$
t_k-t_{k-1}\leq T^{(3\alpha -1)/2}\ \ \text{for all}\ k .
$$
{}For $\gamma$ chosen as above, $\gamma > (3\alpha -1)/2$, and so
$t_k-t_{k-1} << T^\gamma$ for 
large $T$.  Also, $[T/M,T]\subset [t_1,t_K]$, where $K$ is as in
Proposition 2.2.

It follows from these observations, that an appropriate collection
${\Cal T}_{k_1},\ {\Cal T}_{k_2},\dots$ of at most $2T^{1-\gamma} =
2T^{18\beta}$ of these intervals covers $[T/M,T]$.  Applying these ${\Cal
T}_{k_i}$ in (7.25) and summing the probabilities implies that
$$
P\Bigl( \sup_{t\in [T/M,T]} \sum^I_{i=1} \xi^m_t(D^i)\geq
T^{d/4-\beta}\Bigr) \leq 4T^{-2\beta}
$$
for large enough $T$.  This gives (7.9).\hfill //
\vskip .3cm
We now demonstrate Proposition 7.3.  The proposition would be a simple
application of Proposition 2.2 and Chebyshev's inequality if one replaced
the upper bound $T^{-20\beta}$ in (7.22) by a multiple of
$T^{-7\beta}$.  The bound 
$T^{-7\beta}$ is too coarse, however, to apply in the proof of Proposition
7.2, since it needs to be applied over $2T^{18\beta}$ events.

One can get around this problem by covering each of the cubes $\bar D^i$
with disjoint cubes $D^{i,j},\ j=1,\dots ,J$, each of length
$\lfloor T^{\alpha'/2}\rfloor$, where  $\alpha' = 3\alpha -2$.  We do this with
$\cup_jD^{i,j}$ contained in the cube having the same center as $\bar D^i$
but with length $6T^{\alpha/2}$, so that $J\leq (7T^{(\alpha-\alpha')/2})^d =
(7T^{1-\alpha})^d$, and we center each $D^{i,j}$ so that it is a translate
(in ${\Bbb Z}^d$) of $D_{\lfloor T^{\alpha'/2}\rfloor}$.  We will
apply Proposition 2.2 to 
each of these cubes $D^{i,j}$, which gives smaller bounds on the
exceptional probabilities than if we applied it to $\bar D^i$ directly.
We will then show that the fluctuation of $\xi^m_{t_k}(D^{i,j})$ over
$j = 1,\dots ,J$, for fixed $i$, is small enough so that one retains
the improved bounds in (7.22) as well, when one replaces
$\sum_{i,j}\xi^m_{t_k}(D^{i,j})$ by $\sum_i\xi^m_{t_k}(\bar D^i)$.

Before presenting the proof of Proposition 7.3, we first give two
preliminary lemmas.  The first lemma says that the fluctuation in
$\xi_t (D^{i,j})$, between different $D^{i,j}$ with the same $i$, will
typically be small.
\vskip .3cm
\noindent{\bf Lemma 7.4.} {\it Let $d<4$, $\alpha\in [1-2/1,000,\ 1)$, $\beta =
(1-\alpha)/16$ and choose $D^{i,j}$ as above.  Then, for all $M>1$,
$$
P\Bigl( \max_j\xi_t(D^{i,j})-\min_j \xi_t(D^{i,j})\geq
3T^{d(2\alpha'-1)/4-2\beta}\Bigr)\leq 2\exp\{-T^{2\beta}\}
\leqno(7.26)
$$
for sufficiently large $T,\ t\in [T/M,T]$ and all $i$.}
\vskip .3cm
\noindent{\it Proof}. Let $G_1$ denote the nonexceptional set in (7.6), and
fix $i$.  Since $\vert D^{i,j}\vert = \lfloor T^{\alpha'/2}\rfloor^d$ for all
$j$,
$$
\max_j (\xi_0*N_t)(D^{i,j}) -\min_j (\xi_0*N_t)(D^{i,j}) <
T^{d(2\alpha'-1)/4-2\beta}
\leqno(7.27)
$$
on $G_1$ for each $t\in [T/M,T]$.  Let $G_2$ denote the nonexceptional
set in (2.4).  Then, for 
each $j$,
$$
\vert\xi_t(D^{i,j}) - (\xi_0*N_t)(D^{i,j})\vert < T^{d/4-1/80}
\leqno(7.28)
$$
holds on $G_2$.  Since $\alpha\geq 1-2/1,000$, one can check that the
bound in (7.27) is larger than that in (7.28).  So, (7.27) and (7.28) imply
that
$$
\max_j \xi_t(D^{i,j})-\min_j \xi_t(D^{i,j}) < 3T^{d(2\alpha'-1)/4-2\beta}
\leqno(7.29)
$$
on $G_1\cap G_2$.  By (7.6) and (2.4),
$$
P((G_1\cap G_2)^c)\leq\exp\{-T^{2\beta}\} + \exp\{-T^{1/42}\}
\leq 2\exp\{-T^{2\beta}\}.
\leqno(7.30)
$$
for large $T$.  The bound in (7.26) follows from (7.29) and (7.30).\hfill
//
\vskip .3cm
Let $H_i$ denote the set of realizations where, for a given $t$,
$\xi_t(D^{i,j})\geq 0$ holds for all $j$ or $\xi_t(D^{i,j})\leq 0$
holds for all $j$.  The next 
lemma says that, on $H^c_i$, the quantities $\xi^m_t(\bar D^i)$ and
$\sum_j\xi^m_t(D^{i,j})$ will typically be close.
\vskip .3cm
\noindent{\bf Lemma 7.5.} {\it Let $d<4$, $\alpha\in [1-2/1,000,\ 1)$,
$\beta = (1-\alpha)/16$, and choose $\bar D^i$ and $D^{i,j}$ as above.
Then, for all $M>1$,
$$
P\Bigl( \xi^m_t(\bar D^i)-\sum^J_{j=1}\xi^m_t(D^{i,j})\geq
3\cdot 7^dT^{d(2\alpha-1)/4-2\beta};H^c_i\Bigr)
\leq 2\exp\{ -T^{2\beta}\}
\leqno(7.31)
$$
for sufficiently large $T$, $t\in [T/M,T]$, and all $i$.}
\vskip .3cm
\noindent{\it Proof}. On $H^c_i$, $\max_j \xi_t(D^{i,j}) > 0$ and
$\min_j\xi_t(D^{i,j}) < 0$.  So, on $H^c_i$,
$$
\max_j \vert\xi_t(D^{i,j})\vert < \max_j \xi_t(D^{i,j}) -\min_j
\xi_t(D^{i,j}) .
$$
It therefore follows from Lemma 7.4, that
$$
P(\max_j \vert\xi_t(D^{i,j})\vert\geq 3T^{d(2\alpha'-1)/4-2\beta};H^c_i)\leq
2\exp\{-T^{2\beta}\}.
\leqno(7.32)
$$
That is, the numbers of $A$ and $B$ particles are almost the same over each
$D^{i,j}$.

One always has that
$$
\eqalign{\xi^m_t(\bar D^i) &\leq \xi^m_t(\cup_j D^{i,j}) =
\Bigl( \sum_j\xi^A_t(D^{i,j})\Bigr)\wedge \Bigl(
\sum_j\xi^B_t(D^{i,j})\Bigr)\cr
&\leq\sum_j (\xi^A_t(D^{i,j})\vee \xi^B_t(D^{i,j}))=\sum_j
(\xi^m_t(D^{i,j})+\vert\xi_t(D^{i,j})\vert ),\cr}
\leqno(7.33)
$$
since $\bar D^i\subset \cup_j D^{i,j}$ and the sets $D^{i,j}$, $j=1,\dots$,
$J$ are disjoint.  Off of $H_i$ and the exceptional set in (7.32), this is
$$
<\sum_j\xi^m_t(D^{i,j}) + 3JT^{d(2\alpha'-1)/4-2\beta}
< \sum_j\xi^m_t(D^{i,j}) + 3\cdot 7^dT^{d(2\alpha-1)/4-2\beta}.
\leqno(7.34)
$$
Together with (7.32), (7.33) and (7.34) imply (7.31).\hfill //
\vskip .3cm
We now prove Proposition 7.3, and hence complete the proof of Theorem 3.
The argument combines Proposition 2.2 with Lemma 7.5.
\vskip .3cm
\noindent{\it Proof of Proposition 7.3}. Fix $M$, and let $t_k,\
k=1,\dots ,K-1$, be chosen as in Proposition 2.2.  We apply (2.6) to each
$D^{i,j}$, $i=1,\dots ,I$ and $j=1,\dots ,J$, choosing $R_T =
\lfloor T^{\alpha'/2}\rfloor$, and using the translation invariance of
$\xi_t$.  Since $\delta_1(T)\leq T^{(\alpha'-1)/2} = T^{-24\beta}$, one has
$$
E\Bigl[ \sum_{i,j} \xi^m_{t_k} (D^{i,j})\Bigr]\leq
C_1IJT^{d(2\alpha'-1)/4-24\beta}
\leq C_142^dT^{d/4-24\beta},
\leqno(7.35)
$$
for large $T$ and appropriate $C_1$.  By Markov's inequality, this implies
$$
P\Bigl( \sum_{i,j}\xi^m_{t_k}(D^{i,j})\geq C_142^dT^{d/4-2\beta}\Bigr) \leq
T^{-22\beta} .
\leqno(7.36)
$$

We wish to replace $\sum_{i,j}\xi^m_{t_k}(D^{i,j})$ with
$\sum_i\xi^m_{t_k}(\bar D^i)$, in (7.34).  On the set $H_i$ defined above
Lemma 7.5, for a given $t_k$,
$$
\xi^m_{t_k}(\bar D^i)\leq\sum_j\xi^m_{t_k}(D^{i,j})
\leqno(7.37)
$$
clearly holds, since the minimum type over each $D^{i,j},\ j=1,\dots ,J$,
is the same.  On $H^c_i$, we can employ Lemma 7.5 (with the value of $M$
being twice the value chosen here).  Summing the exceptional
probabilities in (7.31) over $i = 1,\dots ,I$, and combining this with
(7.37), gives
$$
\align
P\bigl( \sum_i \xi^m_{t_k}(\bar D^i) -\sum_{i,j}\xi^m_{t_k}(D^{i,j})\geq
3\cdot 42^d\bigr. &\bigl.T^{d/4-2\beta}\bigr)\\
&\leq 2\cdot 6^dT^{d(1-\alpha)/2}\exp\{ -T^{2\beta}\}\tag7.38
\endalign
$$
for large $T$ and $k=1,\dots ,K-1$.  Together with (7.36), (7.38) implies that
$$
P\Bigl( \sum_i\xi^m_{t_k}(\bar D^i)\geq T^{d/4-\beta}\Bigr)
\leq T^{-20\beta} ,
$$
which is (7.22).\hfill //
\vskip .5truein
\noindent{\bf 8. Convergence of $^T\hat\xi_t$ to $(2\lambda)^{1/2}(\Phi
*N_t)$}
\vskip .3cm
Theorem 1 states that $^T\hat\xi_t$ converges to
$(2\lambda)^{1/2}(\Phi *N_t)$ as $T\to\infty$, where $\Phi$ is the mean-0
generalized Gaussian random field with covariance given by (1.3).  
In the mathematical physics and other literature, $\Phi$ is referred to as
white noise.  In this section, we first discuss white noise and its connection
with Brownian sheet.  We then demonstrate Theorem 4, which is a
generalization of Theorem 1.

Brownian sheet is the higher dimensional analog of Brownian motion.
Brownian sheet $W(x)$, with $x = (x_1,\dots ,x_d)\in [0,\infty)^d$, is the
real-valued Gaussian process with mean 0 and covariances
$$
E[W(x)W(y)] =\prod^d_{j=1} (x_j\wedge y_j) .
\leqno(8.1)
$$
A version of this process exists where almost all realizations are
continuous in $x$; we will, from now on, automatically choose this version.
Various more refined sample path properties of Brownian sheet have been
investigated in [OrPr73] and the references given there.

One can extend the domain of $W$ from $[0,\infty)^d$ to ${\Bbb R}^d$.
One can do this by employing $2^d$ independent copies $W^1,\dots ,W^{2d}$,
of $W$, each defined on $[0,\infty)^d$, where each $W^n$ is identified
with a different one of the 
$2^d$ orthants.  Writing $x^{\text{pos}} = (\vert x_1\vert,\dots ,\vert
x_d\vert)$, where $x=(x_1,\dots ,x_d)$, one can extend
$W(x)$ to $x\in {\Bbb R}^d$ by setting $W(x) = W^n(x^{\text{pos}})$, where
$W^n$ is the copy identified with the orthant
containing $x$.  As before, almost all realizations of $W(x)$ will be
continuous in $x$.

Let $D\subset {\Bbb R}^d$ be a finite rectangle.  Denote by $x$ and
$y$ the vertices where all of the coordinates are maximized, respectively,
minimized, and, for $z$ any vertex of $D$, let $\nu (z)$ denote the
number of coordinates $z$ shares with $y$.  We set
$$
\Phi (D) =\sum_z(-1)^{\nu (z)}W(z).
\leqno(8.2)
$$
When $y=0$, one has $\Phi (D) = W(x)$.  The operator $\Phi$ defines a
mean-0 generalized Gaussian random field, with covariance satisfying
$$
E[\Phi (D_1)\Phi (D_2)] = \vert D_1\cap D_2\vert
\leqno(8.3)
$$
for pairs of rectangles $D_1$ and $D_2$.  This is the same
expression as (1.3).  Thus, (8.2) gives a representation for white noise in
terms of Brownian sheet.  One may check (8.3) by decomposing $D_1$ and
$D_2$ into unions of rectangles in the different orthants, and
then writing these as differences of rectangles, with each rectangle having
the origin as a vertex.  One then applies the formula (8.1) to each
such pair.  The 
white noise $\Phi$ given by (8.2) is almost surely continuous in $D$ as its
vertices are varied; we shall henceforth assume this continuity for $\Phi$.

We note that for $W$ defined here, $W(x) = 0$ for any $x\in {\Bbb R}^d$ with
at least one coordinate equal to 0.  Thus, $W(x)$ is ``centered'' at 0.
One can recenter $W(x)$ at any given point $y$ by setting
$$
W^y(x) = W(x)-\sum^d_{j=1}g_j(x) ,
$$
where each $g_j(x),\ j=1,\dots ,d$, is an appropriate random function
which is constant in 
its $j$th coordinate.  (First set $g_1(x) = W(x)$ for each $x$ sharing its
first coordinate with $y$, then set $g_2(x) = W(x)-g_1(x)$ for each $x$
sharing its second coordinate with $y$, etc.)  Replacing $W$ by $W^y$ does
not change the corresponding operator $\Phi$.  ({}For instance, subtracting
$g_j(x)$ from $W(x)$ does not change $\Phi (D)$ in (8.2), since the effect,
on the right side, on pairs of vertices differing only in the $j$th
coordinate, cancels out due to the factor $\nu (z)$.)

There exists a unique process $V$, with domain ${\Bbb Z}^d$ and centered at 0,
which corresponds to $\xi_0$ as $W$ does to $\Phi$.  That is,
$$
\xi_0(D) =\sum_z(-1)^{\nu (z)}V(z)
\leqno(8.4)
$$
for all rectangles $D$ having vertices $z\in {\Bbb Z}^d$.  When $x$ and $y$
are chosen as above (8.2), with $y=0$, one has $\xi_0(D) = V(x)$.  As
in the definition for 
$^T\hat\xi$ in (1.6), we set
$$
^T\hat V(z) = V(T^{1/2}z)/T^{d/4} ,
\leqno(8.5)
$$
for $z\in {\Bbb Z}^d_{T^{1/2}}$.  It then follows that
$$
^T\hat\xi_0(D) =\sum_z {(-1)^{\nu (z)}} ^T\hat V(z)
\leqno(8.6)
$$
for all $D$ having vertices $z\in {\Bbb Z}^d_{T^{1/2}}$.

We want to be able to compare $^T\hat V$ with $W$, when $T\to\infty$.
{}For this, we need to define $V$ at nonlattice points, which we do by
interpolating.  The most natural way is to use the following scheme.
{}For $d=1$, the interpolation between $y$ and $y+1$ will be linear.  For
$d=2$ and $x\in y+(0,1]^2$, with $y = (y_1,y_2)\in {\Bbb Z}^2$ and $x =
(x_1,x_2)$, set
$$
\eqalign{V(x) &= V(y) +
[V(y+(1,0))-V(y)](x_1-y_1)+[V(y+(0,1))-V(y)](x_2-y_2)\cr 
&\ +[V(y+(1,1)) - V(y+(1,0)) - V(y+(0,1))+V(y)](x_1-y_1)(x_2-y_2).\cr}
$$
This interpolation is linear along the sides of $y + [0,1]^2$, and has a
correction term that is proportional to the area of the rectangle
given by the opposing vertices $y$ and $x$.  The interpolation for $d>2$ is
defined analogously, with the new volume term, for each added dimension,
corresponding to the right side of (8.4).

We will also find it useful to
extend $\xi_0(y)$ to all of ${\Bbb R}^d$. We do this by setting
$_e\xi_0(x) =\xi_0(y)$ for 
$x\in y-[0,1)^d$ and $y\in {\Bbb Z}^d$.  One then has, by (8.4),
$$
_e\xi_0(x) = {\partial^dV(x)\over\partial x_1\dots \partial x_d}
$$
for $x$ with noninteger coordinates.  Since $_e\xi_0(y)$ will serve the role
of a density, the ``extension'' $^T_e\hat\xi_0$ to ${\Bbb R}^d$ needs to be
scaled differently than $^T\hat\xi_0$ to be useful.  {}For $x\in {\Bbb
R}^d$, we set 
$$
^T_e\hat\xi_0(x) = T^{d/4}\ _e\xi_0(T^{1/2}x);
\leqno(8.7)
$$
at $x\in {\Bbb Z}^d_{T^{1/2}}$, one has $^T_e\hat\xi_0(x) = T^{d/2}\
^T\hat\xi_0(x)$.  This scaling gives
$$
^T_e\hat\xi_0(x) = {\partial^d\ ^T\hat V(x)\over\partial x_1\dots\partial x_d}
\leqno(8.8)
$$
for $x$ with coordinates not in ${\Bbb Z}_{T^{1/2}}$.

Since $W$ is defined on ${\Bbb R}^d$, convolution with respect to $N_t$ will
be interpreted as an integral over ${\Bbb R}^d$, i.e.,
$$
(W*N_t)(x) =\int_{{\Bbb R}^d} W(x-y)N_t(y)dy,\quad \text{for}\ x\in {\Bbb R}^d
.
\leqno(8.9)
$$
Since the growth of $\vert W(x)\vert$ can be controlled as $\vert
x\vert\to\infty$, the integral is almost surely well defined and finite.
(One can employ estimates similar to those in the proofs of Lemmas 8.2 and
8.3.)  We define $^T\hat V*N_t$ in the same way.  We will also employ the
convolutions $W*N'_t$ and $^T\hat V*N'_t$, where
$$
N'_t(x) \buildrel def.\over = {\partial^dN_t(x)\over \partial
x_1\dots\partial x_d} = (-1)^d\ {x_1\cdots x_d\over t^d} N_t(x) .
\leqno(8.10)
$$
Using (8.8)-(8.10) and integrating by parts in each direction, one can
check that 
$$
(^T\hat V*N'_t)(x) = (^T_e\hat\xi_0*N_t)(x)\quad \text{for}\ x\in {\Bbb R}^d
.
\leqno(8.11)
$$

It follows from (8.2), that
$$
(\Phi *N_t)(D) =\sum_z(-1)^{\nu (z)}(W*N_t)(z),
\leqno(8.12)
$$
where $z$ are the vertices of the rectangle $D$.  Using (8.1), one can
check that $\Phi *N_t$ scales according to
$$
(\Phi *N_t)(D) = M^{-d/4}(\Phi *N_{Mt})(M^{1/2}D),
\leqno(8.13)
$$
for $M>0$.  We write $(\Phi *N_t)(x)$ for
the density of $(\Phi *N_t)(D)$.  Then,
$$
(\Phi *N_t)(x) = (W*N'_t)(x);
\leqno(8.14)
$$
(8.14) can be thought of as a formal $d$-fold integration by parts.  It
follows from (8.14), that
$(\Phi *N_t)(x)$ is continuous in $(t,x)$ for almost all realizations.
One can check that 
since $W$ is a family of Gaussian random variables, so is $\Phi *N$.  A
simple computation shows that, for any $(t_1,x_1)$ and $(t_2,x_2)$,
$$
E[(\Phi *N_{t_1})(x_1)(\Phi *N_{t_2})(x_2)] =\int_{{\Bbb R}^d}
N_{t_1}(x_1-y)N_{t_2}(x_2-y)dy.
\leqno(8.15)
$$
{}For $t_1=t_2=t$, this equals $(4\pi t)^{-d/2}\exp\{-\vert
x_1-x_2\vert^2/4t\}$.  In particular, $\sigma^2(\Phi *N_t)(x) = (4\pi
t)^{-d/2}$. 

The main result in this section is Theorem 4, which is a stronger version
of Theorem 1.  The main tools for demonstrating the theorem are Theorem
$3'$ of Section 7 and Proposition 8.1.  Let $\bar D_M = [-M/2,M/2]^d$, for
$M>0$.  (We will use this notation throughout the remainder of the section.)
Proposition 8.1 states that
$(^T_e\hat\xi_0 *N_t)(x)$, with $(t,x)\in (0,1]\times \bar D_1$, converges
weakly to $(2\lambda)^{1/2}(\Phi *N_t)(x)$ as $T\to\infty$.  Convergence is
with respect to the uniform topology on compact sets of $C((0,1]\times
\bar D_1)$, the space of continuous functions on $(0,1]\times \bar D_1$.
\vskip .3cm
\noindent{\bf Proposition 8.1.} {\it {}For all $d$,
$$
(^T_e\hat\xi_0 *N_.)(\cdot)\Rightarrow (2\lambda)^{1/2}(\Phi
*N_.)(\cdot)\quad \text{as}\ T\to\infty ,
\leqno(8.16)
$$
on $(0,1]\times \bar D_1$.}

Since the demonstration of Proposition 8.1 requires several steps, it will
be postponed until the latter part of the section.  The proposition has two
useful corollaries.  The first will be employed in Section 9.
\vskip .3cm
\noindent{\bf Corollary 8.1.}  {\it {}For all $d$,
$$
T^{d/4}(\xi_0*N_T)(0)\Rightarrow (2\lambda)^{1/2}(4\pi )^{-d/4}Z_{0,1}\quad
\text{as}\ T\to\infty ,
\leqno(8.17)
$$
where $Z_{0,1}$ has the standard normal distribution.}
\vskip .3cm
\noindent{\it Proof}. By the observations immediately above and below
(8.15), $(\Phi *N_1)(0)$ is normally distributed with mean 0 and variance
$(4\pi)^{-d/2}$.  Also, using (8.7), one can check that
$$
(^T_e\hat\xi_0 *N_1)(0) = T^{d/4}(_e\xi_0*N_T)(0).
\leqno(8.18)
$$
So, (8.17) will follow from (8.16) and (8.18) once we show that
$$
T^{d/4}[(\xi_0*N_T)(0) - (_e\xi_0*N_T)(0)]\to 0\quad \text{as}\ T\to\infty .
\leqno(8.19)
$$
(These two convolutions are not identical, since $\xi_0*N_T$ is a sum over
${\Bbb Z}^d$, whereas $_e\xi_0*N_T$ is an integral over ${\Bbb R}^d$.)
Since $_e\xi_0(x) = \xi_0(y)$ for $x\in y-[0,1)^d$ and $y\in {\Bbb Z}^d$,
$(_e\xi_0*N_T)(0)$ is the average of $(\xi_0*N_T)(z)$ over $z\in [0,1)^d$.
By Lemma 7.2, with $\alpha = 1/2$, all of these values are, off of a set of
probability $\exp\{-T^{1/16}\}$, within $T^{-d/4-1/16}$ of
$(\xi_0*N_T)(0)$, and hence so is $(_e\xi_0*N_T)(0)$.  This implies
(8.19).\hfill //
\vskip .3cm
The space $C((0,1]\times D_1)$ admits a complete, separable metric.
Consequently, weak convergence in (8.16) implies the corresponding
convergence in probability, if, for each $T$, $\xi_0$ and $\Phi$ are coupled
appropriately (see e.g., [Bi71], Theorem 3.3).  That is
$$
(^T_e\hat\xi_0*N_t)(x) - (2\lambda)^{1/2}(\Phi *N_t)(x)\to 0\quad \text{in
probability as}\ T\to\infty ,
$$
uniformly in $(t,x)$ on compact sets.
Integration of $(^T_e\hat\xi_0 *N)(x)$ and $(\Phi *N)(x)$ over rectangles
$D\subset D_1$, in the $d$ space variables, immediately produces the
following result.  By $(^T_e\hat\xi *N_t)^\pm (D)$ and $(\Phi *N_t)^\pm
(D)$, we mean the functions $(^T_e\hat\xi *N_t)(x)^\pm$, respectively
$(\Phi *N_t)(x)^\pm$, integrated over $x\in D$.
\vskip .3cm
\noindent{\bf Corollary 8.2.} {\it Fix $d,\ M>1$ and $\epsilon > 0$.
{}For each $T$, there exist copies of $\xi_0$ and $\Phi$, so that
$$
\lim_{T\to\infty} P\Bigl( \sup_{t\in [1/M,1]} \sup_{D\in {\Cal D}_1} \vert
(^T_e\hat\xi_0 *N_t)^-(D)-(2\lambda)^{1/2}(\Phi *N_t)^-(D)\vert
>\epsilon\Bigr) = 0
\leqno(8.20)
$$
and}
$$
\lim_{T\to\infty} P\Bigl( \sup_{t\in [1/M,1]} \sup_{D\in {\Cal D}_1} \mid
(^T_e\hat\xi_0 *N_t)^+(D)-(2\lambda)^{1/2}(\Phi *N_t)^+(D) >\epsilon\Bigr)
= 0.
\leqno(8.21)
$$

Theorem $3'$, in Section 7, states that, for large $T$ in $d<4$,
$^T\hat\xi^A_t(D)$ and $^T\hat\xi^B_t(D)$ are, over $t\in [1/M,M]$ and
$D\in {\Cal D}_M$, uniformly approximated by $\sum_{x\in D\cap
{\Bbb Z}^d_{T^{1/2}}} (^T\hat\xi_0*N_t)(x)^-$, respectively
$\sum_{x\in D\cap {\Bbb Z}^d_{T^{1/2}}} (^T\hat\xi_0 *N_t)(x)^+$.
Putting this together with Corollary 8.2 produces the following uniform
approximations on $\xi^A_t(D)$ and $\xi^B_t(D)$.
\vskip .3cm
\noindent{\bf Theorem 4.} {\it Let $d<4$, and fix $M>1$ and $\epsilon >
0$.  {}For each $T$, there exist copies of $\xi_0$ and $\Phi$, so that
$$
\lim_{T\to\infty} P\Bigl( \sup_{t\in [1/M,M]} \sup_{D\in {\Cal D}_M} \vert
^T\hat\xi^A_t(D)-(2\lambda)^{1/2}(\Phi *N_t)^-(D)\vert >\epsilon\Bigr) = 0
\leqno(8.22)
$$
and}
$$
\lim_{T\to\infty} P\Bigl( \sup_{t\in [1/M,M]} \sup_{D\in {\Cal D}_M}
\vert ^T\hat\xi^B_t(D)-(2\lambda)^{1/2}(\Phi *N_t)^+(D)\vert >\epsilon
\Bigr) = 0 . 
\leqno(8.23)
$$
By fixing $t$ and $D$, one obtains Theorem 1 as a special
case of Theorem 4.  One can also phrase Theorem 4 in terms of weak
convergence, if one wishes.  In (8.22)-(8.23), it suffices to consider
those $D$ with a vertex at the origin; the four quantities in (8.22)-(8.23)
can each be written as a function of the opposite vertex.  The limits can
then be formulated in terms of weak convergence, with respect to the
uniform topology on compact sets, of continuous functions from
$(0,\infty)\times {\Bbb R}^d$ to ${\Bbb R}^2$.

We now demonstrate Theorem 4.  The main estimates that are needed are
supplied by Theorem $3'$ and Corollary 8.2.  One also needs to do some
tedious, but straightforward comparisons to coordinate these estimates.
\vskip .3cm
\noindent{\it Proof of Theorem 4}. Since the arguments are the same for
both parts, we will just show (8.23).  Note that
$$
^T\hat\xi_t(D) = M^{-d/4}\ ^{T/M}\hat\xi_{Mt}(M^{1/2}D)\quad \text{for}\
M>1 .
\leqno(8.24)
$$
Using (8.13) and (8.24), it is enough to show (8.23) for $t\in [1/M,1]$ and
$D\in {\Cal D}_1$. We claim, it suffices to show this for $D'\in
{\Cal D}_1$, with $D'$ having vertices in ${\Bbb Z}^d_{T^{1/2}}$; we denote
the set by ${\Cal D}_{1,T^{1/2}}$.  To see this, note that for any $D\in
{\Cal D}_1$, there is a $D'\in {\Cal D}_{1,T^{1/2}}$, with $D'\cap {\Bbb
Z}^d_{T^{1/2}} = D\cap {\Bbb Z}^d_{T^{1/2}}$, so that the volume of the
symmetric difference $D\Delta D'$ is at most $2d/T^{1/2}$.  Since
$(\Phi *N_t)(x)$ is continuous in $(t,x)$ (for almost all realizations),
it is bounded on $[1/M,1]\times D_1$, and so the same is true for
$(\Phi *N_t)^+(D\Delta D')/\vert D\Delta D'\vert$, for $|D\Delta
D'|>0$.  Replacing $D$ by $D'$, in (8.23), thus changes the second
term by a random multiple of 
$T^{-1/2}$, and leaves the first term unchanged.

The display (8.23), with $D\in {\Cal D}_{1,T^{1/2}}$ replacing $D\in {\Cal
D}_1$, will follow immediately from Theorem $3'$ and Corollary 8.2, once we
have shown that
$$
\align
\sup_{t\in [1/M,1]} \sup_{D\in {\Cal D}_{1,T^{1/2}}} \Bigl|
T^{-d/2}\sum_{y\in D\cap {\Bbb Z}^d_{T^{1/2}}}\Bigr.
&(^T\hat\xi_0 *N_t)(y)^+\\
&\Bigl.-\int_D (^T_e\hat\xi_0 *N_t)(x)^+ dx\Bigr| \to 0\tag8.25
\endalign
$$
in probability as $T\to\infty$.  The reasoning here is similar to that for
(8.19), in the proof of Corollary 8.1.  The convolutions on the left and
on the right are somewhat different, since $^T_e\hat\xi_0$
on the left is defined on ${\Bbb Z}^d_{T^{1/2}}$ (and so $*$ is defined as a
sum), whereas $^T\hat\xi_0$ on the right has been extended to ${\Bbb R}^d$
(and so $*$ is defined as an integral).  Since $_e\xi_0(x) =\xi_0(y)$ for
$x\in y -[0,1)^d$ and $y\in {\Bbb Z}^d$, one can check that $(^T_e\hat\xi_0
*N_t)(x)$ is the average of $(^T\hat\xi_0 *N_t)(z)$ over $z\in
x+[0,T^{-1/2})^d$.  By (7.19) and Lemma 7.2, with $\alpha = 1/2$, all of
these values are, off of a set of probability $\exp\{ -T^{1/16}\}$ (not
depending on $t$ or $x$), within $T^{-1/16}$ of $(^T\hat\xi_0*N_t)(y)$, and
hence so is $(^T_e\hat\xi_0 *N_t)(x)$.  Integration over $D$ produces an
error that is at most $T^{-1/16}$.  This implies (8.25).

\hfill //
\vskip .3cm
We now set out to show Proposition 8.1.  The basic idea is as follows.  By
the invariance principle in Proposition 8.2, $^T\hat V(\cdot)\Rightarrow
(2\lambda)^{1/2}W(\cdot)$ as $T\to\infty$.  Convolution by $N'$ will
be a continuous map if one 
truncates the tail of $N'$, and the error involved in this truncation can be
made as small as desired.  By employing a standard weak convergence result,
one can therefore show that
$$
(^T\hat V*N'_{\cdot})(\cdot)\Rightarrow
(2\lambda)^{1/2}(W*N'_{\cdot})(\cdot)
\leqno(8.26)
$$
on $(0,1]\times\bar D_1$.  On account of (8.11) and (8.14),
this is equivalent to (8.16).

In Proposition 8.2, the domains of $^T\hat V$ and $W$ are each ${\Bbb R}^d$.
Convergence is with respect to the uniform topology on compact sets of
$C({\Bbb R}^d)$.
\vskip .3cm
\noindent{\bf Proposition 8.2.} {\it {}For all $d$,}
$$
^T\hat V(\cdot)\Rightarrow (2\lambda)^{1/2} W(\cdot)\quad \text{as}\
T\to\infty .
\leqno(8.27)
$$

The proof of Proposition 8.2, for general $d$, is similar to the proof
for $d=1$, 
which is a special case of the standard invariance principle.
Rather than go into the proof in detail, we will briefly discuss related
results in [Ku73].  We will also summarize the key steps for $d=1$, in
[Bi68], and will 
indicate how they extend to general $d$.

{}For $d=1$, Proposition 8.2 is just a special case of the invariance
principle, since $\xi_0(x)$, $x\in {\Bbb Z}$, are i.i.d. with mean 0
and variance 
$2\lambda$.  {}For $d=2$, the analog of (8.27), on $[0,1]^2$, is
shown in Theorem 3 of [Ku73] for i.i.d. random
variables with finite variance.  By intersecting $\bar D_M$, $M>0$,
with each of 
the four quadrants and treating each of them separately, the problem in
Proposition 8.2, for $d=2$, reduces to this setting.  The extension of the
invariance principle, to $d\geq 3$, is briefly discussed in [Ku73].
Theorem 3, in [Ku73], is an application of an earlier result in the paper
on the convergence of Banach-valued random variables to the corresponding
Banach-valued Brownian motion.  In the proof of the theorem, the
dimension of the 
parameter space is effectively lowered from 2 to 1, by treating the
evolution of the process along one of the directions as the state of a
corresponding process, whose state space consists of continuous functions
on $[0,1]$ with the uniform topology.  The extension, from $d$ to $d+1$,
involves a similar induction argument.  It is sketched in [Ku73].

If one wishes, one can instead show Proposition 8.2 directly, using
reasoning corresponding to that given in [Bi68] for the one dimensional
case.  The two main steps are as follows.  One first shows convergence of the
joint distributions.  This part proceeds as in $d=1$, with the extension of
the dimension not affecting the argument.  One then needs to show
tightness of the 
sequence of probability measures.  {}For this, one can apply the analog
of Theorem 8.3 of 
[Bi68].  The main condition in Theorem 8.3 is that, for
fixed $\epsilon > 0$, the probability of an oscillation, of size at last
$\epsilon$, occurring over any
interval of length $\delta$, converges to 0 as $\delta\to 0$.  The analog
of this condition in our setting, where such intervals are replaced by
cubes of length $\delta$, is not difficult to show.  Since $\xi_0(x)$ is
the difference of two independent mean-$\lambda$ Poisson distributions at
each $x$, it is symmetric.  So, using the reflection principle (as in Lemma
8.3, below), one can show that the probability of a large fluctuation
occurring in 
a cube is only a fixed multiple of the probability of $^T\hat V$ attaining
a large value at one of the vertices.  One can show this is small by
employing second moments together with Chebyshev's inequality (as in Lemma
8.2).

We will break the work in showing (8.26), and hence Proposition 8.1, into
the following two steps.  Lemma 8.1 is the analog of (8.26), with the
integral associated with $*$ restricted to the domain $\bar D_M$.  Convergence
is with respect to the uniform topology on compact sets for $C((0,1]\times
D_1)$.
\vskip .3cm
\noindent{\bf Lemma 8.1.} {\it {}For all $d$ and $M$,}
$$
\int_{\bar D_M}{^T\hat V}(y)N'_.(\cdot -y)dy\Rightarrow (2\lambda)^{1/2}
\int_{\bar D_M} W(y)N'_.(\cdot
-y)dy\quad \text{as}\ T\to\infty .
\leqno(8.28)
$$

The other step says that the contribution to the integral associated with
$*$ is insignificant off of the set $\bar D_M$.
\vskip .3cm
\noindent{\bf Lemma 8.2.} {\it {}For all $d$ and $\epsilon > 0$,
$$
P\Bigl( \sup_{t\in [0,1]} \sup_{x\in \bar D_1} \int_{\bar D^c_M} \vert ^T\hat
V(y)N'_t(x-y)\vert dy\geq\epsilon \Bigr) \to 0
\leqno(8.29)
$$
uniformly in $T$ as $M\to\infty$.  Similarly,
$$
P\Bigl( \sup_{t\in [0,1]} \sup_{x\in \bar D_1} \int_{\bar D^c_M} \vert
W(y)N'_t(x-y)\vert dy\geq\epsilon \Bigr)\to 0
\leqno(8.30)
$$
as $M\to\infty$.}
\vskip .3cm
The proof of Proposition 8.1 is immediate from Lemmas 8.1 and 8.2.
Together, (8.28)-(8.30) imply that
$$
\int_{{\Bbb R}^d}{^T\hat V}(y)N'_.(\cdot -y)dy\Rightarrow
(2\lambda)^{1/2}\int_{{\Bbb R}^d} 
W(y) N'_.(\cdot -y)dy ,
\leqno(8.31)
$$
where $\Rightarrow$ denotes weak convergence with respect to the uniform
topology on compact subsets of $C((0,1]\times \bar D_1)$.  This is (8.26),
which is equivalent to (8.16).
\vskip .3cm
\noindent{\it Proof of Lemma 8.1}. {}For fixed $M>0$, let $\Xi$ denote the
linear map from the space of continuous functions on $\bar D_M$ to the space of
continuous functions on $(0,1]\times \bar D_1$, which is given by
$$
(\Xi (g))(t,x) = \int_{\bar D_M} g(y)N'_t(x-y)dy.
$$
Denote by $\Xi^L$ the map $\Xi$ where the domain of $\Xi (g)$
is restricted to $[1/L,1]\times \bar D_1$.  We write $\Vert\cdot\Vert_{\bar
D_M}$ and $\Vert\cdot\Vert_{[1/L,1]\times \bar D_1}$ for the uniform
metrics on the spaces 
of continuous functions on $\bar D_M$ and $[1/L,1]\times \bar D_1$,
respectively. 
Since
$$
\Vert\Xi^L(g)\Vert_{[1/L,1]\times \bar D_1}\leq C(L)\Vert
g\Vert_{\bar D_M}
$$
for all $g$ and appropriate $C(L)$, the map $\Xi^{L}$ is
continuous.  Hence, so is $\Xi$.

We know from Proposition 8.2, that $^T\hat V(\cdot) \Rightarrow
(2\lambda)^{1/2}W(\cdot)$ as
$T\to\infty$.  Since $\Xi$ is continuous, it follows from a standard
result on weak convergence, that
$$
(\Xi (^T\hat V))(\cdot ,\cdot)\Rightarrow (2\lambda)^{1/2}(\Xi (W))(\cdot
,\cdot)\quad \text{as}\ T\to\infty .
\leqno(8.32)
$$
(See, e.g., Theorem 5.1 of [Bi68].)  The limit (8.32) is equivalent to
(8.28).\hfill //
\vskip .3cm
In order to demonstrate Lemma 8.2, we first need the following bounds on
$\sup_{\vert y\vert\leq j}\vert{^T\hat V}(y)\vert$, $j\in {\Bbb Z}^+$,
and $\sup_{\vert y\vert\leq r} \vert W(y)\vert,\ r\in {\Bbb R}^+$.  Both
bounds are repeated applications of the reflection principle, and employ
Chebyshev's inequality with the 2nd moments of
$^T\hat V$ and $W$.  Here, ${\bold 1} = (1,\dots ,1)\in {\Bbb R}^d$ and
$\vert\cdot\vert_\infty$ denotes the max norm on ${\Bbb R}^d$.
\vskip .3cm
\noindent{\bf Lemma 8.3.} {\it {}For all $d,\ \epsilon > 0$ and $j\in {\Bbb
Z}^+$,
$$
P\Bigl( \sup_{\vert y\vert_\infty\leq j}\vert ^T\hat V(y)\vert\geq\epsilon
\Bigr) \leq 2\cdot 4^dP\bigl({^T\hat V}({\bold 1}j)\geq\epsilon\bigr) .
\leqno(8.33)
$$
Similarly, for all $r>0$,}
$$
P\Bigl( \sup_{\vert y\vert_\infty\leq r} \vert W(y)\vert\geq\epsilon\Bigr) \leq
2\cdot 4^dP(W({\bold 1}r)\geq\epsilon).
\leqno(8.34)
$$
\vskip .3cm
\noindent{\it Proof}. Since the proofs are similar, we demonstrate just
(8.34).  We first note that it is enough to show
$$
P\Bigl( \sup_{y\in H_r} W(y)\geq\epsilon\Bigr)\leq 2^dP\Bigl( W({\bold
1}r)\geq\epsilon\Bigr) ,
\leqno(8.35)
$$
where $H_r$ is the set of $y$ with $\vert y\vert_\infty\leq r$ and having
nonnegative coordinates.  (The different orthants contribute an additional
factor $2^d$, and the absolute value contributes the factor 2.)

In order to show (8.35), we repeatedly apply the reflection principle.  Let
$H^1_r$ denote the subset of points $y\in H_r$, $y = (y_1,\dots ,y_d)$,
with $y_1 = r$.  We will show that
$$
P\Bigl( \sup_{y\in H_r}W(y)\geq\epsilon\Bigr) \leq 2P\Bigl( \sup_{y\in
H^1_r}W(y)\geq\epsilon\Bigr) .
\leqno(8.36)
$$
To obtain (8.36), set
$$
Y_1 = \inf\{y_1:W(y) =\epsilon\ \ \text{for some}\ y\in H_r\}\wedge r .
\leqno(8.37)
$$
Also, let $Y = (Y_1,\dots ,Y_d)$ denote the smallest point in $H_r$ at
which this occurs (ordering $y_1$ first, then $y_2,\dots$, down through
$y_d$).  One has $W(Y) =\epsilon$ unless $Y_1 = r$.  Set $Y' =(r,Y_2,\dots ,
Y_d)$.  When the first coordinate is varied, with the other
coordinates remaining fixed, 
the increments of $W(y)$ are independent, and so it follows by symmetry that
$$
P(W(Y')\geq W(Y))\geq 1/2 .
\leqno(8.38)
$$
This implies (8.36).

Proceeding inductively, one can continue in the same manner as above.
Starting on $H^{i-1}_r$, the $d-i+1$ dimensional face, where the first
$i-1$ coordinates are all fixed and equal $r$, one can reflect, in the $i$th
direction, to obtain
$$
P\Bigl( \sup_{y\in H^{i-1}_r} W(y)\geq\epsilon \Bigr) \leq 2P
\Bigl( \sup_{y\in H^i_r} W(y)\geq\epsilon \Bigr) .
\leqno(8.39)
$$
Continuing until $i=d$, where $H^d_r =\{{\bold 1} r\}$, one obtains (8.35)
after putting the inequalities together.

We point out that we are implicitly employing the strong Markov property
here.  To justify its application for $i=1$, for example, let $W_1(y_1)$,
$y_1\in [0,r]$, denote the process whose value at $y_1$ is the map $W^1$ from
$[0,r]^{d-1}$ to ${\Bbb R}$, with
$$
W^1(y_2,\dots ,y_d) =W(y_1,\dots ,y_d),
$$
i.e., $W_1(y_1)$ is given by the ``slice'' of $W$ taken where the first
coordinate is $y_1$.  Assign the uniform topology on $C([0,r]^{d-1})$ to
these states.  Then, one can check that $W_1(y_1)$ is Feller continuous,
and is hence strong Markov.  The justification for the other steps is
analogous.\hfill //
\vskip .3cm
Employing Lemma 8.3, we now show Lemma 8.2.  This will complete the proof
of Proposition 8.1.
\vskip .3cm
\noindent{\it Proof of Lemma 8.2}.  Since the proofs of (8.29) and (8.30)
are similar, we will do just (8.30).  We first note that $W({\bold 1}j)$
is normally distributed with 2nd moment $j^d$.  So, by Chebyshev's
inequality,
$$
P(W({\bold 1}j)\geq e^j)\leq j^d/e^{2j} .
\leqno(8.40)
$$
Set $E_j =\{y:\vert y\vert_\infty\in (j-1,j]\}$.  It follows from (8.34)
and (8.40), that
$$
P\Bigl( \sup_{y\in E_j} \vert W(y)\vert\geq e^j\Bigr) \leq 2(4j)^d/e^{2j} .
\leqno(8.41)
$$

{}For $x\in \bar D_1$, $y\in E_j$ and $j\geq 4$, one has $\vert
x-y\vert_\infty\geq j/2$.  It is therefore easy to check that, for large
enough $j$,
$$
\sup_{t\in [0,1]} \sup_{x\in \bar D_1\atop y\in E_j} \vert N'_t(x-y)\vert <
e^{-j^2/9} .
\leqno(8.42)
$$
Together with (8.41), (8.42) implies that
$$
P\Bigl( \sup_{t\in [0,1]} \sup_{x\in \bar D_1} \int_{E_j} \vert
W(y)N'_t(x-y)\vert 
dy\geq (2j)^de^{j-j^2/9}\Bigr) \leq 2(4j)^d/e^{2j}
\leqno(8.43)
$$
for large enough $j$.  Since $\cup^\infty_{j=\lfloor M\rfloor/2} E_j\supset
\bar D^c_M$, it follows from (8.43), that
$$
P\Bigl( \sup_{t\in [0,1]} \sup_{x\in \bar D_1} \int_{\bar D^c_M} \vert
W(y)N'_t(x-y)\vert dy\geq e^{-M^2/40}\Bigr) \leq 2^{-M} ,
$$
for large enough $M$.  This clearly implies (8.30).\hfill //
\vskip .5truein
\centerline{\bf 9. Local Behavior of $\xi_t$}
\vskip .3cm
In this section, we are interested in the local (or microscopic)
behavior of $\xi_t$ for 
large $t$, after space has been appropriately scaled.  On account of (1.8)
(or (1.2)), the scaling given in (1.10), by $\check \xi_t(E) =
\xi_t(t^{1/4}E)$, is the right scaling.  The goal here is to
demonstrate Theorem 2, which 
states that $\check \xi_t$ converges to a mixture of Poisson random fields.

We will divide the work needed for Theorem 2 into two main steps.  In
Proposition 9.1, we break the evolution of $\xi_r,\ r\in [0,t]$, into
the time intervals 
$[0,s]$ and $[s,t]$, where $t = s+s^\alpha$ and $\alpha > 0$.  (Later on, we
will choose $\alpha$ to be slightly less than 1.)  We examine there the
behavior of $_s\tilde\eta_r$, which will be a slight variant of the process
$_s\eta_r$ defined in Section 2, where the annihilation between particles is
suppressed starting at time $s$. The interval $[s,t]$ has been chosen so
that it is (1) long enough so that particles will mix locally to form
Poisson random fields, but (2) short enough so that the density
changes insignificantly.  In Proposition 9.2, we restore the annihilation
between particles over $[s,t]$.  On account of (2), $\xi_t$ and
$_s\tilde\eta_t$ will typically be the same locally. It will follow
from Proposition 9.2, that the Laplace functionals of $\xi_t$ converge to
the desired limits, which implies Theorem 2.

In order to show Propositions 9.1 and 9.2, we employ the following two
lemmas.  {}For Lemma 9.1, we partition ${\Bbb R}^d$ by cubes $D^1,D^2,\dots
,$ each having length $\lfloor s^\beta\rfloor$, where $\beta\in
(0,\alpha /2)$.  We then 
have $\vert D^i\vert =\lfloor s^\beta\rfloor^d$ for each $i$; the
exact choice of the 
translates does not matter.  (Later on, $\beta$ will be slightly less than
1/2, with $\beta =\alpha -1/2$.)  The following bound limits the local
fluctuations of $K_{s^\alpha}(x)$ as $x$ varies.  (Recall that $K_t(x)$ is
the random walk kernel introduced in Section 2.)
\vskip .3cm
\noindent{\bf Lemma 9.1.} {\it Fix $d$.  {}For $\alpha > 0$, $\beta\in
(0,\alpha /2)$ and large $s$,
$$
\sum^\infty_{i=1} \max_{y,y'\in D^i} \vert
K_{s^\alpha}(x-y)-K_{s^\alpha}(x-y')\vert\leq C_{41}s^{-\alpha
/2-(d-1)\beta}
\leqno(9.1)
$$
holds for all $x$ and appropriate $C_{41}$.}
\vskip .3cm
\noindent{\it Proof}. It follows without difficulty from (4.13), that for
large $s$,
$$
\sum^\infty_{i=1} \max_{y,y'\in D^i} \vert
N_{s^\alpha}(x-y)-N_{s^\alpha}(x-y')\vert\leq C_{42}s^{-\alpha
/2-(d-1)\beta}
\leqno(9.2)
$$
holds for all $x$ and appropriate $C_{42}$.  It also follows immediately
from (4.8), that
$$
\sum^\infty_{i=1} \max_{y\in D^i} \vert N_{s^\alpha}(x-y) - K_{s^\alpha}
(x-y)\vert\leq C_{11}s^{-\alpha /2-d\beta}
\leqno(9.3)
$$
for all $x$.  Together, (9.2) and (9.3) imply (9.1).\hfill //
\vskip .3cm
We will employ Lemma 9.1 in Proposition 9.1, in the form of the following
corollary.  We need the following terminology.  Set
$$
m^i(x) = \min_{y\in D^i}K_{s^\alpha}(x-y),\quad M^i(x) = \max_{y\in D^i}
K_{s^\alpha} (x-y),
\leqno(9.4)
$$
where $D^i,\ i=1,2,\dots$ are given above.  Also, let $I$ denote the
smallest set of indices of these cubes so that $D_{2s^\gamma}\subset
\cup_{i\in I} D^i$, where $\gamma\in (\alpha /2,1]$ is fixed.  (Later on,
$\gamma$ will be slightly less than 1/2.)
\vskip .3cm
\noindent{\bf Corollary 9.1.} {\it Fix $d$.  {}For $\alpha > 0$, $\beta\in
(\alpha /4,\alpha /2),\ \gamma\in (\alpha /2,1]$ and large $s$,
$$
(1-C_{43}s^{-\alpha /2+\beta})s^{-d\beta} \leq\sum_{i\in I}
m^i(x)\leq\sum^\infty_{i=1} M^i(x)\leq (1+C_{43}s^{-\alpha
/2+\beta})s^{-d\beta}
\leqno(9.5)
$$
holds for all $x\in D_{s^\gamma}$ and appropriate $C_{43}$.}
\vskip .3cm
\noindent{\it Proof.} We consider the lower bound.  Let $A^i(x)$ denote the
average of $K_{s^\alpha}(x-y)$ over $y\in D^i$.  By (9.1),
$$
\sum^\infty_{i=1} (A^i(x)-m^i(x))\leq C_{41}s^{-\alpha /2-(d-1)\beta}
\leqno(9.6)
$$
holds for large $s$.  On the other hand, since $\vert D^i\vert\leq s^{d\beta}$
for all $i$,
$$
\sum^\infty_{i=1} A^i(x)\geq s^{-d\beta}\sum_{y\in {\Bbb Z}^d} K_{s^\alpha}
(y) = s^{-d\beta} .
\leqno(9.7)
$$
Also, using a simple large deviations estimate, one can check that
$$
\sum_{i\not\in I} A^i(x)\leq 2s^{-d\beta}\sum_{y\not\in D_{s^\gamma}}
K_{s^\alpha} (y)\leq \exp\{ -C_{44}s^{2\gamma -\alpha}\}
$$
for large $s$, $x\in D_{s^\gamma}$ and appropriate $C_{44}$.  Since
$\gamma > \alpha /2$, 
this last term goes to 0 quickly as $s\to\infty$.  Together, the above three
estimates imply the lower bound in (9.5).  Only (9.6) and (9.7) are needed
for the upper bound, which follows in the same manner with the inequalities
reversed.  (The upper bound holds for all $x$.)\hfill //
\vskip .3cm
The process $_s\tilde\eta_r$ alluded to earlier is the same as $_s\eta_r$,
except that, at time $s$, one kills all of the particles outside
$\cup_{i\in I}D^i$, where $D^i$ and $I$ are specified before Corollary 9.1.
Over $(s,t]$, the process evolves without interaction between particles.
The following lemma says that this modification will typically not affect
the configuration of particles in $D_{s^\gamma}$, at time $t$, which
contains the regions we are interested in.
\vskip .3cm
\noindent{\bf Lemma 9.2.}  {\it Fix $d$.  {}For $\alpha > 0$ and $\gamma\in
(\alpha /2,1]$,
$$
P(_s\tilde\eta_t^A(x)\ne {_s\eta^A_t(x)}\ \text{for some}\ x\in
D_{s^\gamma})\leq\exp\{-C_{45}s^{2\gamma-\alpha}\}
\leqno(9.8)
$$
and
$$
P(_s\tilde\eta^B_t(x)\ne {_s\eta^B_t(x)}\ \text{for some}\ x\in
D_{s^\gamma})\leq\exp\{-C_{45}s^{2\gamma-\alpha}\}
\leqno(9.9)
$$
for large $s$ and appropriate $C_{45}$.}
\vskip .3cm
\noindent{\it Proof.}  Since $D_{2s^\gamma}\subset \cup_{i\in I}D^i$,
it suffices, for each case, to calculate an upper bound on the
expected number of particles in $D^c_{2s^\gamma}$ at time $s$, which are in
$D_{s^\gamma}$ at time $t$, for the process $\eta_r$.  One can then apply
Markov's inequality.  The configurations of $A$ and $B$ particles, at time
$s$, are dominated by Poisson random fields with intensity $\lambda$.  So,
the argument reduces to elementary large deviation estimates on the
probability of a particle moving distance greater than $s^\gamma$ over the
time interval $[s,s+s^\alpha]$.  The estimates required here are similar to
those in Lemma 7.3.\hfill //
\vskip .3cm
Proposition 9.1 provides information on the behavior near $0$ of
$_s\tilde\eta_t$. (The replacement of $_s\eta_t$ by $_s\tilde\eta_t$
simplifies the reasoning somewhat.)  The main tools in the proof of
Proposition 9.1 are Corollaries 7.2 and 9.1.
Corollary 7.2 allows us to approximate $\xi^A_s(D^i)$ and $\xi^B_s(D^i)$ by
$\lfloor s^\beta\rfloor^d (\xi_0*N_s)(0)^-$ and
$\lfloor s^\beta\rfloor^d(\xi_0*N_s)(0)^+$ for all $i\in I$, since
both $\beta$ and 
$\gamma$ will be slightly less than 1/2.  On account of Corollary 9.1, if
one ignores annihilations over $(s,t]$, the probabilities of such
particles being at a given site $x\in D_{s^\gamma}$ at time $t$, do
not depend much on 
their exact locations within $D^i$ at time $s$.  Together with some
approximation, this behavior will imply (9.10) and (9.11) as $t\to\infty$.
\vskip .3cm
\noindent{\bf Proposition 9.1.} {\it Assume that $d<4$, and set $\alpha =
1-10^{-5}$ and $\gamma = 1/2-10^{-6}$.  Then, for $f\in C^+_c({\Bbb R}^d)$,}
$$
\align
E\Bigl[\exp\Bigl\{&-\sum_{x\in {\Bbb Z}^d} f(x/t^{1/4})
_s\tilde\eta^A_t (x)\Bigr\}\mid {\Cal F}_0\Bigr] \\
&- \exp\Bigl\{t^{d/4}(\xi_0*N_t)(0)^-\int_{{\Bbb R}^d}
(e^{-f(x)}-1)dx\Bigr\} \\
&\to 0\quad \text{in probability as}\ t\to\infty\tag9.10
\endalign
$$
{\it and}
$$
\align
E\Bigl[\exp\Bigl\{&-\sum_{x\in {\Bbb Z}^d} f(x/t^{1/4}) _s\tilde\eta^B_t
(x)\Bigr\}\mid {\Cal F}_0\Bigr] \\
&- \exp\Bigl\{ t^{d/4}(\xi_0*N_t)(0)^+ \int_{{\Bbb R}^d}
(e^{-f(x)}-1)dx\Bigr\} \\
&\to 0\quad \text{in probability as}\ t\to\infty .\tag9.11
\endalign
$$
\vskip .3cm
\noindent{\it Proof}. We will demonstrate just (9.11), since the argument
for (9.10) is the same.  Set $\beta = 1/2-10^{-5}$, and let $D^1,D^2,\dots$
denote the cubes of length $\lfloor s^\beta\rfloor$, and $I$ the set of
indices that were introduced above.  Then, $D^I\buildrel def.\over =
\cup_{i\in I} D^i\subset D_{3s^\gamma}$.  Also, set $\epsilon = 10^{-6}$,
and let $H_s$ denote the set of realizations where
$$
\vert\xi^B_s(D^i) -\lfloor s^\beta\rfloor^d(\xi_0*N_s)(0)^+\vert < s^{d\beta
-(d+\epsilon /2)/4}
\leqno(9.12)
$$
for all $i\in I$.  Since $\gamma < 1/2$, it follows from Corollary 7.2,
that
$$
P(H_s)\to 1\quad \text{as}\ s\to\infty .
\leqno(9.13)
$$

Using (9.13), we first obtain upper bounds for the left side of (9.11).
Since the particles of $_s\tilde\eta_r$ execute independent random walks
over $(s,t]$, and $_s\tilde\eta_s = \xi_s$ on $D^I$,
$$
\align
E\Bigl[ \exp\Bigl\{ -\sum_{x\in {\Bbb Z}^d} &f(x/t^{1/4}) _s\tilde\eta^B_t
(x)\Bigr\}\mid {\Cal F}_s\Bigr] \\
(9.14)\hskip .6in &= \prod_{y\in D^I} \Bigl[ \sum_{x\in {\Bbb Z}^d} \exp\{
-f(x/t^{1/4})\}K_{s^\alpha}(x-y)\Bigr]^{\xi^B_s(y)} \\
&= \prod_{y\in D^I} \Bigl[ 1+\sum_{x\in {\Bbb Z}^d} (\exp\{
-f(x/t^{1/4})\}-1)K_{s^\alpha} 
(x-y)\Bigr]^{\xi^B_s(y)}.
\endalign
$$
Define $m^i(x)$ as in (9.4) and set $Z_s = [\lfloor s^\beta\rfloor^d
(\xi_0*N_s)(0) -s^{d\beta -(d+\epsilon /2)/4}]^+$.  On $H_s$, $Z_s$ is a lower
bound of $\xi^B_s(D^i)$, for all $i\in I$.  Grouping all $B$ particles
for each $D^i$ together, one can therefore check that, on $H_s$, (9.14) is
$$
\align
&\leq \prod_{i\in I} \Bigl[ 1+\sum_{x\in {\Bbb Z}^d} (\exp\{
-f(x/t^{1/4})\}-1)m^i(x)\Bigr]^{Z_s} \\
&\leq\prod_{i\in I}\exp\Bigl\{ Z_s\sum_{x\in {\Bbb Z}^d}
(\exp\{ -f(x/t^{1/4})\}-1)m^i(x)\Bigr\} \\
&= \exp\Bigl\{ Z_s\sum_{x\in {\Bbb Z}^d} \Bigl[
(\exp\{-f(x/t^{1/4})\}-1)\sum_{i\in I} m^i(x)\Bigr]\Bigr\}.\tag9.15
\endalign
$$
By applying the lower bound for $\sum_{i\in I}m^i(x)$ in Corollary 9.1,
with $x\in D_{s^\gamma}$, one obtains the upper bound
$$
\exp\Bigl\{ (1-C_{43}s^{-\alpha /2+\beta})s^{-d\beta} Z_s\sum_{x\in {\Bbb
Z}^d} (\exp\{ -f(x/t^{1/4})\}-1)\Bigr\}
$$
for large $t$. (Since $f(\cdot)$ has compact support, the values of
$m^i(x)$, for $x\not\in D_{s^\gamma}$, do not matter.) Substituting in for
$Z_s$, one can check that this is at most
$$
\align
\exp\Bigl\{ (1-C_{46}s^{-\alpha /2+\beta})&[(\xi_0*N_s)(0) -
s^{-(d+\epsilon /2)/4}]^+ \\
&\times\sum_{x\in {\Bbb Z}^d} (\exp\{
-f(x/t^{1/4})\}-1)\Bigr\} ,\tag9.16
\endalign
$$
for appropriate $C_{46}$.

Since $f$ is continuous and has compact support,
$$
t^{-d/4}\sum_{x\in {\Bbb Z}^d} (\exp\{-f(x/t^{1/4})\}-1)\to \int_{{\Bbb R}^d}
(e^{-f(x)}-1)dx\quad \text{as}\quad t\to\infty .
\leqno(9.17)
$$
One has $\beta < \alpha /2$ and $\epsilon > 0$, and so one can use this to
write (9.16) as
$$
\exp\Bigl\{ [c_{1,s}t^{d/4}(\xi_0*N_s)(0) +c_{2,s}]^+\int_{{\Bbb R}^d}
(e^{-f(x)}-1)dx\Bigr\} ,
\leqno(9.18)
$$
where $c_{1,s}\to 1$ and $c_{2,s}\to 0$ as $s\to\infty$.  Since $f(x)\geq
0$ for all $x$, (9.18) is asymptotically equivalent to the expression
obtained by dropping the terms $c_{1,s}$ and $c_{2,s}$.  So, combining
(9.13)-(9.18), one sees that
$$
\align
\Bigl[ E\Bigl[ \exp\Bigl\{ &-\sum_x f(x/t^{1/4}) _s\tilde\eta^B_t
(x)\Bigr\} \mid {\Cal F}_s\Bigr]\Bigr.\\
& - \Bigl.\exp\Bigl\{ t^{d/4}(\xi_0*N_s)(0)^+\int_{{\Bbb
R}^d}(e^{-f(x)}-1)dx\Bigr\}\Bigr]^+\\
&\to 0\quad \text{in probability as}\ t\to\infty.\tag9.19
\endalign
$$
Taking the conditional expectation of the left side of (9.19), with respect
to ${\Cal F}_0$, produces the same expression as in (9.19), but with ${\Cal
F}_0$ replacing ${\Cal F}_s$.  Moreover, it follows from Lemma 7.1, that
$$
P(\vert (\xi_0*N_t)(0)^+ -(\xi_0*N_s)(0)^+\vert\geq
t^{-(d+1-\alpha)/4})\leq 2\exp\{ -C_{33}t^{(1-\alpha)/2}\}
\leqno(9.20)
$$
for large $t$; note that $\alpha < 1$.  Together, (9.19)-(9.20) imply that
$$
\align
\Bigl[ E\Bigl[ \exp\Bigl\{ &-\sum_x f(x/t^{1/4}) _s\tilde\eta^B_t
(x)\Bigr\} \mid {\Cal F}_0\Bigr]\Bigr.\\
&-\exp\Bigl.\Bigl\{ t^{d/4}(\xi_0*N_t)(0)^+\int_{{\Bbb R}^d}
(e^{-f(x)}-1)dx\Bigr\} \Bigr]^+\\
&\to 0\quad \text{in probability as}\ t\to\infty.\tag9.21
\endalign
$$
This is the desired upper bound for the left side of (9.11).

We still need to show the analog of (9.21), but with $[\ \cdot\ ]^-$ replacing
$[\ \cdot\ ]^+$ on the left side of (9.21).  The argument for this direction is
essentially the same as before.  We define $M^i(x)$ as in (9.4), and set
$Z'_s = [\lfloor s^\beta\rfloor^d (\xi_0*N_s)(0) + s^{d\beta -(d+\epsilon
/2)/4}]^+$.
Reasoning as in (9.14) through the first line of (9.15), one obtains that,
on $H_s$,
$$
\align
&E\Bigl[ \exp\Bigl\{ -\sum_{x\in {\Bbb Z}^d} f(x/t^{1/4})
_s\tilde\eta^B_t (x)\Bigr\} \mid {\Cal F}_s\Bigr] \\
&\geq\prod_{i\in I} \Bigl[ 1+\sum_{x\in {\Bbb Z}^d}
(\exp\{-f(x/t^{1/4})\}-1) M^i(x)\Bigr]^{Z'_s}.\tag9.22
\endalign
$$
Note that the process $_s\tilde\eta_t$, rather than $_s\eta_t$, is needed
for (9.22), because the above product is restricted to $i\in I$.
Since $f$ has compact support, one can check, using a standard 
version of the local central limit theorem, that
$$
\sum_{x\in {\Bbb Z}^d} (\exp\{ -f(x/t^{1/4})\} - 1)M^i(x)\leq
C_{47}s^{d(1-2\alpha)/4} 
$$
for large $t$ and appropriate $C_{47}$.  The right side of (9.22) is
therefore at least
$$
\exp\Bigl\{ (1+C_{47}s^{d(1-2\alpha)/4})Z'_s\sum_{x\in {\Bbb Z}^d} \Bigl[
(\exp\{-f(x/t^{1/4})\} -1)\sum_{i\in I} M^i(x)\Bigr]\Bigr\} ,
$$
which is the analog of the last line in (9.15).  Since $\alpha > 1/2$, the
term $C_{47}s^{d(1-2\alpha)/4}$ is negligible.

{}From here on, the arguments leading to (9.21) can be copied, with the upper
bound for $\sum^\infty_{i=1} M^i(x)$ in Corollary 9.1, and Lemma 7.1 being
applied.  In place of (9.21), one obtains
$$
\align
\Bigl[ E\Bigl[ \exp\Bigl\{ &-\sum_{x\in {\Bbb Z}^d} f(x/t^{1/4})
_s\tilde\eta^B_t(x)\Bigr\}\mid {\Cal F}_0\Bigr]\Bigr. \\
&-\exp\Bigl.\Bigl\{ t^{d/4}(\xi_0*N_t)(0)^+\int_{{\Bbb R}^d}
(e^{-f(x)}-1)dx\Bigr\}\Bigr]^-\\
&\to 0\quad \text{in probability as}\ t\to\infty.\tag9.23
\endalign
$$
Together, (9.21) and (9.23) imply (9.11).\hfill //
\vskip .3cm
In Proposition 9.2, we replace $_s\tilde\eta_t$, in (9.10)-(9.11), with
$\xi_t$; we also examine the joint behavior of $\xi^A_t$ and $\xi^B_t$.
In addition to Proposition 9.1, we employ Lemma 9.2, which allows
us to compare $_s\tilde\eta_t$ with $_s\eta_t$.  On account of (1.8), the
decrease in the density $\rho (t)$ is smooth, and so comparison of
$_s\eta_t$ with $\xi_t$ is also not difficult; the reasoning for this
follows [Ar81].  Together, these results will imply (9.25).
\vskip .3cm
\noindent{\bf Proposition 9.2.} {\it Assume that $d<4$.  Then, for $f =
(f_1,f_2)$, with $f_i\in C^+_c({\Bbb R}^d)$,}
$$
\align
&E\Bigl[ \exp\Bigl\{ -\sum_{x\in {\Bbb Z}^d} (f_1(x/t^{1/4})\xi^A_t(x) +
f_2(x/t^{1/4})\xi^B_t(x))\Bigr\}\mid {\Cal F}_0\Bigr] \\
&-\exp\Bigl\{ t^{d/4}\Bigl[ (\xi_0*N_t)(0)^- \int_{{\Bbb R}^d}
(e^{-f_1(x)}-1)dx + (\xi_0*N_t)(0)^+\Bigr.\Bigr.\\
&\Bigl.\Bigl.\times\int_{{\Bbb R}^d} (e^{-f_2(x)}-1)dx\Bigr]\Bigr\}
\to 0\quad \text{in probability as}\ t\to\infty.\tag9.25
\endalign
$$
\vskip .3cm
\noindent{\it Proof.} We first compare $_s\tilde\eta_t$ and $\xi_t$.
Recall that $t=s+s^\alpha$; as in Proposition 9.1, we set $\alpha =
1-10^{-5}$.  It therefore follows from (1.8), that
$$
t^{d/4}(\rho (s)-\rho (t))\leq C_{48}t^{-10^{-5}}
\leqno(9.26)
$$
for large $t$ and appropriate $C_{48}$.  Consequently, for given $M>0$,
$$
E[_s\eta^B_t (D_{Mt^{1/4}})] - E[\xi^B_t(D_{Mt^{1/4}})]\leq
2C_{48}M^dt^{-10^{-5}} .
\leqno(9.27)
$$
The particles of $\xi_t$ form a subset of those of $_s\eta_t$.  Therefore,
by (9.27) and Markov's inequality,
$$
P(_s\eta^B_t(x)\ne \xi^B_t(x)\quad \text{for some}\ x\in D_{Mt^{1/4}})\to
0\quad \text{as}\ t\to\infty .
\leqno(9.28)
$$
The analogous limit holds for $A$ particles as well.  It follows from this and
Lemma 9.2, that
$$
\align
P(_s\tilde\eta^A_t(x) &\ne \xi^A_t(x)\quad \text{for some}\ x\in
D_{Mt^{1/4}})\to 0\quad \text{as}\ t\to\infty ,\\
P(_s\tilde\eta^B_t(x) &\ne \xi^B_t(x)\quad \text{for some}\ x\in
D_{Mt^{1/4}})\to 0\quad \text{as}\ t\to\infty.\tag9.29
\endalign
$$

We now derive (9.25) from (9.10)-(9.11).  Let $G_t$ denote the event where
$(\xi_0*N_t)(0)\geq 0$.  It follows from (9.10)-(9.11) that, for $f_i\in
C^+_c({\Bbb R}^d)$,
$$
\align
1_{G_t}\sum_{x\in {\Bbb Z}^d} f_1(x/t^{1/4})\
_s\tilde\eta^A_t(x)\to 0\quad &\text{in probability as}\ t\to\infty,\\
1_{G^c_t}\sum_{x\in {\Bbb Z}^d} f_2(x/t^{1/4})\ _s\tilde\eta^B_t(x)\to 0\quad
&\text{in probability as}\ t\to\infty,\tag9.30
\endalign
$$
where $1_G$ denotes the indicator function of the event $G$.
Consequently, by (9.29),
$$
\align
1_{G_t}\sum_{x\in {\Bbb Z}^d} f_1(x/t^{1/4})\xi^A_t(x)\to 0\quad
&\text{in probability as}\ t\to\infty,\\
1_{G^c_t}\sum_{x\in {\Bbb Z}^d} f_2(x/t^{1/4})\xi^B_t(x)\to 0\quad
&\text{in probability as}\ t\to\infty.\tag9.31
\endalign
$$
So, in order to demonstrate (9.25), it suffices to show the analogous
limit,
$$
\align
&E\bigl[ \exp\bigl\{ -1_{G^c_t}\sum_{x\in {\Bbb Z}^d}
f_1(x/t^{1/4})\xi^A_t(x)-1_{G_t}\sum_{x\in {\Bbb
Z}^d}f_2(x/t^{1/4})\xi^B_t(x)\bigr\}\mid {\Cal F}_0\bigr] \\
&-\exp\bigl\{ t^{d/4}\bigl[ (\xi_0*N_t)(0)^-\int_{{\Bbb R}^d}
(e^{-f_1(x)}-1)dx + 
(\xi_0*N_t)(0)^+\bigr.\bigr. \\
&\bigl.\bigl.\times\int_{{\Bbb R}^d} (e^{-f_2(x)}-1)dx\bigr]\bigr\}\to 0
\quad \text{in probability as}\ t\to\infty.\tag9.32
\endalign
$$
On $G_t$, the left side of (9.32) reduces to the left side of (9.11), if
$f_2$ is replaced by $f$ and $\xi^B_t(x)$ by $_s\tilde\eta^B_t(x)$;
similarly, on $G^c_t$, the left side of (9.32) reduces to the left
side of (9.10).  So, (9.32) follows from (9.10)-(9.11) and (9.29).
This demonstrates the proposition.\hfill //
\vskip .3cm
We now demonstrate Theorem 2. We know from Corollary 8.1, that
$$
t^{d/4}(\xi_0*N_t)(0) \Rightarrow b_dZ_{0,1}\quad \text{as}\ t\to\infty ,
\leqno(9.33)
$$
where $Z_{0,1}$ has a standard normal distribution, and $b_d =
(2\lambda)^{1/2}(4\pi)^{-d/4}$.  Taking expectations in (9.25), and
substituting in (9.33) implies that
$$
\align
E &\Bigl[ \exp\Bigl\{ -\sum_{x\in {\Bbb Z}^d} (f_1(x/t^{1/4})\xi^A_t(x) +
f_2(x/t^{1/4})\xi^B_t(x))\Bigr\}\Bigr] \\
&\to E\Bigl[ \exp\Bigl\{ b_d\Bigl( Z^-_{0,1} \int_{{\Bbb R}^d}
(e^{-f_1(x)}-1)dx + Z^+_{0,1} 
\int_{{\Bbb R}^d} (e^{-f_2(x)}-1)dx\Bigr)\Bigr\}\Bigr]\tag9.34
\endalign
$$
as $t\to\infty$, for $f_i\in C^+_c({\Bbb R}^d)$.  One can rescale $\xi_t$
as in (1.10), setting $\check \xi_t(E) =\xi_t(t^{1/4}E)$.  One can also
rewrite the left side of (9.34), viewing $\check \xi^A_t$ and
$\check \xi^B_t$ as random measures on ${\Bbb R}^d$.  Doing this, one
can rephrase 
(9.34) as
$$
\align
E &\Bigl[ \exp\Bigl\{-\int_{{\Bbb R}^d} f_1(x)\check \xi^A_t(dx) -\int_{{\Bbb
R}^d} f_2(x)\check \xi^B_t(dx)\Bigr\}\Bigr] \\
&\to E\Bigl[ \exp\Bigl\{ b_d\Bigl( Z^-_{0,1} \int_{{\Bbb R}^d}
(e^{-f_1(x)} -1)dx +Z^+_{0,1} 
\int_{{\Bbb R}^d} (e^{-f_2(x)}-1)dx\Bigr)\Bigr\}\Bigr]\tag9.35
\endalign
$$
as $t\to\infty$.  The right side of (9.35) is the Laplace functional of a
convex combination of Poisson random fields with two types of particles,
where the intensities are given by $b_dZ^-_{0,1}$ and $b_dZ^+_{0,1}$.
Letting $F$ denote the distribution function of $b_dZ_{0,1}$, we can write
this random field as ${\Cal P}_F$, as in (1.11).  It
follows from (9.35), that the pair $(\check\xi^A_t,\check\xi^B_t)$ converges
weakly to ${\Cal P}_F$, on the Borel measures on ${\Bbb R}^d$ with
finite mass on compact subsets.  That is,
$$
(\check\xi^A_t,\check\xi^B_t)\Rightarrow {\Cal P}_F\quad \text{as}\
t\to\infty .
\leqno(9.36)
$$
The limit in (9.36) is the same as that in (1.12).  This completes the proof
of Theorem 2.
\vskip .3cm
\noindent {\bf REFERENCES}
\vskip .3cm
\noindent{[Ar79]} Arratia, R. (1979). {\it Coalescing Brownian Motions on
the Line}. Ph.D. thesis, University of Wisconsin at Madison.
\vskip .3cm
\noindent [Ar81] Arratia, R. (1981).  Limiting point processes for
rescalings of coalescing and annihilating random walks on ${\Bbb Z}^d$.
{\it Ann. Probab.} $\bold{9}$ 909-936.
\vskip .3cm
\noindent [BhRa86] Bhattacharya, R.N. and Rao, R.R. (1986).
{\it Normal Approximation and Asymptotic Expansions}. Krieger, Malabar.
\vskip .3cm
\noindent [Bi68] Billingsley, P. (1968).  {\it Convergence of Probability
Measures}. Wiley, New York.
\vskip .3cm
\noindent [Bi71]  Billingsley, P. (1971).
{\it Weak Convergence of Measures: Applications in Probability}.
Regional Conference Series in Applied Mathematics. SIAM, Philadelpha.
\vskip .3cm
\noindent [BrGr80] Bramson, M. and Griffeath, D. (1980).  Asymptotics for
interacting particle systems on ${\Bbb Z}^d$. {\it Z. Wahrsch. verw. Geb.}
$\bold{53}$ 183-196.
\vskip .3cm
\noindent [BrLe91a] Bramson, M. and Lebowitz, J.L. (1991).
Asymptotic behavior of densities for two-particle annihilating
random walks. {\it J. Stat. Phys.} $\bold{62}$ 297-372.
\vskip .3cm
\noindent [BrLe91b] Bramson, M. and Lebowitz, J.L. (1991).
Spatial structure in diffusion limited two-particle reactions.
{\it J. Stat. Phys.} $\bold{65}$ 941-952.
\vskip .3cm
\noindent [BrLe99]  Bramson, M. and Lebowitz, J.L. (1999).
Spatial structure in high dimensions for diffusion limited
two-particle reactions.  In preparation.
\vskip .3cm
\noindent [Ch74]  Chung, K.L. (1974).  {\it A Course in Probability Theory}.
Academic Press, New York.
\vskip .3cm
\noindent [KaRe85] Kang, K. and Redner, S. (1985).
Fluctuation-dominated kinetics in diffusion-controlled reactions.
{\it Phys. Rev. A} $\bold{32}$ 435-447.
\vskip .3cm
\noindent [KeVa98] Kesten, H. and van den Berg, J. (1998).  Asymptotic
density in a coalescing random walk model.  Preprint.
\vskip .3cm
\noindent [Ku73] Kuelbs, J. (1973).
The invariance principle for Banach space valued random variables.
{\it J. Mult. Analysis} $\bold{3}$ 161-172.
\vskip .3cm
\noindent [LeCa95] Lee, B.P. and Cardy, J. (1995). Renormalization group
study of the $A+B\to\emptyset$ diffusion-limited reaction. {\it J. Stat.
Phys.} $\bold{80}$ 971-1007.
\vskip .3cm
\noindent [LeCa97] Lee, B.P. and Cardy, J. (1997). Erratum: Renormalization
group study of the $A+B\to\emptyset$ diffusion-limited reaction. {\it J.
Stat. Phys.} $\bold{87}$ 951-954.
\vskip .3cm
\noindent [OrPr73] Orey, S. and Pruitt, W. (1973).
Sample functions of the N-parameter Wiener process.
{\it Ann. Probab.} $\bold{1}$ 138-163.
\vskip .3cm
\noindent [OvZe83] Ovchinnikov, A.A. and Zeldovich, Ya.B. (1978).
Role of density fluctuations in bimolecular reaction kinetics.
{\it Chem. Phys.} $\bold{28}$ 215-218.
\vskip .3cm
\noindent [Pe75] Petrov, V.V. (1975). {\it Sums of Independent Random
Variables}.  Springer, New York.
\vskip .3cm
\noindent [ToWi83] Toussaint, D. and Wilczek, F. (1983).
Particle-antiparticle annihilation in diffusive motion.
{\it J. Chem. Phys.} $\bold{78}$ 2642-2647.
\vskip .5truein
\noindent MAURY BRAMSON

\noindent SCHOOL OF MATHEMATICS

\noindent UNIVERSITY OF MINNESOTA

\noindent MINNEAPOLIS, MINNESOTA 55455

\noindent E-MAIL - bramson\@math.umn.edu

\vskip .3in

\noindent JOEL L. LEBOWITZ

\noindent DEPARTMENTS OF MATHEMATICS AND PHYSICS

\noindent RUTGERS UNIVERSITY

\noindent NEW BRUNSWICK, NEW JERSEY  08903

\noindent E-MAIL - lebowitz\@sakharov.rutgers.edu

\end